\documentclass[12pt]{article}
\usepackage{fullpage}
\usepackage[compress]{cite}
\usepackage[centertags]{amsmath}
\numberwithin{equation}{section}
\usepackage{color}
\usepackage{amsbsy}
\usepackage{amsfonts}
\usepackage{amssymb}
\usepackage{graphicx}
\usepackage{bm}
\usepackage{url}
\usepackage{hyperref}
\usepackage[ddmmyy,24hr]{datetime}
\usepackage{subfigure}
\usepackage{multirow}
\newcommand{\nua}[1]{\ensuremath{\rlap
 {\kern-2.5pt\ensuremath
 {\overset{\scriptscriptstyle(-)}{\phantom{\nu}}}}
 {\ensuremath{{\nu}_{#1}}}}}
\newcommand{\re}[1]{\ensuremath{\Re\mathfrak{e}\!\left[#1\right]}}
\newcommand{\im}[1]{\ensuremath{\Im\mathfrak{m}\!\left[#1\right]}}

\RequirePackage{ifthen}

\newcommand{\nHe}[1]{\ensuremath{\ifthenelse{#1>9} {{}^{#1}_{\;\:2}} {{}^{#1}_2} \textrm{He}}}
\newcommand{\nLi}[1]{\ensuremath{\ifthenelse{#1>9} {{}^{#1}_{\;\:3}} {{}^{#1}_3} \textrm{Li}}}
\newcommand{\nBe}[1]{\ensuremath{\ifthenelse{#1>9} {{}^{#1}_{\;\:4}} {{}^{#1}_4} \textrm{Be}}}
\newcommand{\nB} [1]{\ensuremath{\ifthenelse{#1>9} {{}^{#1}_{\;\:5}} {{}^{#1}_5} \textrm{B}}}
\newcommand{\nC} [1]{\ensuremath{\ifthenelse{#1>9} {{}^{#1}_{\;\:6}} {{}^{#1}_6} \textrm{C}}}
\newcommand{\nN} [1]{\ensuremath{\ifthenelse{#1>9} {{}^{#1}_{\;\:7}} {{}^{#1}_7} \textrm{N}}}
\newcommand{\nO} [1]{\ensuremath{\ifthenelse{#1>9} {{}^{#1}_{\;\:8}} {{}^{#1}_8} \textrm{O}}}
\newcommand{\nF} [1]{\ensuremath{\ifthenelse{#1>9} {{}^{#1}_{\;\:9}} {{}^{#1}_9} \textrm{F}}}

\newcommand{\nCa}[1]{\ensuremath{{}^{#1}_{20} \textrm{Ca}}}

\newcommand{\nTi}[1]{\ensuremath{{}^{#1}_{22} \textrm{Ti}}}

\newcommand{\nZn}[1]{\ensuremath{{}^{#1}_{30} \textrm{Zn}}}

\newcommand{\nGe}[1]{\ensuremath{{}^{#1}_{32} \textrm{Ge}}}
\newcommand{\nAs}[1]{\ensuremath{{}^{#1}_{33} \textrm{As}}}
\newcommand{\nSe}[1]{\ensuremath{{}^{#1}_{34} \textrm{Se}}}

\newcommand{\nKr}[1]{\ensuremath{\ifthenelse{#1>99} {{}^{#1}_{\;\:36}} {{}^{#1}_{36}} \textrm{Kr}}}
\newcommand{\nRb}[1]{\ensuremath{\ifthenelse{#1>99} {{}^{#1}_{\;\:37}} {{}^{#1}_{37}} \textrm{Rb}}}
\newcommand{\nSr}[1]{\ensuremath{\ifthenelse{#1>99} {{}^{#1}_{\;\:38}} {{}^{#1}_{38}} \textrm{Sr}}}
\newcommand{\nY} [1]{\ensuremath{\ifthenelse{#1>99} {{}^{#1}_{\;\:39}} {{}^{#1}_{39}} \textrm{Y}}}
\newcommand{\nZr}[1]{\ensuremath{\ifthenelse{#1>99} {{}^{#1}_{\;\:40}} {{}^{#1}_{40}} \textrm{Zr}}}
\newcommand{\nNb}[1]{\ensuremath{\ifthenelse{#1>99} {{}^{#1}_{\;\:41}} {{}^{#1}_{41}} \textrm{Nb}}}
\newcommand{\nMo}[1]{\ensuremath{\ifthenelse{#1>99} {{}^{#1}_{\;\:42}} {{}^{#1}_{42}} \textrm{Mo}}}
\newcommand{\nTc}[1]{\ensuremath{\ifthenelse{#1>99} {{}^{#1}_{\;\:43}} {{}^{#1}_{43}} \textrm{Tc}}}
\newcommand{\nRu}[1]{\ensuremath{\ifthenelse{#1>99} {{}^{#1}_{\;\:44}} {{}^{#1}_{44}} \textrm{Ru}}}
\newcommand{\nRh}[1]{\ensuremath{\ifthenelse{#1>99} {{}^{#1}_{\;\:45}} {{}^{#1}_{45}} \textrm{Rh}}}
\newcommand{\nPd}[1]{\ensuremath{\ifthenelse{#1>99} {{}^{#1}_{\;\:46}} {{}^{#1}_{46}} \textrm{Pd}}}
\newcommand{\nAg}[1]{\ensuremath{\ifthenelse{#1>99} {{}^{#1}_{\;\:47}} {{}^{#1}_{47}} \textrm{Ag}}}
\newcommand{\nCd}[1]{\ensuremath{\ifthenelse{#1>99} {{}^{#1}_{\;\:48}} {{}^{#1}_{48}} \textrm{Cd}}}
\newcommand{\nIn}[1]{\ensuremath{\ifthenelse{#1>99} {{}^{#1}_{\;\:49}} {{}^{#1}_{49}} \textrm{In}}}
\newcommand{\nSn}[1]{\ensuremath{\ifthenelse{#1>99} {{}^{#1}_{\;\:50}} {{}^{#1}_{50}} \textrm{Sn}}}
\newcommand{\nSb}[1]{\ensuremath{\ifthenelse{#1>99} {{}^{#1}_{\;\:51}} {{}^{#1}_{51}} \textrm{Sb}}}
\newcommand{\nTe}[1]{\ensuremath{\ifthenelse{#1>99} {{}^{#1}_{\;\:52}} {{}^{#1}_{52}} \textrm{Te}}}
\newcommand{\nI} [1]{\ensuremath{\ifthenelse{#1>99} {{}^{#1}_{\;\:53}} {{}^{#1}_{53}} \textrm{I}}}
\newcommand{\nXe}[1]{\ensuremath{\ifthenelse{#1>99} {{}^{#1}_{\;\:54}} {{}^{#1}_{54}} \textrm{Xe}}}
\newcommand{\nCs}[1]{\ensuremath{\ifthenelse{#1>99} {{}^{#1}_{\;\:55}} {{}^{#1}_{55}} \textrm{Cs}}}
\newcommand{\nBa}[1]{\ensuremath{\ifthenelse{#1>99} {{}^{#1}_{\;\:56}} {{}^{#1}_{56}} \textrm{Ba}}}
\newcommand{\nLa}[1]{\ensuremath{\ifthenelse{#1>99} {{}^{#1}_{\;\:57}} {{}^{#1}_{57}} \textrm{La}}}
\newcommand{\nCe}[1]{\ensuremath{\ifthenelse{#1>99} {{}^{#1}_{\;\:58}} {{}^{#1}_{58}} \textrm{Ce}}}
\newcommand{\nPr}[1]{\ensuremath{\ifthenelse{#1>99} {{}^{#1}_{\;\:59}} {{}^{#1}_{59}} \textrm{Pr}}}
\newcommand{\nNd}[1]{\ensuremath{\ifthenelse{#1>99} {{}^{#1}_{\;\:60}} {{}^{#1}_{60}} \textrm{Nd}}}
\newcommand{\nPm}[1]{\ensuremath{\ifthenelse{#1>99} {{}^{#1}_{\;\:61}} {{}^{#1}_{61}} \textrm{Pm}}}
\newcommand{\nSm}[1]{\ensuremath{\ifthenelse{#1>99} {{}^{#1}_{\;\:62}} {{}^{#1}_{62}} \textrm{Sm}}}
\newcommand{\nEu}[1]{\ensuremath{\ifthenelse{#1>99} {{}^{#1}_{\;\:63}} {{}^{#1}_{63}} \textrm{Eu}}}
\newcommand{\nGd}[1]{\ensuremath{\ifthenelse{#1>99} {{}^{#1}_{\;\:64}} {{}^{#1}_{64}} \textrm{Gd}}}
\newcommand{\nTb}[1]{\ensuremath{\ifthenelse{#1>99} {{}^{#1}_{\;\:65}} {{}^{#1}_{65}} \textrm{Tb}}}
\newcommand{\nDy}[1]{\ensuremath{\ifthenelse{#1>99} {{}^{#1}_{\;\:66}} {{}^{#1}_{66}} \textrm{Dy}}}
\newcommand{\nHo}[1]{\ensuremath{\ifthenelse{#1>99} {{}^{#1}_{\;\:67}} {{}^{#1}_{67}} \textrm{Ho}}}
\newcommand{\nEr}[1]{\ensuremath{\ifthenelse{#1>99} {{}^{#1}_{\;\:68}} {{}^{#1}_{68}} \textrm{Er}}}
\newcommand{\nTm}[1]{\ensuremath{\ifthenelse{#1>99} {{}^{#1}_{\;\:69}} {{}^{#1}_{69}} \textrm{Tm}}}
\newcommand{\nYb}[1]{\ensuremath{\ifthenelse{#1>99} {{}^{#1}_{\;\:70}} {{}^{#1}_{70}} \textrm{Yb}}}
\newcommand{\nLu}[1]{\ensuremath{\ifthenelse{#1>99} {{}^{#1}_{\;\:71}} {{}^{#1}_{71}} \textrm{Lu}}}
\newcommand{\nHf}[1]{\ensuremath{\ifthenelse{#1>99} {{}^{#1}_{\;\:72}} {{}^{#1}_{72}} \textrm{Hf}}}
\newcommand{\nTa}[1]{\ensuremath{\ifthenelse{#1>99} {{}^{#1}_{\;\:73}} {{}^{#1}_{73}} \textrm{Ta}}}
\newcommand{\nW} [1]{\ensuremath{\ifthenelse{#1>99} {{}^{#1}_{\;\:74}} {{}^{#1}_{74}} \textrm{W}}}
\newcommand{\nRe}[1]{\ensuremath{\ifthenelse{#1>99} {{}^{#1}_{\;\:75}} {{}^{#1}_{75}} \textrm{Re}}}
\newcommand{\nOs}[1]{\ensuremath{\ifthenelse{#1>99} {{}^{#1}_{\;\:76}} {{}^{#1}_{76}} \textrm{Os}}}
\newcommand{\nIr}[1]{\ensuremath{\ifthenelse{#1>99} {{}^{#1}_{\;\:77}} {{}^{#1}_{77}} \textrm{Ir}}}
\newcommand{\nPt}[1]{\ensuremath{\ifthenelse{#1>99} {{}^{#1}_{\;\:78}} {{}^{#1}_{78}} \textrm{Pt}}}
\newcommand{\nAu}[1]{\ensuremath{\ifthenelse{#1>99} {{}^{#1}_{\;\:79}} {{}^{#1}_{79}} \textrm{Au}}}
\newcommand{\nHg}[1]{\ensuremath{\ifthenelse{#1>99} {{}^{#1}_{\;\:80}} {{}^{#1}_{80}} \textrm{Hg}}}
\newcommand{\nTl}[1]{\ensuremath{\ifthenelse{#1>99} {{}^{#1}_{\;\:81}} {{}^{#1}_{81}} \textrm{Tl}}}
\newcommand{\nPb}[1]{\ensuremath{\ifthenelse{#1>99} {{}^{#1}_{\;\:82}} {{}^{#1}_{82}} \textrm{Pb}}}
\newcommand{\nBi}[1]{\ensuremath{\ifthenelse{#1>99} {{}^{#1}_{\;\:83}} {{}^{#1}_{83}} \textrm{Bi}}}
\newcommand{\nPo}[1]{\ensuremath{\ifthenelse{#1>99} {{}^{#1}_{\;\:84}} {{}^{#1}_{84}} \textrm{Po}}}
\newcommand{\nAt}[1]{\ensuremath{\ifthenelse{#1>99} {{}^{#1}_{\;\:85}} {{}^{#1}_{85}} \textrm{At}}}
\newcommand{\nRn}[1]{\ensuremath{\ifthenelse{#1>99} {{}^{#1}_{\;\:86}} {{}^{#1}_{86}} \textrm{Rn}}}
\newcommand{\nFr}[1]{\ensuremath{\ifthenelse{#1>99} {{}^{#1}_{\;\:87}} {{}^{#1}_{87}} \textrm{Fr}}}
\newcommand{\nRa}[1]{\ensuremath{\ifthenelse{#1>99} {{}^{#1}_{\;\:88}} {{}^{#1}_{88}} \textrm{Ra}}}
\newcommand{\nAc}[1]{\ensuremath{\ifthenelse{#1>99} {{}^{#1}_{\;\:89}} {{}^{#1}_{89}} \textrm{Ac}}}
\newcommand{\nTh}[1]{\ensuremath{\ifthenelse{#1>99} {{}^{#1}_{\;\:90}} {{}^{#1}_{90}} \textrm{Th}}}
\newcommand{\nPa}[1]{\ensuremath{\ifthenelse{#1>99} {{}^{#1}_{\;\:91}} {{}^{#1}_{91}} \textrm{Pa}}}
\newcommand{\nU} [1]{\ensuremath{\ifthenelse{#1>99} {{}^{#1}_{\;\:92}} {{}^{#1}_{92}} \textrm{U}}}
\newcommand{\nNp}[1]{\ensuremath{\ifthenelse{#1>99} {{}^{#1}_{\;\:93}} {{}^{#1}_{93}} \textrm{Np}}}
\newcommand{\nPu}[1]{\ensuremath{\ifthenelse{#1>99} {{}^{#1}_{\;\:94}} {{}^{#1}_{94}} \textrm{Pu}}}
\newcommand{\nAm}[1]{\ensuremath{\ifthenelse{#1>99} {{}^{#1}_{\;\:95}} {{}^{#1}_{95}} \textrm{Am}}}
\newcommand{\nCm}[1]{\ensuremath{\ifthenelse{#1>99} {{}^{#1}_{\;\:96}} {{}^{#1}_{96}} \textrm{Cm}}}
\newcommand{\nBk}[1]{\ensuremath{\ifthenelse{#1>99} {{}^{#1}_{\;\:97}} {{}^{#1}_{97}} \textrm{Bk}}}
\newcommand{\nCf}[1]{\ensuremath{\ifthenelse{#1>99} {{}^{#1}_{\;\:98}} {{}^{#1}_{98}} \textrm{Cf}}}
\newcommand{\nEs}[1]{\ensuremath{\ifthenelse{#1>99} {{}^{#1}_{\;\:99}} {{}^{#1}_{99}} \textrm{Es}}}

\begin{document}

\title{\textbf{Neutrinoless Double-Beta Decay: a Probe of Physics Beyond the Standard Model}}

\author{
\textbf{S.M. Bilenky}\\
{\small\textit{Joint Institute for Nuclear Research, Dubna, R-141980, Russia,}}\\
{\small\textit{and}}\\
{\small\textit{TRIUMF, 4004 Wesbrook Mall, Vancouver, BC V6T 2A3, Canada}}
\and
\textbf{C. Giunti}\\
{\small\textit{INFN, Sezione di Torino, Via P. Giuria 1, I--10125 Torino, Italy}}
}

\date{\textit{International Journal of Modern Physics A} \textbf{30}, 1530001 (2015)}

\maketitle

\begin{abstract}
In the Standard Model
the total lepton number is conserved.
Thus,
neutrinoless double-$\beta$ decay,
in which the total lepton number is violated by two units,
is a probe of physics beyond the Standard Model.
In this review we consider the basic mechanism of neutrinoless double-$\beta$ decay
induced by light Majorana neutrino masses.
After a brief summary of the present status of our knowledge of
neutrino masses and mixing
and
an introduction to the seesaw mechanism for the generation of
light Majorana neutrino masses,
in this review we discuss
the theory and phenomenology of neutrinoless double-$\beta$ decay.
We present the basic elements of the theory of
neutrinoless double-$\beta$ decay,
our view of the present status of the challenging problem of the calculation of the nuclear matrix element of the process
and a summary of the experimental results.
\end{abstract}


\tableofcontents


\section{Introduction}
\label{bb1}

The experimental evidence of neutrino oscillations
is one of the most important recent discoveries in particle physics.
Model-independent evidences of neutrino oscillations
have been obtained
in 1998 by the
atmospheric neutrino experiment Super-Kamiokande
\cite{Fukuda:1998mi},
in 2002 by the
solar neutrino experiment SNO
\cite{Ahmad:2002jz},
and in 2004 by the
reactor neutrino experiment KamLAND
\cite{Araki:2004mb}.

The existence of neutrino oscillations implies that neutrinos are massive particles
and that the three flavor neutrinos
$\nu_{e}$,
$\nu_{\mu}$,
$\nu_{\tau}$
are mixtures of neutrinos with definite masses
$\nu_{i}$
(with $i=1,2,\ldots$).
The phenomenon of neutrino oscillations was studied in several
atmospheric
(Kamiokande \cite{Fukuda:1994mc},
IMB \cite{Becker-Szendy:1992hq},
Super-Kamiokande \cite{Fukuda:1998mi},
Soudan-2 \cite{Sanchez:2003rb},
MACRO \cite{Ambrosio:2003yz},
MINOS \cite{Adamson:2012gt},
ANTARES \cite{AdrianMartinez:2012ph},
IceCube \cite{Aartsen:2013jza}),
solar
(Homestake \cite{Cleveland:1998nv},
GALLEX/GNO \cite{Altmann:2005ix},
SAGE \cite{Abdurashitov:2002nt},
Super-Kamiokande \cite{Renshaw:2013dzu},
SNO \cite{Aharmim:2011vm},
Borexino \cite{Bellini:2014uqa},
KamLAND \cite{Gando:2014wjd}),
reactor
(KamLAND \cite{Gando:2013nba},
Daya Bay \cite{An:2013zwz},
Double Chooz \cite{Abe:2013sxa},
RENO \cite{Ahn:2012nd})
and accelerator
(K2K \cite{Ahn:2006zza},
MINOS \cite{Adamson:2013whj,Adamson:2013ue},
T2K \cite{Abe:2013hdq,Abe:2014ugx})
experiments.
These experiments fully confirmed the existence of neutrino oscillations
in different channels
($\nua{e}\to\nua{e}$,
$\nu_{e}\to\nu_{\mu,\tau}$,
$\nua{\mu}\to\nua{\mu}$,
$\nu_{\mu}\to\nu_{e}$,
$\nu_{\mu}\to\nu_{\tau}$).

Now that we know that neutrinos are massive,
one of the most fundamental open problems which must be investigated by experiments
is the determination of the nature of neutrinos with definite mass:
are they four-component Dirac particles
possessing a conserved lepton number $L$
or
two-component truly neutral
(no electric charge and no lepton number)
Majorana particles?

Neutrino oscillation experiments do not allow to
answer this fundamental question,
because in the neutrino oscillation transitions
$\nua{l}\to\nua{l'}$
the total lepton number $L$ is conserved.
Therefore,
neutrino oscillations do not depend on
the Majorana phases which enter in the neutrino mixing matrix
if massive neutrinos are Majorana particles
\cite{Bilenky:1980cx,Doi:1980yb,Langacker:1986jv,Giunti:2010ec}.

In order to probe the nature of massive neutrinos
we need to study processes in which the total lepton number $L$
is not conserved.
The highest sensitivity to small Majorana neutrino masses can be reached
in experiments on the search of the
$L$-violating neutrinoless double-$\beta$ decay process
\begin{equation}
\beta\beta_{0\nu}:
\quad
{}^{A}_{Z}\text{X} \to {}^{\phantom{+2}A}_{Z+2}\text{X} + 2 e^{-}
,
\label{bb0-}
\end{equation}
where
${}^{A}_{Z}\text{X}$
is a nucleus with atomic number $Z$ and atomic mass $A$.

\begin{figure}
\begin{center}
\begin{minipage}[r]{0.49\linewidth}
\begin{center}
\subfigure[]{
\includegraphics*[width=\linewidth]{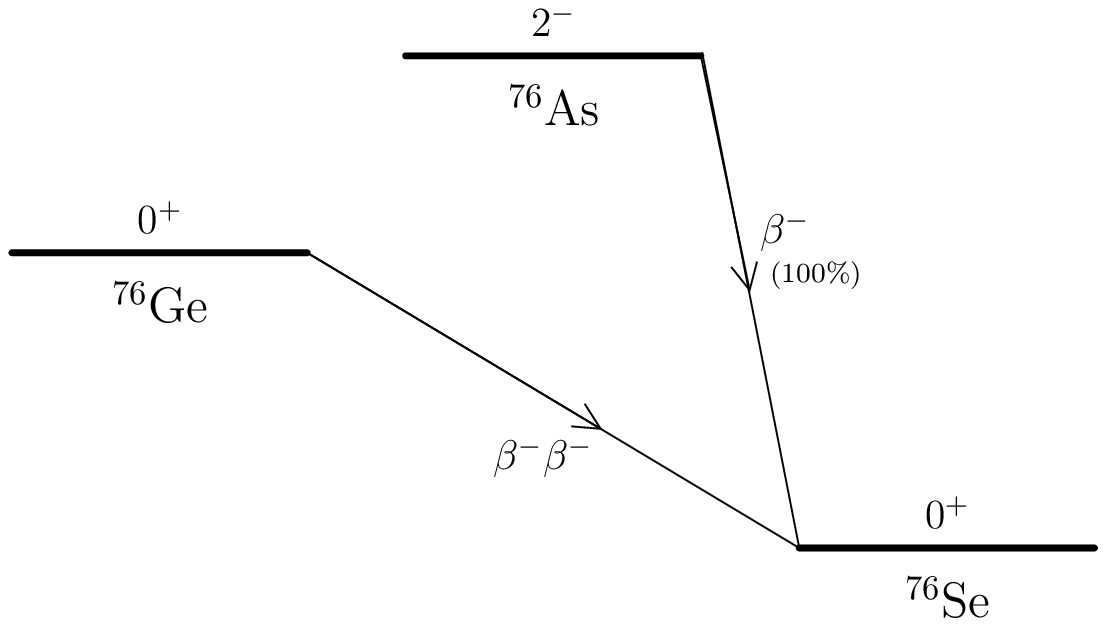}
\label{fig:ge76}
}
\end{center}
\end{minipage}
\hfill
\begin{minipage}[l]{0.49\linewidth}
\begin{center}
\subfigure[]{
\includegraphics*[width=\linewidth]{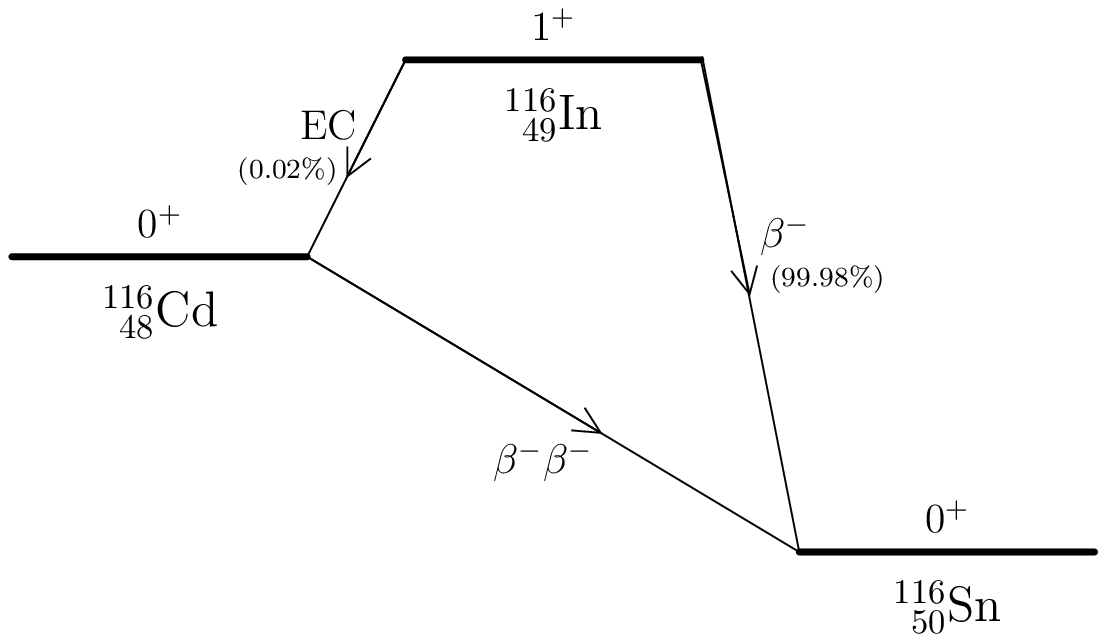}
\label{fig:cd116}
}
\end{center}
\end{minipage}
\end{center}
\caption{ \label{fig:ge76-cd116}
Schematic illustration of the ground-state energy level structure
and decays
of the
$\protect\nGe{76}$,
$\protect\nAs{76}$,
$\protect\nSe{76}$
\subref{fig:ge76}
and
$\protect\nCd{116}$,
$\protect\nIn{116}$,
$\protect\nSn{116}$
\subref{fig:cd116}
isobar nuclei.
}
\end{figure}

The two-neutrino double-$\beta$ decay process
\begin{equation}
\beta\beta_{2\nu}:
\quad
{}^{A}_{Z}\text{X} \to {}^{\phantom{+2}A}_{Z+2}\text{X} + 2 e^{-} + 2 \bar\nu_{e}
\label{bb2-}
\end{equation}
is allowed by the Standard Model
for some even-even nuclei
for which the single $\beta$ decay or electron capture is forbidden
(see the illustrations in Fig.~\ref{fig:ge76-cd116}).
They are extremely rare processes
of the second order of perturbation theory in the Fermi constant $G_{F}$.
As will be explained in Section~\ref{bb4},
the neutrinoless double-$\beta$ decay process (\ref{bb0-})
is further suppressed by the proportionality
of the transition amplitude to the effective Majorana mass
\begin{equation}\label{effMaj}
m_{\beta\beta} = \sum_{i} U^{2}_{ei} m_{i}
,
\end{equation}
where
$m_{i}$
are the small masses of the Majorana neutrinos $\nu_{i}$
and
$U_{ei}$
are the elements of the neutrino mixing matrix
(see Eq.~(\ref{mixing})).

Up to now,
$\beta\beta_{0\nu}$ decay has not been observed\footnote{
The claim of observation of
$\beta\beta_{0\nu}$ decay of $\nGe{76}$
presented in Ref.~\citen{KlapdorKleingrothaus:2004wj}
is strongly disfavored by the recent results of the GERDA experiment
\cite{Agostini:2013mzu}.
The more elaborated analysis of the same data of Ref.~\citen{KlapdorKleingrothaus:2004wj}
presented in Ref.~\citen{KlapdorKleingrothaus:2006ff}
has been severely criticized in Ref.~\citen{Schwingenheuer:2012zs}.
}.
The most sensitive experiments
obtained for $|m_{\beta\beta}|$
upper limits in the range $0.2 - 0.6 \, \text{eV}$
(depending on the theoretical calculation of the nuclear matrix elements discussed in Section~\ref{bb5}),
which correspond to $\beta\beta_{0\nu}$ decay half-lives
of the order of $10^{25} \, \text{y}$.
These impressive results have been reached
using detectors with large masses
located underground,
constructed with radiopure materials
and
with high energy resolution
in order to reach very low background levels in the energy region
of the $\beta\beta_{0\nu}$ decay signal.

The possible value of the effective Majorana mass in Eq.~(\ref{effMaj})
depends on the normal or inverted character of the neutrino mass spectrum
(see Fig.~\ref{fig:mass}).
The next generation of
$\beta\beta_{0\nu}$ decay experiments
will probe the inverted mass hierarchy region,
in which
$|m_{\beta\beta}| \sim (2 - 5) \times 10^{-2} \, \text{eV}$
(see Fig.~\ref{fig:mbb}).

In the Standard Model
the total lepton number $L$ is conserved
because
it is not possible to construct $L$-violating Lagrangian terms
with products of fields with total energy dimension smaller or equal to four.
This constraint is required by renormalizability.
However,
if the Standard Model is a low-energy effective theory,
the high-energy physics beyond the Standard Model
can generate effective low-energy non-renormalizable
Lagrangian terms with products of fields with total energy dimension larger than four
\cite{Weinberg:1979sa}.
Since the effective Lagrangian term of a dimension-$N$ field product
with $N>4$
is suppressed by a coefficient $\Lambda^{4-N}$,
where $\Lambda$ is the scale of new physics beyond the Standard Model,
in practice only the terms with field products with small dimensionalities can be observable.
It is remarkable that there is only one dimension-5 field product
that can be constructed with Standard Model fields
and this term violates the total lepton number $L$.
After spontaneous electroweak symmetry breaking,
this term generates small Majorana neutrino masses and neutrino mixing.
If this mechanism is realized in nature,
the small Majorana neutrino masses and neutrino mixing
are the most accessible low-energy phenomena generated by
the physics beyond the Standard Model.

However,
the small neutrino masses discovered in neutrino oscillation experiments
can have a Standard Model origin if
$L$ is conserved and
neutrinos are Dirac particles
with masses generated by the Standard Model Higgs mechanism
with extremely small Yukawa couplings
(smaller than about $10^{-11}$).
In this case there is no $\beta\beta_{0\nu}$ decay.

Therefore, it is very important to search for
$L$-violating processes as $\beta\beta_{0\nu}$ decay,
which can reveal the Majorana nature of massive neutrinos
and open a window on the physics beyond the Standard Model.

The $L$-violating five-dimensional field product
which generates small Majorana neutrino masses
can be induced by the exchange of heavy virtual Majorana leptons
between two lepton-Higgs vertices.
Thus,
the observation of $\beta\beta_{0\nu}$ decay
would be an indication in favor of the existence of heavy Majorana leptons.
The CP-violating decays of these particles in the early Universe
could be the origin of
the baryon asymmetry of the Universe
through the leptogenesis mechanism
(see Refs.~\citen{DiBari:2012fz,Fong:2013wr}).

In this review we discuss the theory and phenomenology of neutrinoless double-$\beta$ decay
induced by light Majorana neutrino masses
(see also the old but very instructive reviews in
Refs.~\citen{Haxton:1984am,Doi:1985dx}
and the recent reviews in
Refs.~\citen{GomezCadenas:2011it,Schwingenheuer:2012zs,Vergados:2012xy,Rodejohann:2012xd,Deppisch:2012nb,Giuliani:2012zu,Cremonesi:2013vla}).
In Section~\ref{bb2}
we briefly summarize the present status of neutrino masses and mixing.
In Section~\ref{bb3}
we discuss the effective Lagrangian approach
and the seesaw mechanism of generation of small Majorana neutrino masses
and neutrino mixing.
In Section~\ref{bb4}
we derive the rate of $\beta\beta_{0\nu}$ decay
and in Section~\ref{bb5}
we discuss the problem of the calculation of the nuclear matrix elements
of different $\beta\beta_{0\nu}$ decays.
In Section~\ref{bb6}
we discuss the phenomenological implications
of our knowledge of the neutrino oscillation parameters
for the effective Majorana mass $m_{\beta\beta}$ in $\beta\beta_{0\nu}$ decay,
with emphasis on the differences between the normal and inverted hierarchies.
In Section~\ref{bb7} we present the results of recent $\beta\beta_{0\nu}$ decay
experiments and we briefly discuss the general strategies of future projects.
Finally, we draw our conclusions in Section~\ref{bb8}.
\section{Experimental status of neutrino masses and mixing}
\label{bb2}

Atmospheric, solar, reactor and accelerator neutrino oscillation experiments
proved that neutrinos are massive and mixed particles
(see Refs.~\citen{Giunti:2007ry,Bilenky:2010zza,Xing:2011zza}).
Neutrino oscillations
\cite{Pontecorvo:1957cp,Pontecorvo:1957qd,Maki:1962mu,Pontecorvo:1968fh,Gribov:1968kq}
is a quantum-mechanical phenomenon
due to the fact the left-handed neutrino flavor fields $\nu_{lL}(x)$
($l=e,\mu,\tau$)
are superpositions of left-handed neutrino fields
$\nu_{iL}(x)$ with masses $m_{i}$
according to the mixing relation
\begin{equation}\label{mixing}
\nu_{lL}(x) = \sum_{i} U_{li} \, \nu_{iL}(x) \qquad (l=e,\mu,\tau)
.
\end{equation}
In the standard framework of three-neutrino mixing,
$i=1,2,3$
and
$U$ is
the 3$\times$3 unitary PMNS
(Pontecorvo, Maki, Nakagawa, Sakata)
mixing matrix
which has the standard parameterization
\begin{equation}
U
=
\begin{pmatrix}
c_{12}
c_{13}
&
s_{12}
c_{13}
&
s_{13}
e^{-i\delta}
\\
-
s_{12}
c_{23}
-
c_{12}
s_{23}
s_{13}
e^{i\delta}
&
c_{12}
c_{23}
-
s_{12}
s_{23}
s_{13}
e^{i\delta}
&
s_{23}
c_{13}
\\
s_{12}
s_{23}
-
c_{12}
c_{23}
s_{13}
e^{i\delta}
&
-
c_{12}
s_{23}
-
s_{12}
c_{23}
s_{13}
e^{i\delta}
&
c_{23}
c_{13}
\end{pmatrix}
\text{diag}\!\left(
e^{i\lambda_1},
e^{i\lambda_2},
1
\right)
,
\label{U}
\end{equation}
where
$ c_{ab} \equiv \cos\vartheta_{ab} $
and
$ s_{ab} \equiv \sin\vartheta_{ab} $,
with the three mixing angles
$\vartheta_{12}$,
$\vartheta_{23}$,
$\vartheta_{13}$.
The Dirac CP-violating phase $\delta$
can generate CP violation in neutrino oscillations,
whereas the Majorana CP-violating phases
$\lambda_1$ and $\lambda_2$
contribute to processes like
neutrinoless double-$\beta$ decay
in which the total lepton number is violated
(see Section~\ref{bb6}).

A flavor neutrino $\nu_{l}$,
which is produced in CC weak processes together with a charged lepton $l^{+}$
is described by the mixed state
\begin{equation}\label{state}
|\nu_{l}\rangle=\sum_{i} U^{*}_{li} \, |\nu_{i}\rangle
,
\end{equation}
where $|\nu_{i}\rangle $ is the state of a Majorana or Dirac neutrino with mass $m_{i}$.

The probability of the transition $\nu_{l}\to \nu_{l'}$ in vacuum is
given by the standard expression
(see Refs.~\citen{Giunti:2007ry,Bilenky:2010zza,Xing:2011zza})
\begin{align}
P_{\nu_{l}\to\nu_{l'}}
=
\null & \null
\delta_{ll'}
-
4
\sum_{i>k}
\re{
U_{li}^{*}
\,
U_{l'i}
\,
U_{lk}
\,
U_{l'k}^{*}
}
\,
\sin^{2}\left( \frac{\Delta{m}^{2}_{ki} L}{4E} \right)
\nonumber
\\
\null & \null
\phantom{
\delta_{ll'}
}
+
2
\sum_{i>k}
\im{
U_{li}^{*}
\,
U_{l'i}
\,
U_{lk}
\,
U_{l'k}^{*}
}
\,
\sin\!\left( \frac{\Delta{m}^{2}_{ki} L}{2E} \right)
.
\label{oscprob}
\end{align}
Here $\Delta{m}^{2}_{ki}= m_{i}^2 - m_{k}^2$,
$L$ is the distance between the neutrino
detector and the neutrino source, and $E$ is the neutrino energy.
Since the oscillation probabilities of antineutrinos are given by the exchange
$U \leftrightarrows U^*$,
if the mixing matrix is complex
($U \neq U^*$)
the last term in Eq.~(\ref{oscprob})
describes CP violation
($P_{\nu_{l}\to\nu_{l'}} \neq P_{\bar\nu_{l}\to\bar\nu_{l'}}$)
in appearance experiments ($l \neq l'$).
In the standard parameterization (\ref{U}) of three-neutrino mixing,
$
\im{
U_{li}^{*}
\,
U_{l'i}
\,
U_{lk}
\,
U_{l'k}^{*}
}
=
\pm
c_{12} s_{12} c_{23} s_{23} c_{13}^2 s_{13} \sin\delta
$,
where the sign depends on the values of the flavor and mass indices.

It is also useful to express the transition probability as
\cite{Bilenky:2012zp}
\begin{align}
P_{\nua{l}\to\nua{l'}}
=
\null & \null
\delta_{ll'}
-
4
\sum_{i \neq p}
|U_{li}|^2
\left( \delta_{ll'} -  |U_{l'i}|^2 \right)
\,
\sin^{2}\Delta_{pi}
\nonumber
\\
\null & \null
+
8
\sum_{\stackrel{i>k}{i,k \neq p}}
\Big\{
\re{
U_{li}^{*}
\,
U_{l'i}
\,
U_{lk}
\,
U_{l'k}^{*}
}
\,
\cos\!\left( \Delta_{pi} - \Delta_{pk} \right)
\sin\Delta_{pi}
\sin\Delta_{pk}
\nonumber
\\
\null & \null
\phantom{
8
\sum_{\stackrel{i>k}{i,k \neq p}}
}
\pm
\im{
U_{li}^{*}
\,
U_{l'i}
\,
U_{lk}
\,
U_{l'k}^{*}
}
\,
\sin\!\left( \Delta_{pi} - \Delta_{pk} \right)
\sin\Delta_{pi}
\sin\Delta_{pk}
\Big\}
,
\label{oscprobalt}
\end{align}
where
$\Delta_{pi} \equiv \Delta{m}^{2}_{pi}L/4E$
and $p$ is an arbitrary fixed index,
which can be chosen in the most convenient way
depending on the case under consideration.
In the case of three-neutrino mixing,
there is only one interference term in Eq.~(\ref{oscprobalt}),
because for any choice of $p$ there is only one possibility for $i$ and $k$
such that $i>k$.

Neutrino oscillations have been measured in different channels
and the oscillations in each channel are mainly due to
one squared-mass difference and one of the three mixing angles in
the standard parameterization (\ref{U}) of three-neutrino mixing.
These measurements are described\footnote{
In the following we adopt the traditional classification of
terrestrial neutrino oscillation experiments
according to their baseline $L$ relative to the average energy $E$,
which determines the maximal sensitivity to the squared-mass difference
$\Delta{m}^2$
for which
$\Delta{m}^2 L / E \sim 1$.
Historically the first terrestrial neutrino oscillation experiments
have been short-baseline experiments,
which are sensitive to $\Delta{m}^2 \gtrsim 3 \times 10^{-2} \, \text{eV}^2$,
which imply
$L \lesssim 100 \, \text{m}$
for reactor experiments with
$E \sim 3 \, \text{MeV}$
and
$L \lesssim 30 \, \text{km}$
for accelerator experiments with
$E \sim 1 \, \text{GeV}$.
We call
long-baseline experiments those which have maximal sensitivity to
$3 \times 10^{-4} \lesssim \Delta{m}^2 \lesssim 3 \times 10^{-2} \, \text{eV}^2$
and
very-long-baseline experiments those which have maximal sensitivity to
$\Delta{m}^2 \lesssim 3 \times 10^{-4} \, \text{eV}^2$.
}
in Subsections~\ref{sub:VLBL}--\ref{sub:LBLnumu}.
In Subsection~\ref{sub:global} we review the results
of a recent global analysis of neutrino oscillation data
in the framework of three-neutrino mixing.
In Subsection~\ref{sub:absolute} we discuss briefly the problem of
the determination of the absolute scale of neutrino masses.


\subsection{Very-long-baseline $\protect\nua{e}$ experiments}
\label{sub:VLBL}

Solar neutrino experiments
(Homestake \cite{Cleveland:1998nv},
GALLEX/GNO \cite{Altmann:2005ix},
SAGE \cite{Abdurashitov:2002nt},
Super-Kamiokande \cite{Renshaw:2013dzu},
SNO \cite{Aharmim:2011vm},
Borexino \cite{Bellini:2014uqa},
KamLAND \cite{Gando:2014wjd};
see the recent reviews in Refs.~\citen{Antonelli:2012qu,Robertson:2012ib,Antonelli:2013nua,Bellini:2013wra})
measured $\nu_{e}$ disappearance due to $\nu_{e} \to \nu_{\mu}, \nu_{\tau}$
transitions generated by the solar squared-mass difference
$
\Delta m^2_{\text{S}}
\simeq
7 \times 10^{-5} \, \text{eV}^2
$
and a mixing angle
$
\sin^2 \vartheta_{\text{S}}
\simeq
0.3
$.
The measured solar neutrino energy spectrum ranges from about 0.2 MeV
to about 15 MeV
and the $\nu_{e} \to \nu_{\mu}, \nu_{\tau}$ transitions are
mainly due to averaged oscillations in vacuum for $E \lesssim 1.2 \, \text{MeV}$
and to matter effects
\cite{Wolfenstein:1978ue,Mikheev:1986gs}
for larger energies.

The very-long-baseline KamLAND experiment
\cite{Gando:2013nba}
confirmed these oscillations by observing the disappearance
of reactor $\bar\nu_{e}$
with $\langle E \rangle \simeq 3.6 \, \text{MeV}$
and
$\langle L \rangle \simeq 180 \, \text{km}$.

In the framework of three-neutrino mixing,
it is convenient to choose the numbering of the massive neutrinos
in order to have
\begin{equation}
\Delta{m}^{2}_{\text{S}}
=
\Delta{m}^{2}_{12}
,
\qquad
\vartheta_{\text{S}}
=
\vartheta_{12}
.
\label{dms}
\end{equation}

\subsection{Long-baseline $\protect\nua{\mu}$ disappearance experiments}
\label{sub:LBL}

Atmospheric neutrino experiments
(Kamiokande \cite{Fukuda:1994mc},
IMB \cite{Becker-Szendy:1992hq},
Super-Kamiokande \cite{Fukuda:1998mi},
Soudan-2 \cite{Sanchez:2003rb},
MACRO \cite{Ambrosio:2003yz},
MINOS \cite{Adamson:2012gt},
ANTARES \cite{AdrianMartinez:2012ph},
IceCube \cite{Aartsen:2013jza})
measured $\nu_{\mu}$ and $\bar\nu_{\mu}$ disappearance
due to oscillations generated by the atmospheric squared-mass difference
$
\Delta m^2_{\text{A}}
\simeq
2.3 \times 10^{-3} \, \text{eV}^2
$
and a mixing angle
$
\sin^2 \vartheta_{\text{A}}
\simeq
0.5
$.
The detectable energy spectrum of atmospheric neutrinos is very wide,
ranging from about 100 MeV to about 100 TeV,
and the source-detector distance varies from about 20 km for downward-going neutrinos
to about $1.3 \times 10^4 \, \text{km}$ for upward-going neutrinos.
The most precise determination of
$\Delta m^2_{\text{A}}$
and
$\sin^2 \vartheta_{\text{A}}$
has been obtained
in the high-statistics Super-Kamiokande atmospheric neutrino experiment
\cite{Abe:2011ph}
through the measurement of events contained in the detector volume,
which correspond to neutrino energies between
about 100 MeV and 10 GeV.

The disappearance of $\nu_{\mu}$ and $\bar\nu_{\mu}$
due to oscillations generated by
$\Delta m^2_{\text{A}}$
have been confirmed by the following accelerator long-baseline experiment:
the K2K experiment \cite{Ahn:2006zza}
measured the disappearance
of $\nu_{\mu}$
with
$\langle E \rangle \simeq 1.3 \, \text{GeV}$
at
$L \simeq 250 \, \text{km}$ (KEK--Kamioka);
the MINOS experiment \cite{Adamson:2013whj}
observed the disappearance
of $\nu_{\mu}$ and $\bar\nu_{\mu}$
with
$\langle E \rangle \simeq 3 \, \text{GeV}$
at
$L \simeq 735 \, \text{km}$ (Fermilab--Soudan);
the T2K experiment \cite{Abe:2014ugx}
measurement the disappearance
of $\nu_{\mu}$
at
$L \simeq 295 \, \text{km}$ (Tokai--Kamioka)
with a narrow-band off-axis beam peaked at
$\langle E \rangle \simeq 0.6 \, \text{GeV}$.

The Super-Kamiokande atmospheric neutrino data indicate that the
disappearance of $\nu_{\mu}$ and $\bar\nu_{\mu}$
is predominantly due to
$\nu_{\mu}\to\nu_{\tau}$ and $\bar\nu_{\mu}\to\bar\nu_{\tau}$ transitions,
respectively,
with a statistical significance of $3.8\sigma$
\cite{Abe:2012jj}.
This oscillation channel is confirmed at
$4.2\sigma$ by the observation of four
$\nu_{\mu}\to\nu_{\tau}$ candidate events
in the OPERA long-baseline accelerator experiment
\cite{Agafonova:2014ptn}
which was exposed to the CNGS (CERN--Gran Sasso) beam
with
$\langle E \rangle \simeq 13 \, \text{GeV}$
and
$L \simeq 730 \, \text{km}$.

\begin{figure}
\begin{center}
\begin{minipage}[r]{0.49\linewidth}
\begin{center}
\subfigure[]{
\includegraphics*[width=\linewidth]{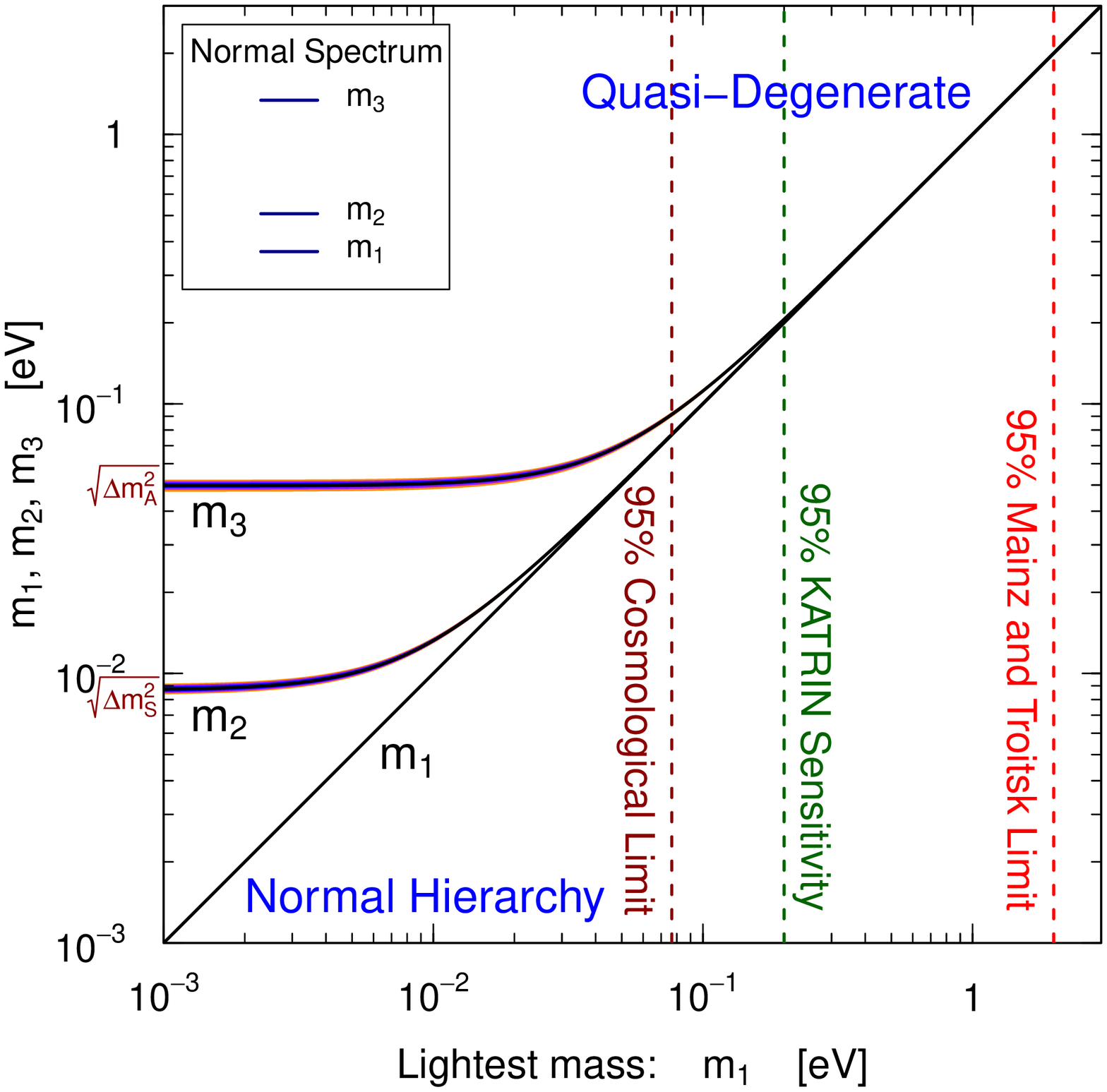}
\label{fig:mas-nor}
}
\end{center}
\end{minipage}
\hfill
\begin{minipage}[l]{0.49\linewidth}
\begin{center}
\subfigure[]{
\includegraphics*[width=\linewidth]{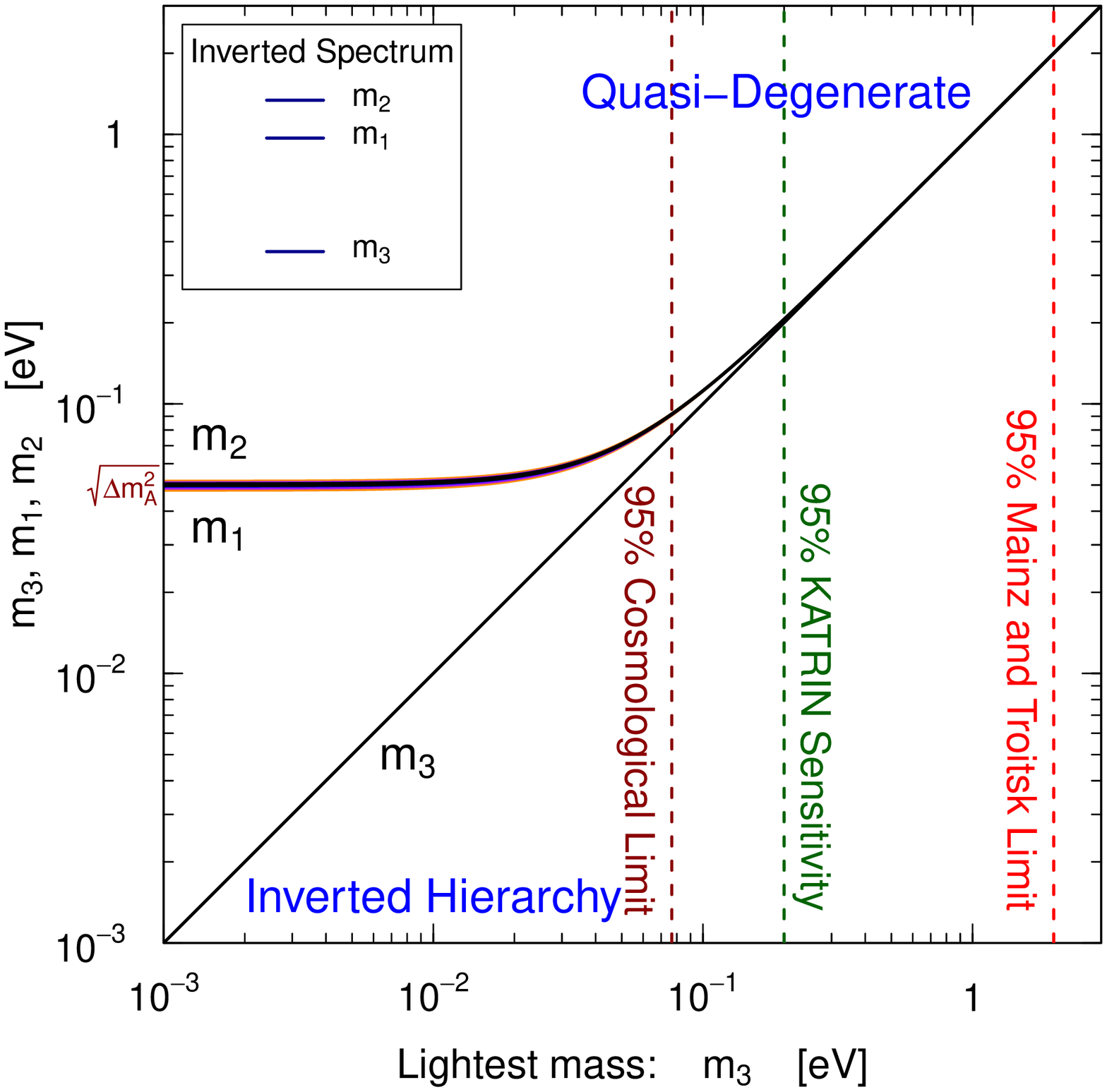}
\label{fig:mas-inv}
}
\end{center}
\end{minipage}
\end{center}
\caption{\label{fig:mass}
\subref{fig:mas-nor}
Values of the neutrino masses as functions of the lightest mass $m_{1}$ in the normal mass spectrum
obtained with the squared-mass differences in Tab.~\ref{tab:global}.
\subref{fig:mas-inv}
Corresponding
values of the neutrino masses as functions of the lightest mass $m_{3}$ in the inverted mass spectrum.
}
\end{figure}

In the framework of three-neutrino mixing,
with the convention (\ref{dms})
and taking into account that
$\Delta{m}^{2}_{\text{S}} \ll \Delta{m}^{2}_{\text{A}}$,
we have
\begin{equation}
\Delta{m}^{2}_{\text{A}}
=
\frac{1}{2}
\left| \Delta{m}^{2}_{13} + \Delta{m}^{2}_{23} \right|
,
\qquad
\vartheta_{\text{A}}
=
\vartheta_{\text{23}}
.
\label{dma}
\end{equation}
The absolute value in the definition of $\Delta{m}^{2}_{\text{A}}$
is necessary,
because
there are the two possible spectra for the neutrino masses
illustrated schematically in the insets of the two corresponding
panels in Fig.~\ref{fig:mass}:

\begin{description}

\item[The normal mass spectrum (NS):]
\begin{equation}
m_{1}<m_{2}<m_{3}
,
\quad \text{with} \quad
\Delta{m}^{2}_{12} \ll \Delta{m}^{2}_{23}
.
\label{NS}
\end{equation}

\item[The inverted mass spectrum (IS):]
\begin{equation}
m_{3}<m_{1}<m_{2}
,
\quad \text{with} \quad
\Delta{m}^{2}_{12} \ll |\Delta{m}^{2}_{13}|
.
\label{IS}
\end{equation}

\end{description}

The two spectra differ by the sign of
$\Delta{m}^{2}_{13}$
and
$\Delta{m}^{2}_{23}$,
which is positive in the normal spectrum
and negative in the inverted spectrum.

\subsection{Long-baseline $\bar\nu_{e}$ disappearance}
\label{sub:LBLnue}

In the beginning of 2012 the Daya Bay experiment
\cite{An:2012eh}
measured the disappearance of reactor $\bar\nu_{e}$
with
$\langle E \rangle \simeq 3.6 \, \text{MeV}$
by comparing the event rate measured at near detectors
located at 512~m and 561~m from the reactors
and that measured at far detectors
at a distance of 1579~m from the reactors.
This disappearance is due to neutrino oscillations generated by
$\Delta m^2_{\text{A}}$
and
the mixing angle $\vartheta_{13}$\footnote{
The value of $\vartheta_{13}$
was previously constrained by the negative results of
the Chooz \cite{Apollonio:2002gd}
and
Palo-Verde \cite{Boehm:2001ik}
long-baseline reactor antineutrino experiments.
}.
The current Daya Bay determination of the value of
$\vartheta_{13}$
is \cite{An:2013zwz}
\begin{equation}
\sin^{2}2\vartheta_{13}
=
0.090 {}^{+0.008}_{-0.009}
.
\label{t13}
\end{equation}
This result has been confirmed, with less precision, by the
Double Chooz \cite{Abe:2013sxa} ($L \simeq 1050 \, \text{m}$)
and
RENO \cite{Ahn:2012nd}
($L_{\text{near}} \simeq 294 \, \text{m}$
and
$L_{\text{far}} \simeq 1383 \, \text{m}$)
reactor experiments.

\subsection{Long-baseline $\nu_{\mu} \to \nu_{e}$ transitions}
\label{sub:LBLnumu}

In 2011 the T2K experiment
\cite{Abe:2011sj}
reported a first $2.5\sigma$ indication of long-baseline
$\nu_{\mu}\to\nu_{e}$
transitions
generated by
$\Delta m^2_{\text{A}}$
and
the mixing angle $\vartheta_{13}$.
Recently,
the T2K collaboration reported a convincing
$7.3\sigma$
observation of
$\nu_{\mu}\to\nu_{e}$
transitions
through the measurement of
28
$\nu_{e}$
events
with an expected background of
$4.92 \pm 0.55$
events
\cite{Abe:2013hdq}.
Assuming a negligible CP-violating phase $\delta$,
they obtained
$\sin^{2}2\vartheta_{13} = 0.140 {}^{+0.038}_{-0.032}$
for the normal spectrum
and
$\sin^{2}2\vartheta_{13} = 0.170 {}^{+0.045}_{-0.037}$
for the inverted spectrum,
which are in reasonable agreement with the more precise Daya Bay value in Eq.~(\ref{t13}).
The T2K observation of long-baseline
$\nu_{\mu}\to\nu_{e}$
transitions is supported by a less precise measurement of
the MINOS experiment
\cite{Adamson:2013ue}.


\subsection{Global analysis}
\label{sub:global}

The three-neutrino mixing parameters can be determined with good precision
with a global fit of neutrino oscillation data.
In Tab.~\ref{tab:global}
we report the results of the latest global fit presented in Ref.~\citen{Capozzi:2013csa},
which agree, within the uncertainties,
with the NuFIT-v2.0
\cite{NuFIT-2014}
update of the global analysis presented in Ref.~\citen{GonzalezGarcia:2012sz}.
One can see that all the oscillation parameters are determined with precisions between
about 3\% and 11\%.
The largest uncertainty is that of $\vartheta_{23}$,
which is known to be close to maximal ($\pi/4$),
but it is not known if it is smaller or larger than $\pi/4$.
For the Dirac CP-violating phase $\delta$,
there is an indication in favor of $\delta \approx 3\pi/2$,
which would give maximal CP violation,
but at $3\sigma$ all the values of $\delta$ are allowed,
including the CP-conserving values $\delta=0,\pi$.

\begin{table}[b!]
\caption{\label{tab:global}
Values of the neutrino mixing parameters obtained with
global analysis of neutrino oscillation data
presented in Ref.~\citen{Capozzi:2013csa}
in the framework of three-neutrino mixing
with a normal spectrum (NS) or an inverted spectrum (IS).
The relative uncertainty (rel. unc.) has been obtained from the
$3\sigma$ range divided by 6.}
\centering
{
\begin{tabular}{lcccccc}
Parameter & Spectrum & Best fit & $1\sigma$ range & $2\sigma$ range & $3\sigma$ range & rel. unc. \\
\hline
$\Delta{m}^2_{\text{S}}/10^{-5}\,\text{eV}^2 $ & & 7.54 & 7.32 -- 7.80 & 7.15 -- 8.00 & 6.99 -- 8.18 & 3\% \\
\hline
$\sin^2 \theta_{12}/10^{-1}$ & & 3.08 & 2.91 -- 3.25 & 2.75 -- 3.42 & 2.59 -- 3.59 & 5\% \\
\hline
\multirow{2}{*}{$\Delta{m}^2_{\text{A}}/10^{-3}\,\text{eV}^2$}
& NS & 2.44 & 2.38 -- 2.52 & 2.30 -- 2.59 & 2.22 -- 2.66 & 3\% \\
& IS & 2.40 & 2.33 -- 2.47 & 2.25 -- 2.54 & 2.17 -- 2.61 & 3\% \\
\hline
\multirow{2}{*}{$\sin^2 \theta_{23}/10^{-1}$}
& NS & 4.25 & 3.98 -- 4.54 & 3.76 -- 5.06 & 3.57 -- 6.41 & 11\% \\
& IS & 4.37 & 4.08 -- 6.10 & 3.84 -- 6.37 & 3.63 -- 6.59 & 11\% \\
\hline
\multirow{2}{*}{$\sin^2 \theta_{13}/10^{-2}$}
& NS & 2.34 & 2.16 -- 2.56 & 1.97 -- 2.76 & 1.77 -- 2.97 & 9\% \\
& IS & 2.39 & 2.18 -- 2.60 & 1.98 -- 2.80 & 1.78 -- 3.00 & 9\% \\
\hline
\end{tabular}
}
\end{table}

\subsection{Absolute scale of neutrino masses}
\label{sub:absolute}

The determination of the absolute value of neutrino masses
is an open problem
which cannot be resolved by neutrino oscillations,
that depend only on the differences of the squares of the neutrino masses.
However,
the measurement in neutrino oscillation experiments of the
neutrino squared-mass differences allows us to constraint the allowed patterns of neutrino masses.
A convenient way to see the allowed patterns of neutrino masses
is to plot the values of the masses as functions of the unknown lightest mass
$m_{\text{min}}$,
as shown in Fig.~\ref{fig:mass},
where
we used the squared-mass differences in Tab.~\ref{tab:global}
with:

\begin{description}

\item[Normal mass spectrum (NS):] $m_{\text{min}}=m_{1}$,
\begin{equation}
m_{2} = \sqrt{ m_{\text{min}}^2 + \Delta{m}^{2}_{\text{S}} }
,
\quad
m_{3} = \sqrt{ m_{\text{min}}^2 + \Delta{m}^{2}_{\text{A}} + \Delta{m}^{2}_{\text{S}}/2 }
.
\label{mNS}
\end{equation}

\item[Inverted mass spectrum (IS):] $m_{\text{min}}=m_{3}$,
\begin{equation}
\null\hspace{-1cm}\null
m_{1} = \sqrt{ m_{\text{min}}^2 + \Delta{m}^{2}_{\text{A}} - \Delta{m}^{2}_{\text{S}}/2 }
,
\quad
m_{2} = \sqrt{ m_{\text{min}}^2 + \Delta{m}^{2}_{\text{A}} + \Delta{m}^{2}_{\text{S}}/2 }
.
\label{mIS}
\end{equation}

\end{description}

Figure~\ref{fig:mass} shows that there are three extreme possibilities:

\begin{description}

\item[A normal hierarchy:]
$m_{1} \ll m_{2} \ll m_{3}$.
In this case
\begin{equation}
m_{1}
\ll
m_{2} \simeq \sqrt{\Delta{m}^{2}_{\text{S}}} \approx 9 \times 10^{-3} \, \text{eV}
,
\quad
m_{3} \simeq \sqrt{\Delta{m}^{2}_{\text{A}}} \approx 5 \times 10^{-2} \, \text{eV}
.
\label{mNH}
\end{equation}

\item[An inverted hierarchy:]
$m_{3} \ll m_{1} \lesssim m_{2}$.
In this case
\begin{equation}
m_{3}
\ll
m_{1}
\lesssim
m_{2}
\simeq \sqrt{\Delta{m}^{2}_{\text{A}}} \approx 5 \times 10^{-2} \, \text{eV}
.
\label{mIH}
\end{equation}

\item[Quasi-degenerate spectra:]
$m_{1} \lesssim m_{2} \lesssim m_{3} \simeq m_{\nu}^{\text{QD}}$ in the normal scheme
and
$m_{3} \lesssim m_{1} \lesssim m_{2} \simeq m_{\nu}^{\text{QD}}$ in the inverted scheme,
with
\begin{equation}
m_{\nu}^{\text{QD}}
\gg
\sqrt{\Delta{m}^{2}_{\text{A}}} \approx 5 \times 10^{-2} \, \text{eV}
.
\label{mQD}
\end{equation}

\end{description}

Besides neutrinoless double-$\beta$ decay,
which is the topic of this review and will be discussed in depth in the following Sections,
there are two main sources of information
on the absolute scale of neutrino masses:


\noindent
\textbf{Beta decay.}
The end-point part of the spectrum of electrons emitted in $\beta$ decay is affected by neutrino masses through the phase-space factor.
Hence,
the $\beta$ decay information on neutrino masses
is very robust.
Tritium $\beta$-decay experiments obtained the most stringent model-independent bounds
on the neutrino masses by limiting the effective electron neutrino mass
(see the recent reviews in Refs.~\citen{Drexlin:2013lha,Weinheimer:2013hya})
\begin{equation}
m_{\beta}^2
=
\sum_{k=1}^{3} |U_{ek}|^{2} m^{2}_{k}
.
\label{trimass}
\end{equation}
The most stringent limits have been obtained
in the Mainz \cite{hep-ex/0412056} and Troitsk \cite{1108.5034} experiments:
\begin{equation}
m_{\beta} \leq
\left\{
\begin{array}{ll} \displaystyle
2.3 \, \text{eV}
&
\quad
(\text{Mainz})
,
\\ \displaystyle
2.05 \, \text{eV}
&
\quad
(\text{Troitsk})
,
\end{array}
\right.
\label{mb}
\end{equation}
at 95\% CL.
The KATRIN experiment \cite{Fraenkle:2011uu},
which is scheduled to start data taking in 2016,
is expected to have a sensitivity to
$m_{\beta}$ of about 0.2 eV.
An approximate Mainz and Troitsk upper bound of 2 eV for the lightest mass
and the sensitivity of KATRIN
are depicted in Fig.~\ref{fig:mass}.

\noindent
\textbf{Cosmology.}
Since light massive neutrinos constitute hot dark matter,
cosmological data give information on the sum of neutrino masses
(see Refs.~\citen{Wong:2011ip,Lesgourgues:2012uu,Lesgourgues-Mangano-Miele-Pastor-2013}).
The analysis of cosmological data in the framework of the standard
Cold Dark Matter model with a cosmological constant ($\Lambda$CDM)
disfavors neutrino masses larger than some fraction of eV,
because free-streaming neutrinos suppress
small-scale clustering.
The value of the upper bound on the sum of neutrino masses
depends on model assumptions and on the considered data set.
Assuming a spatially flat Universe,
the recent results of the Planck experiment
\cite{Ade:2013zuv}
combined with other cosmological data gave,
with 95\% Bayesian probability,
the upper bound
\begin{equation}
\sum_{k=1}^{3} m_{k} < 0.23 \, \text{eV}
.
\label{cosmo}
\end{equation}
The corresponding upper bound for the lightest mass is depicted in Fig.~\ref{fig:mass}.

\section{Theory of neutrino masses and mixing}
\label{bb3}

Several mechanisms of neutrino mass generation have been proposed in the literature.
In the prevailing opinion, the most plausible and the most viable is the seesaw mechanism
\cite{Minkowski:1977sc,Yanagida:1979as,GellMann:1980vs,Glashow:1979nm,Mohapatra:1980ia}.
The most general approach to the seesaw mechanism is based on the effective Lagrangian formalism \cite{Weinberg:1979sa}
discussed in Subsection~\ref{sub:ela}.
In Subsection~\ref{sub:MD} we discuss in details the simplest ``type I'' implementation of the seesaw mechanism
based on the existence of Majorana and Dirac mass terms which can be generated in the framework of GUT models.

\subsection{Effective Lagrangian approach}
\label{sub:ela}

Let us start by remarking that the left-handed and right-handed quark and charged lepton fields and the left-handed neutrino fields
$\nu_{lL}$ ($l=e, \mu, \tau$) are Standard Model (SM) fields.
What about right-handed neutrino fields?
If $\nu_{lR}$ are also considered to be SM fields,
the neutrino masses (as the quark and charged lepton masses) can be generated by the standard Higgs mechanism via the Yukawa interaction
\begin{equation}\label{Yukawa}
\mathcal{L}^{Y}_{I}
=
-
\sqrt{2}
\sum_{l'l} \overline{L_{l'L}} \, Y_{l'l} \, \nu_{lR} \, \widetilde{\Phi}
+
\mathrm{h.c.}
.
\end{equation}
Here
\begin{eqnarray}\label{doublets}
L_{lL}
=
\left(
\begin{array}{c}
\nu_{lL}
\\
l_L
\end{array}
\right)
,
\qquad
\Phi
=
\left(
\begin{array}{c}
\Phi^{(+)}
\\
\Phi^{(0)}
\end{array}
\right)
\end{eqnarray}
are the lepton and Higgs doublets, $\widetilde{\Phi}=i\tau_{2}\Phi^{*}$
(here $\tau_{2}$ is the second Pauli matrix),
and $Y_{l'l}$ are dimensionless Yukawa coupling constants.

After spontaneous breaking of the electroweak symmetry, we have
\begin{eqnarray}\label{break}
\widetilde{\Phi}
=
\frac{1}{\sqrt{2}}
\left(
\begin{array}{c}
v + H
\\
0
\end{array}
\right)
,
\end{eqnarray}
where $H$ is the field of the physical Higgs boson.
From Eqs.~(\ref{Yukawa}) and (\ref{break}), we obtain the Dirac mass term
\begin{equation}\label{Dmass}
\mathcal{L}^{\mathrm{D}}
=
-
\sum_{l',l} \overline{\nu_{l'L}} \, M_{l'l}^{\mathrm{D}} \, \nu_{lR}
+\mathrm{h.c.}
.
\end{equation}
Here
\begin{equation}\label{Dmass1}
M_{l'l}^{\mathrm{D}} = v \, Y_{l'l}
,
\end{equation}
where $v=(\sqrt{2}G_{F})^{-1/2}\simeq 246$ GeV is the vacuum expectation value of the Higgs field.

After the diagonalization of the matrix $Y$
($Y = U y V^{\dag}$,
where $U$ and $V$ are unitary matrices and $y$ is a diagonal matrix),
we obtain the Dirac neutrino masses
\begin{equation}\label{Dmass2}
m_{i} = v \, y_{i}.
\end{equation}
If we assume a normal hierarchy of neutrino masses,
the value of the largest neutrino mass is
$m_{3}\simeq\sqrt{\Delta m_{\text{A}}^{2}}\simeq 5 \times 10^{-2}$ eV
($\Delta m_{\text{A}}^{2}$ is the atmospheric squared-mass difference in Eq.~(\ref{dma})).
Then, for the corresponding Yukawa coupling we find
\begin{equation}\label{Yukawa1}
y_{3} \simeq 2 \times 10^{-13}.
\end{equation}
The constants $y_{1,2}$ are much smaller.

Comparing the value of $y_{3}$
with the values of the Yukawa couplings of the quarks and the charged lepton of the third generation,
$y_{t} \simeq 0.7$,
$y_{b} \simeq 2 \times 10^{-2}$,
$y_{\tau} \simeq 7 \times 10^{-3}$,
it is clear that the neutrino Yukawa couplings are unnaturally small.

The necessity of extremely small values of the neutrino Yukawa coupling constants is commonly considered as a strong argument against a SM origin of the neutrino masses.

Thus, we are naturally led to consider\footnote{
In the simplest case of two neutrinos,
the Majorana mass term was considered for the first time in Ref.~\citen{Gribov:1968kq}.
}
the Majorana mass term
\begin{equation}\label{Mjmass}
\mathcal{L}_{L}^{\mathrm{M}}
=
-
\frac{1}{2}
\,
\sum_{l',l} \overline{\nu_{l'L}} \, M^{L}_{l'l} \, \nu_{lL}^{c}
+
\mathrm{h.c.}
,
\end{equation}
which is the only possible mass term which can be build from the SM neutrino fields $\nu_{lL}$,
but is forbidden by the symmetries of the SM.
Here
$\nu_{lL}^{c} = (\nu_{lL})^{c} = C \, \overline{\nu_{lL}}^{T}$,
where $C$ is the matrix of charge conjugation,
which satisfies the conditions
\begin{equation}\label{Cconj}
C \, \gamma^{T}_{\alpha} \, C^{-1} = - \gamma_{\alpha}
,
\quad C^{T}= -C
,
\end{equation}
where $\gamma_{\alpha}$,
with $\alpha=0,1,2,3$,
are the Dirac matrices.
It follows from the Fermi-Dirac statistics that $M^{L}$ is a symmetric matrix.

After the diagonalization of the matrix $M^{L}$
($M^{L}=UmU^{T}$,
where $U^{\dag}U=1$,
$m_{ik}=m_{i}\delta_{ik}$,
$m_{i}>0$),
we have
\begin{equation}\label{Mjmass1}
\mathcal{L}_{L}^{\mathrm{M}}
=
-
\frac{1}{2}
\,
\sum^{3}_{i=1} m_{i} \overline{\nu_{i}} \, \nu_{i}
.
\end{equation}
Here $\nu_{i}=\nu^{c}_{i}$ is the field of the Majorana neutrino with mass $m_{i}$.

The Majorana mass term in Eq.~(\ref{Mjmass}) violates the total lepton number $L$
and can be generated only in the framework of a theory beyond the SM.
In general,
the effects of high-energy physics beyond the SM can be described by
adding to the SM Lagrangian effective nonrenormalizable Lagrangian terms
which are invariant under the electroweak $SU(2)_{L} \times U(1)_{Y}$ transformations \cite{Weinberg:1979sa,Wilczek:1979hc}.

In order to generate the neutrino masses,
we need to build an effective Lagrangian term which is quadratic in the lepton fields.
The product
$\overline{L_{lL}}\widetilde{\Phi}$ is $SU(2)_{L} \times U(1)_{Y}$ invariant and has dimension $M^{5/2}$.
Taking into account that the Lagrangian must have dimension $M^4$,
we have\footnote{
Let us notice that for the dimensional arguments used here it is important that the Higgs boson is not a composite particle and
that there is a Higgs field having dimension $M$.
The recent discovery of the Higgs boson at CERN \cite{Aad:2012tfa,Chatrchyan:2012ufa} confirms this assumption.
}
\begin{equation}\label{effective}
\mathcal{L}_{I}^{\mathrm{eff}}
=
-
\frac{1}{\Lambda}
\sum_{l'l} \overline{L_{l'L}} \, \widetilde{\Phi} \, \overline{Y}_{l'l} \, \widetilde{\Phi}^{T} \, L^{c}_{lL}
+
\mathrm{h.c.}
,
\end{equation}
where the parameter $\Lambda \gg v$ has the dimension of mass.
It characterizes the high-energy scale at which the total lepton number $L$ is violated.

After spontaneous breaking of the electroweak symmetry, from Eqs.~(\ref{break}) and (\ref{effective}) we obtain the Majorana mass term
in Eq.~(\ref{Mjmass}),
with the mass matrix $M^{\mathrm{L}}$ given by
\begin{equation}\label{Mjmass2}
M^{\mathrm{L}}
=
\frac{v^{2}}{\Lambda} \, \overline{Y}
.
\end{equation}
The neutrino masses are obtained from the diagonalization of $\overline{Y}$
($\overline{Y}^{L} = U \overline{y} U^{T}$, where $U^{\dag}U=1$, $\overline{y}_{ik}=\overline{y}_{i}\delta_{ik}$, $\overline{y}_{i}>0$):
\begin{equation}\label{numass}
m_{i}
=
\frac{v^{2}}{\Lambda} \, \overline{y}_{i}
.
\end{equation}
Thus, if the Majorana neutrino masses are generated by a high-energy mechanism beyond the SM,
the value of the neutrino mass
$m_{i}$ is a product of a ``standard fermion mass'' $v \, \overline{y}_{i}$ and a suppression factor
which is given by the ratio of the electroweak scale $v$ and a high-energy scale $\Lambda$ of new total lepton number violating physics.
This suppression is called ``seesaw mechanism''.

In order to estimate $\Lambda$,
we consider the case of a normal hierarchy of neutrino masses.
In this case, $m_{3} \simeq \sqrt{\Delta m^{2}_{\text{A}}} \simeq 5 \times 10^{-2}$ eV.
Assuming that the parameter $\overline{y}_{3}$ is of the order of one,
we find $\Lambda \sim 10^{15}$ GeV.
Hence, the smallness of the Majorana neutrino masses could mean that $L$ is violated at a very large scale.

The violation of the total lepton number can be connected with the existence of heavy Majorana leptons which interact with the SM particles \cite{Weinberg:1980bf}.
Let us assume that heavy Majorana leptons $N_{i}$ ($i=1,\ldots,n$),
which are singlets of the SM $SU(2)_{L} \times U(1)_{Y}$ symmetry group,
have the Yukawa lepton number violating interaction
\begin{equation}\label{HeavyMj}
\mathcal{L}_{I}^{Y}
=
-
\sqrt{2}
\sum _{l,i} y_{li} \overline{L_{lL}} \, \widetilde{\Phi} \, N_{iR}
+
\mathrm{h.c.}
.
\end{equation}
Here $N_{i}=N^{c}_{i}$ is the field of a Majorana lepton with mass $M_{i} \gg v$ and $y_{li}$ are dimensionless Yukawa coupling constants.

In the second order of perturbation theory,
at electroweak energies the interaction in Eq.~(\ref{HeavyMj}) generates the effective Lagrangian in Eq.~(\ref{effective}),
with the constant $\overline{Y}_{l'l} / \Lambda$ given by
\begin{equation}\label{HeavyMj1}
\frac{\overline{Y}_{l'l}}{\Lambda}=\sum_{i}y_{l'i}\frac{1}{M_{i}} y_{li}.
\end{equation}
From this relation it is clear that the scale of the new lepton number violating physics is determined by the masses of the heavy Majorana leptons.

Summarizing,
if the neutrino masses are generated by the interaction of the lepton and Higgs doublets with heavy Majorana leptons:
\begin{enumerate}
\item Neutrinos with definite masses are Majorana particles.
\item Neutrino masses are much smaller than the masses of charged leptons and quarks.
They are of the order of a ``standard fermion mass'' multiplied by a suppression factor which is given by the ratio of
the Higgs vacuum expectation value $v$ and the high-energy scale of total lepton number violation
(seesaw mechanism).
\item The number of light massive neutrinos is equal to the number of lepton generations (three)\footnote{In the scheme of neutrino masses and mixing based on the effective Lagrangian approach there are no light sterile neutrinos.
However, there are some indications in favor of transitions of active flavor neutrinos into sterile states (see Ref.~\citen{1311.1335}).
Numerous experiments, now in preparation, with the aim of checking the existence of light sterile neutrinos will confirm or disprove
in a few years the scenario that we have considered here.}.
\end{enumerate}

In the approach considered here,
the observation of neutrinoless double-$\beta$ decay would imply the existence of heavy Majorana leptons which induce Majorana neutrino masses.
The scale of the masses of the heavy leptons
is determined by the parameter $\Lambda \sim 10^{15}$ GeV.
Such heavy particles cannot be produced in laboratory.
It is however very interesting that
the CP-violating decays of such particles in the early Universe could be the origin of the baryon asymmetry of the Universe
through the leptogenesis mechanism
(see, for example, Refs.~\citen{Davidson:2008bu,Fong:2013wr}).

The seesaw mechanism based on the interaction in Eq.~(\ref{HeavyMj}) is called type I seesaw.
The effective Lagrangian in Eq.~(\ref{effective}) can also be generated by the Lagrangian of interaction of
the lepton and Higgs doublets with a
heavy scalar boson triplet (type II seesaw)
or with a
heavy lepton triplet (type III seesaw)
(see, for example, Ref.~\citen{Mohapatra:2006gs}).

\subsection{Approach based on Majorana and Dirac neutrino mass term}
\label{sub:MD}

The simplest implementation of the seesaw mechanism of neutrino mass generation
is the so-called ``type I'' seesaw,
which is based on Majorana and Dirac mass terms
that can be generated in the framework of GUT models
\cite{Minkowski:1977sc,Yanagida:1979as,GellMann:1980vs,Glashow:1979nm,Mohapatra:1980ia}.

From three active flavor left-handed fields $\nu_{lL}$ ($l=e,\mu, \tau$) and $N_{s}$ sterile right-handed fields $\nu_{sR}$ ($s=s_{1},\ldots,s_{N_{s}}$),
one can build the general mass term
(see, for example, Refs.~\citen{Bilenky:1976yj,Bilenky:1987ty})
\begin{equation}\label{MDmass}
\mathcal{L}^{\mathrm{M+D}}
=
-
\frac{1}{2}
\,
\sum_{l',l}
\overline{\nu_{l'L}} \, M^{L}_{l'l} \, \nu^{c}_{lL}
-
\sum_{l,s}
\overline{\nu_{lL}} \, \overline{M}^{D}_{ls} \, \nu_{sR}
-
\frac{1}{2}
\,
\sum_{s',s}
\overline{\nu^{c}_{s'R}} \, \overline{M}^{R}_{s's} \, \nu_{sR}
+
\mathrm{h.c.}
,
\end{equation}
where $\overline{M}^{D}$ is a complex matrix and $M^{L}$ and $\overline{M}^{R}$ are complex symmetric matrices.

The left-handed flavor fields $\nu_{lL}$ enter also in the leptonic charged current $j_{\alpha}^{CC}$ and in the neutrino neutral current $j_{\alpha}^{NC}$:
\begin{equation}
j_{\alpha}^{CC} = 2 \sum_{l=e,\mu,\tau} \overline{l_{L}} \, \gamma_{\alpha} \, \nu_{lL}
,
\qquad
j_{\alpha}^{NC} = \sum_{l=e,\mu,\tau} \overline{\nu_{lL}} \, \gamma_{\alpha} \, \nu_{lL}
.
\label{CC-NC}
\end{equation}
On the other hand,
the right-handed sterile fields $\nu_{sR}$ enter only in
the Majorana and Dirac mass term.
Therefore,
we can always choose the right-handed fields in such a way that $\overline{M}^{R}$ is a diagonal matrix.
In fact,
we can make the unitary transformation
$\nu_{sR} = \sum_{i} V_{si} N_{iR}$
(with $V^{\dag}V=1$)
such that
$V^{T} \overline{M}^{R} V = M^{R}$,
with
$M^{R}_{ik}=M_{i}\delta_{ik}$
and
$M_{i}>0$.
Then,
the Majorana and Dirac mass term takes the form
\begin{equation}\label{MDmass1}
\mathcal{L}^{\mathrm{M+D}}
=
-
\frac{1}{2} \, \sum_{l',l} \overline{\nu_{l'L}} \, M^{L}_{l'l} \, \nu^{c}_{lL}
-
\sum_{li} \overline{\nu_{lL}} \, M^{D}_{li} \, N_{iR}
-
\frac{1}{2} \, \sum_{i} M_{i} \, \overline{N^{c}_{iR}} \, N_{iR}
+
\mathrm{h.c.}
,
\end{equation}
where
$M^{D} = \overline{M}^{D} V$
and
$N_{iR} = \sum_{s} V_{si}^{*} \nu_{sR}$.

Let us first consider the simplest case of one left-handed field $\nu_{L}$ and one right-handed field
$\nu_{R}$. The Majorana and Dirac mass term has the form
\begin{eqnarray}\label{2MDmassterm}
\mathcal{L}^{\mathrm{M+D}}
&=&
-
\frac{1}{2}
\,
m_{L} \overline{\nu_{L}} \, \nu_{L}^{c}
-
m_{D} \overline{\nu_{L}} \, \nu_{R}
-
\frac{1}{2}
\,
m_{R} \overline{\nu^{c}_{R}} \, \nu_{R}
+
\mathrm{h.c.}
\nonumber
\\
&=&
-
\frac{1}{2}
\,
\overline{n_{L}} \, M^{\mathrm{M+D}} \, n^{c}_{L}
+
\mathrm{h.c.}
.
\end{eqnarray}
Here $m_{L}, m_{D}, m_{R}$ are real parameters and
\begin{eqnarray}\label{2DMDmatrix1}
M^{\rm{M+D}}=\left(
\begin{array}{cc}
m_{L}&m_{D}\\
m_{D}&m_{R}
\end{array}
\right)
,
\qquad
n_{L}={\nu_{L}\choose \nu^{c}_{R}}
.
\end{eqnarray}
The real, symmetric 2$\times$2 mass matrix $M^{\rm{M+D}}$ can be diagonalized with the transformation
\begin{equation}\label{2diagonalization}
M^{\rm{M+D}}= O\,m'\,O^{T}.
\end{equation}
Here
\begin{eqnarray}\label{2orthogonal}
O=\left(
\begin{array}{cc}
\cos\theta&\sin\theta\\
-\sin\theta&\cos\theta
\end{array}
\right)
,
\end{eqnarray}
and $m'_{ik}=m'_{i}\delta_{ik}$, where
\begin{equation}\label{2eigenvalues}
m'_{1,2}= \frac{1}{2}\,(m_{R}+m_{L}) \mp
\frac{1}{2}\,\sqrt{(m_{R}-m_{L})^{2} + 4\,m_{D}^{2}}.
\end{equation}
are the eigenvalues of the matrix $M^{\rm{M+D}}$.
From Eqs.~(\ref{2diagonalization}), (\ref{2orthogonal})
and (\ref{2eigenvalues}), we find
\begin{equation}\label{2mixangle}
\tan 2\theta
=
\frac{2m_{D}}{m_{R}-m_{L}}
,
\quad
\cos 2\theta
=
\frac{m_{R}-m_{L}}{\sqrt{(m_{R}-m_{L})^{2} + 4 m_{D}^{2}}}
.
\end{equation}
The eigenvalues $m'_{1,2}$ can be positive or negative.
Let us write
\begin{equation}\label{2eigenvalues1}
m'_{i} = m_{i} \, \eta_{i}
,
\end{equation}
where $m_{i}=|m'_{i}|$ and $\eta_{i}=\pm 1$.
From Eqs.~(\ref{2diagonalization}) and (\ref{2eigenvalues1}) we obtain
\begin{equation}\label{2diagonalization1}
M^{\rm{M+D}} = U \, m \, U^{T}
,
\end{equation}
where
\begin{equation}\label{2unitar}
U = O \, \sqrt{\eta}
\end{equation}
is a unitary matrix. From Eqs.~(\ref{2diagonalization1}) and (\ref{2unitar}) it follows that
\begin{eqnarray}
\nu_{L}&=& \cos\theta \sqrt{\eta_{1}}\,\nu_{1 L} + \sin\theta \sqrt{\eta_{2}}\,\nu_{2 L}
\nonumber\\
\nu_{R}^{c}&=&- \sin\theta \sqrt{\eta_{1}}\,\nu_{1 L} + \cos\theta \sqrt{\eta_{2}}\,\nu_{2 L}\label{2mixing}
\end{eqnarray}
where $\nu_{i}=\nu^{c}_{i}$ is the field of the Majorana neutrino with mass $m_{i}$.

The standard seesaw mechanism \cite{Minkowski:1977sc,Yanagida:1979as,GellMann:1980vs,Glashow:1979nm,Mohapatra:1980ia} is based on the
following assumptions:
\begin{enumerate}
\item There is no left-handed Majorana mass term in the Lagrangian
($m_{L}=0$).
\item The mass $m_{R}$ in the right-handed Majorana mass term,
which is the source of violation of the total lepton number $L$, is much larger than
$m_{D}$:
\begin{equation}\label{seesawineq}
m_{R}\gg m_{D}
.
\end{equation}
\item The Dirac mass term is generated by the Higgs
mechanism ($m_{D}$ is of the order of the mass of a charged
lepton or quark).
\end{enumerate}
For $m_{L}=0$ we have
\begin{equation}\label{2mix}
m_{1,2}
=
\frac{1}{2}
\left(
m_{R} \mp \sqrt{m_{R}^{2} + 4\,m_{D}^{2}}
\right)
,
\quad
\tan 2\theta = \frac{2m_{D}}{m_{R}}
.
\end{equation}
Thus, the Dirac mass $m_{D}$ ensures mixing ($\theta\neq 0$) and a nonzero light neutrino mass $m_{1}$.
From Eqs.~(\ref{seesawineq}) and (\ref{2mix}) we find
\begin{equation}\label{2mix1}
m_{1} \simeq \frac{m^{2}_{D}}{m_{R}}
,
\quad
m_{2} \simeq m_{R}
,
\quad
\theta \simeq \frac{m_{D}}{m_{R}} \ll 1
,
\quad
\eta_{1} = -1
,
\quad
\eta_{2} =1
.
\end{equation}
Thus,
the seesaw mechanism implies that in the mass spectrum there is a light Majorana neutrino with mass
$m_{1} \simeq m^{2}_{D} / m_{R} \ll m_{D}$ and a heavy neutral Majorana lepton with mass $m_{2} \simeq m_{R} \gg m_{D}$.
Neglecting the small mixing of light and heavy Majorana leptons we have
$\nu_{L}\simeq i\nu_{1L}$,
$\nu^{c}_{R}\simeq \nu_{2L}$.

Let us consider now the case of three active flavor left-handed fields $\nu_{lL}$ and
$N_{s}$ sterile singlet right-handed fields $\nu_{sR}$.
In this case,
the seesaw mixing matrix has the form
\begin{eqnarray}\label{3mixingmatrix}
M^{\mathrm{seesaw}}
=
\left(\begin{array}{cc}
0&M^{D}\\
(M^{D})^{T}&M^{R}\end{array}\right)
,
\end{eqnarray}
where $M^{D}$ is a complex $3\times3$ matrix
and
$M^{R}_{ik} = M_{i} \delta_{ik}$
with
$M_{i} \gg |M^{D}_{ik}|$.

Let us introduce the matrix $\widetilde{M}$ through the transformation
\begin{equation}\label{3diagonalization}
V^{T}\,M^{\rm{seesaw}} \, V = \widetilde{M}
,
\end{equation}
where $V$ is an unitary matrix.
We will now show that it is possible to choose $V$ in order to obtain a matrix $\widetilde{M}$ which has a block-diagonal form.

Notice that in the case of one generation up to
terms linear in $m_{D} / m_{R} \ll 1$ we have
\begin{equation}
U^{(2)}
\simeq
\left(
\begin{array}{cc}
1&m_{D} / m_{R}\\
-m_{D} / m_{R}& 1
\end{array}
\right)
.
\label{1mixingmatrix}
\end{equation}
Then, let us consider the matrix
\begin{equation}
V
\simeq
\left(
\begin{array}{cc}
1&A\\
-A^{\dag}& 1
\end{array}
\right)
.
\label{2mixingmatrix}
\end{equation}
with $A \ll 1$.
Up to terms linear in $A$ the matrix $V$ is unitary.
In the same linear approximation, the non-diagonal block element of the symmetric matrix $\widetilde{M}$ is given by
$(M^{D})^{T} - M^{R} \, A^{\dag}$.
Thus, if we choose
\begin{equation}\label{matrixA}
A^{\dag} = (M^{R})^{-1} \, (M^{D})^{T}
,
\end{equation}
at the leading order in $A$
the matrix $\widetilde{M}$ takes the block-diagonal form
\begin{eqnarray}
\widetilde{M}
\simeq
\left(\begin{array}{cc}
- M^{D} \, (M^{R})^{-1} \, (M^{D})^{T} & 0\\
0 & M^{R}
\end{array}\right)
.
\label{3massmatrix}
\end{eqnarray}
Thus,
for the left-handed Majorana neutrino mass term we obtain the expression
\begin{equation}\label{numassterm}
\mathcal{L}^{\mathrm{M}}
\simeq
-
\frac{1}{2}
\sum_{l`l}
\overline{\nu_{l'L}} \, M^{L}_{l'l} \, \nu^{c}_{lL}
+
\mathrm{h.c.}
,
\end{equation}
with the matrix $M^{L}$ given by the seesaw relation
\begin{equation}\label{1seesaw}
M^{L}= - M^{D} \, (M^{R})^{-1} \, (M^{D})^{T}
.
\end{equation}
On the other hand,
for the mass term of heavy Majorana leptons we find
\begin{equation}\label{Rmassterm}
\mathcal{L}^{R}
\simeq
-
\frac{1}{2}
\sum_{i} M_{i} \, \overline{N_{i}} \, N_{i}
,
\end{equation}
where $N_{i}=N_{iR}+N^{c}_{iR}=N^{c}_{i}$ is the field of a heavy Majorana lepton with mass $M_{i}$.

Equation~(\ref{numassterm}) is the mass term of the three light
Majorana neutrinos and Eq.~(\ref{Rmassterm}) is the mass term of
the heavy Majorana right-handed leptons.
Thus, in the case of the Dirac and Majorana
mass term with the seesaw mass matrix in Eq.~(\ref{3mixingmatrix})
in the spectrum of masses there are
\begin{itemize}
\item Three light Majorana neutrino masses.
\item $N_{s}$ heavy Majorana masses, which characterize the scale of violation of the total lepton number.
\end{itemize}

The symmetric mass matrix $M^{L}$ of the light neutrinos
can be diagonalized with the transformation
\begin{equation}\label{diag}
M^{L} = U \, m \, U^{T}
,
\end{equation}
with a unitary matrix $U$
such that
$m_{ik}=m_{i}\delta_{ik}$,
where $m_{i}$ are the neutrino masses.
Then,
$U$ is the effective neutrino mixing matrix
neglecting the seesaw-suppressed mixing with the heavy Majorana leptons.
From Eqs.~(\ref{1seesaw}) and (\ref{diag}),
the neutrino masses are given by
\begin{equation}\label{2seesaw}
m_{i}=-\sum_{k}\frac{(U^{\dag}M^{D})^{2}_{ik}}{M_{k}}.
\end{equation}
From this relation it follows that the neutrino masses are much smaller than the masses of quarks and leptons which characterize the matrix $M^{D}$.
\section{Theory of $\beta\beta_{0\nu}$ decay}
\label{bb4}

In this Section we present the basic elements of the phenomenological theory of
neutrinoless double-$\beta$ ($\beta\beta_{0\nu}$) decay
of even-even nuclei (see Refs.~\citen{Doi:1985dx,Bilenky:1987ty}).

In the following derivation we assume that

\begin{itemize}

\item
The interaction Lagrangian is the Charged Current Lagrangian of the Standard Model
\begin{equation}\label{CC1}
\mathcal{L}_{I} (x) = -\frac{g}{2\,\sqrt{2}}\,
j^{CC}_{\alpha}(x) \, W^{\alpha}(x) + \mathrm{h.c.}
.
\end{equation}
Here
\begin{equation}\label{CC2}
j^{CC}_{\alpha}(x) = 2 \sum_{l=e,\mu,\tau} \overline{\nu_{lL}}(x)
\gamma_{\alpha} l_{L}(x) + j_{\alpha}(x)
,
\end{equation}
where $j_{\alpha}(x)$ is the hadronic charged current.

\item
We have the standard three-neutrino mixing in Eq.~(\ref{mixing}) between
the left-handed neutrino flavor fields $\nu_{lL}(x)$
($l=e,\mu,\tau$)
and three left-handed massive neutrino fields
$\nu_{iL}(x)$ with masses $m_{i}$
($i=1,2,3$).
The massive neutrino fields
$\nu_{i}(x) = \nu_{iL}(x) + C \overline{\nu_{iL}}^{T}(x)$
satisfy the Majorana condition
\begin{equation}\label{Mjcond}
\nu_{i}(x)
=
\nu^{c}_{i}(x)
=
C \, \overline{\nu_{i}}^{T}(x)
.
\end{equation}

\end{itemize}

From Eq.~(\ref{CC1}) it follows that the effective Hamiltonian of
$\beta$ decay is given by
\begin{equation}\label{effham}
{\mathcal{H}}_{I}(x)
=
\frac{G_{F}\cos\vartheta_{C}}{\sqrt{2}}
\,
2
\,
\overline{e_{L}}(x)
\,
\gamma_{\alpha}
\,
\nu_{eL}(x)
\,
j^{\alpha}(x) + \mathrm{h.c.}
\end{equation}
Here
$G_{F}$
is the Fermi constant
(with
$G_{F}/\sqrt{2}=g^{2}/8M^{2}_{W}$),
$\vartheta_{C}$ is the Cabibbo angle,
and $j^{\alpha}(x)$ is the $\Delta S=0$ hadronic charged current.

\begin{figure}[t!]
\begin{minipage}[t]{0.49\textwidth}
\begin{center}
\includegraphics*[width=0.5\linewidth]{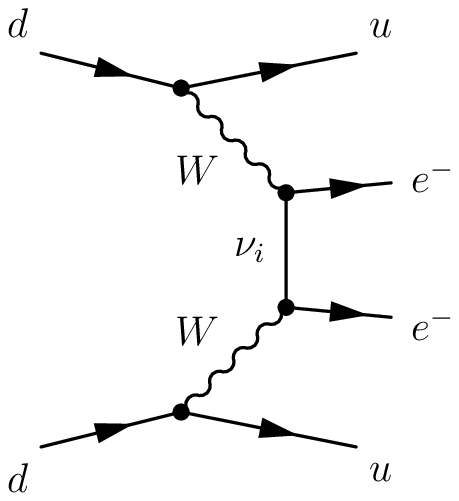}
\end{center}
\caption{ \label{fig:feybb}
Feynman diagram of the transition
$d d \to u u e^{-} e^{-}$
which induces
$\beta\beta_{0\nu}$ decay.
}
\end{minipage}
\hfill
\begin{minipage}[t]{0.49\textwidth}
\begin{center}
\includegraphics*[width=0.85\textwidth]{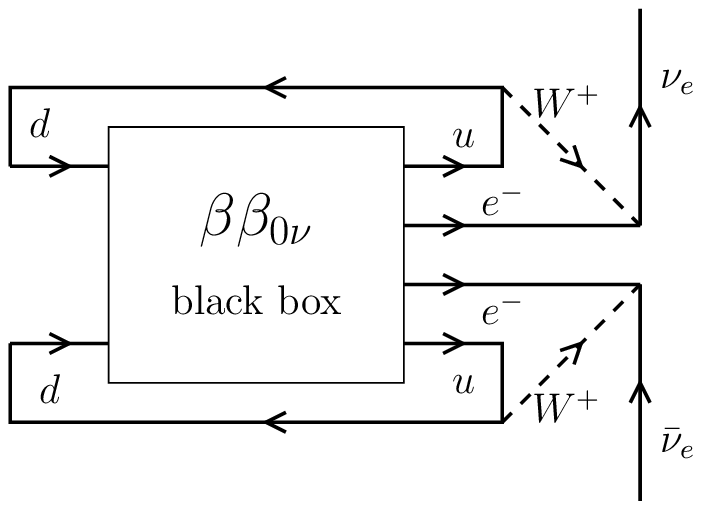}
\end{center}
\caption{ \label{fig:blackbox}
$ \bar\nu_{e} \to \nu_{e} $
transition diagram through a $\beta\beta_{0\nu}$ black box
\cite{Schechter:1982bd}.
}
\end{minipage}
\end{figure}

The Feynman diagram of the neutrinoless double-$\beta$ decay process is presented in Fig.~\ref{fig:feybb}.
One can see that it is a second order process in
$G_{F}$, with the propagation of virtual massive neutrinos.
The matrix element of the process is given by
\begin{align}
\langle f|S^{2}|i \rangle
=
\null & \null
4\frac{(-i)^{2}}{2!}
\left(\frac{G_{F}\cos\vartheta_{C}}{\sqrt{2}}\right)^{2}
N_{p_1}
N_{p_2}
\int
d^{4}x_{1}
d^{4}x_{2}
\,
\overline{u_{L}}(p_1)
e^{ip_{1} \cdot x_{1}}
\gamma_{\alpha}
\nonumber
\\
\null & \null
\times
\langle 0|T(\nu_{eL}(x_{1})\nu^{T}_{eL}(x_{2})|0\rangle
\gamma^{T}_{\beta}
\overline{u_{L}}^{T}(p_2)
e^{ip_{2} \cdot x_{2}}
\langle N_{f}|T(J^{\alpha}(x_{1})J^{\beta}(x_{2}))|N_{i} \rangle
\nonumber
\\
\null & \null
-
(p_{1}\rightleftarrows p_{2})
.
\label{Smatelem}
\end{align}
Here $p_{1}$ and $p_{2}$ are the electron momenta, $J^{\alpha}(x)$
is the weak charged current in the Heisenberg representation\footnote{
Thus, strong interactions are taken into account in Eq.~(\ref{Smatelem}).
},
$N_{i}$ and $N_{f}$ are the states of the initial and final nuclei
with respective four-momenta
$P_{i}=(E_{i}, \vec{p}_{i})$
and
$P_{f}=(E_{f}, \vec{p}_{f})$, and $N_{p}=1/(2\pi)^{3/2}\sqrt{2p^{0}}$ is the standard
normalization factor.

Let us consider the neutrino propagator.
Taking into account the Majorana condition in Eq.~(\ref{Mjcond}), we have
\begin{align}
\langle 0|T(\nu_{eL}(x_{1})
\nu^{T}_{eL}(x_{2})|0\rangle
=
\null & \null
-\frac{1-\gamma_{5}}{2}\sum_{i}U^{2}_{ei}\langle
0|T(\nu_{i}(x_{1})\bar\nu_{i}(x_{2}))|0\rangle
\,
\frac{1-\gamma_{5}}{2}
\,
C
\nonumber
\\
=
\null & \null
-\frac{i}{(2\pi)^{4}}\sum_{i}
\int
d^{4}q
\,
e^{-iq \cdot (x_{1}-x_{2})}\frac{U^{2}_{ei}m_{i}}{q^{2}-m^{2}_{i}}
\,
\frac{1-\gamma_{5}}{2}
\,
C
.
\label{nupropag1}
\end{align}
Thus,
the neutrino propagator is proportional to $m_{i}$.
It is obvious from Eq.~(\ref{nupropag1}) that this is connected with the fact that only left-handed
neutrino fields enter into the Hamiltonian of weak interactions.
In the case of massless neutrinos ($m_{i}=0$, with $i=1,2,3$), in accordance with the theorem on the equivalence of the theories with massless
Majorana and Dirac neutrinos (see Refs.~\citen{Ryan-Okubo-NCS-2-234-1964,Case:1957zza}), the matrix element of neutrinoless double-$\beta$ decay is equal to zero.

Let us consider the second term with $p_{1}\rightleftarrows p_{2}$
of the matrix element in Eq.~(\ref{Smatelem}).
We have
\begin{align}
\overline{u_{L}}(p_1)
\gamma_{\alpha}
(1-\gamma_{5})
\gamma_{\beta}
C \overline{u_{L}}^{T}(p_2)
=
\null & \null
\overline{u_{L}}(p_2)
C^{T}
\gamma^{T}_{\beta}
(1-\gamma^{T}_{5})
\gamma^{T}_{\alpha}
\overline{u}^{T}_{L}(p_1)
\nonumber
\\
\null & \null
-
\overline{u_{L}}(p_2)
\gamma_{\beta}
(1-\gamma_{5})
\gamma_{\alpha}C
\overline{u}^{T}_{L}(p_1)
.
\label{relation}
\end{align}
and
\begin{equation}\label{relation1}
T(J^{\beta}(x_{2})J^{\alpha}(x_{1}))=T(J^{\alpha}(x_{1})J^{\beta}(x_{2}))
.
\end{equation}
From Eqs.~(\ref{relation}) and (\ref{relation1}),
it follows that the second term of the matrix element in Eq.~(\ref{Smatelem}) is equal to the first one.
Hence, we obtain
\begin{align}
\langle f|S^{2}|i \rangle
=
\null & \null
4 \left (\frac{G_{F}\cos\vartheta_{C}}{\sqrt{2}}\right)^{2}N_{p_1}N_{p_2}
\int
d^{4}x_{1}
d^{4}x_{2}
\,
\overline{u_{L}}(p_1)
e^{ip_{1} \cdot x_{1}}
\gamma_{\alpha}
\nonumber
\\
\null & \null
\times
\frac{i}{(2\pi)^{4}}
\sum_{i}
U^{2}_{ei}
m_{i}
\int
d^{4}q
\,
\frac{e^{-iq \cdot (x_{1}-x_{2})}}{ p^{2}-m^{2}_{i}}
\nonumber
\\
\null & \null
\times
\frac{1-\gamma_{5}}{2}
\,
\gamma_{\beta}C
\overline{u_{L}}^{T}(p_2)e^{ip_{2} \cdot x_{2}}
\langle N_{f}|T(J^{\alpha}(x_{1})J^{\beta}(x_{2}))|N_{i} \rangle
.
\label{Smatelem1}
\end{align}
The most interesting nuclear transitions are listed in Tab.~\ref{tab:G0n}.
The calculation of the nuclear part of the matrix element of
the $\beta\beta_{0\nu}$ decay is a complicated nuclear problem
which is reviewed in the following Section~\ref{bb5}. In
such calculation different approximations are used.
In the following part of this Section we derive the form of
the matrix element of the $\beta\beta_{0\nu}$ decay which
is appropriate for the approximate nuclear calculations.

Let us perform in Eq.~(\ref{Smatelem1}) the integration over the
time variables $x^{0}_{2}$ and $x^{0}_{1}$.
We have 
\begin{equation}\label{integration}
\int^{\infty}_{-\infty} dx^{0}_{1} \int^{\infty}_{-\infty}
\ldots dx^{0}_{2}= \int^{\infty}_{-\infty} dx^{0}_{1}\left[\int_{-\infty}^{x^{0}_{1}}
\ldots dx^{0}_{2}+\int_{x^{0}_{1}}^{\infty}
\ldots dx^{0}_{2}\right]
.
\end{equation}
After the integration over $q^{0}$ in the neutrino propagator, we find in the region $x^{0}_{1}>x^{0}_{2}$\footnote{
It is assumed that in the propagator $m^{2}_{i} \to m^{2}_{i}-i\epsilon$.
}
\begin{equation}\label{nupropag4}
\frac{i}{(2\pi)^{4}}
\int\frac{e^{-iq \cdot (x_{1}-x_{2})}} {
q^{2}-m^{2}_{i}}d^{4}q=\frac{1}{(2\pi)^{3}}
\int\frac{e^{-iq^{0}_{i}(x^{0}_{1}-x^{0}_{2})+i\vec{q} \cdot
(\vec{x}_{1}-\vec{x}_{2})}}{2q_{i}^{0}}d^{3}q
,
\end{equation}
where
\begin{equation}\label{energy}
q_{i}^{0}=\sqrt{\vec{q}^{2}+m^{2}_{i}}
.
\end{equation}
In the region $x^{0}_{1}<x^{0}_{2}$ we have
\begin{equation}\label{nupropag5}
\frac{i}{(2\pi)^{4}}
\int
\frac{e^{-iq \cdot (x_{1}-x_{2})}}{q^{2}-m^{2}_{i}}
d^{4}q
=
\frac{1}{(2\pi)^{3}}
\int\frac{e^{-iq^{0}_{i}(x^{0}_{2}-x^{0}_{1})+i\vec{q} \cdot (\vec{x}_{2}-\vec{x}_{1})}}{2q_{i}^{0}}d^{3}q
.
\end{equation}
Furthermore, from the invariance under translations in time we have
\begin{equation}\label{Heisen1}
J^{\alpha}(x)=e^{iHx^{0}}J^{\alpha}(\vec{x})e^{-iHx^{0}},
\end{equation}
where $H$ is the total Hamiltonian and $J^{\alpha}(\vec{x})=J^{\alpha}(0,\vec{x})$.
From this
relation we find at $x^{0}_{1}>x^{0}_{2}$
\begin{align}
\langle N_{f}|T(J^{\alpha}(x_{1})
\null & \null
J^{\beta}(x_{2}))|N_{i} \rangle
=
\langle N_{f}|J^{\alpha}(x_{1})J^{\beta}(x_{2})|N_{i}
\rangle
\nonumber
\\
=
\null & \null
\sum_{n}
e^{i(E_{f}-E_{n})x^{0}_{1}}
e^{i(E_{n}-E_{i})x^{0}_{2}}
\langle N_{f}|J^{\alpha}(\vec{x}_{1})|N_{n}\rangle
\langle N_{n}|J^{\beta}(\vec{x}_{2}))|N_{i}\rangle
,
\label{phases}
\end{align}
where $|N_{n}\rangle$ is the vector of the state of the intermediate
nucleus with four-momentum $P_{n}=(E_{n}, \vec{p_{n}})$ and the sum
is over the total system of states $|N_{n}\rangle$.

In the region $x^{0}_{1}<x^{0}_{2}$ we obtain
\begin{align}
\langle N_{f}|T(J^{\alpha}(x_{1})
\null & \null
J^{\beta}(x_{2}))|N_{i} \rangle
=
\langle N_{f}|J^{\alpha}(x_{2})J^{\beta}(x_{1})|N_{i}
\rangle
\nonumber
\\
=
\null & \null
\sum_{n}
e^{i(E_{f}-E_{n})x^{0}_{2}}
e^{i(E_{n}-E_{i})x^{0}_{1}}
\langle N_{f}|J^{\alpha}(\vec{x}_{2})|N_{n}\rangle
\langle N_{n}|J^{\beta}(\vec{x}_{1}))|N_{i}\rangle
.
\label{phases1}
\end{align}
From Eqs.~(\ref{nupropag4}) and (\ref{phases}) we find
\begin{eqnarray}\label{phases3}
&&\int^{\infty}_{-\infty} dx^{0}_{1}\int_{-\infty}^{x^{0}_{1}}dx^{0}_{2}\langle
N_{f}|J^{\alpha}(\vec{x}_{1}) J^{\beta}(\vec{x}_{2})|N_{i} \rangle
e^{i(p^{0}_{1}x^{0}_{1}+p^{0}_{2}x^{0}_{2})} e^{iq_{i}^{0}(x^{0}_{2}-x^{0}_{1})}
\nonumber\\
&&=-i\sum_{n} \frac{\langle N_{f}|J^{\alpha}(\vec{x}_{1})|N_{n}\rangle\langle N_{n}|
J^{\beta}(\vec{x}_{2}))|N_{i}}{E_{n}+p^{0}_{2}+q^{0}_{i}-E_{i}-i\epsilon}
\,
2\pi\delta(E_{f}+p^{0}_{1}+p^{0}_{2}-E_{i})
.
\end{eqnarray}
Analogously, from (\ref{nupropag5}) and (\ref{phases1}) we obtain the following relation
\begin{eqnarray}\label{phases4}
&&\int^{\infty}_{-\infty} dx^{0}_{1}\int^{\infty}_{x^{0}_{1}}dx^{0}_{2}\langle
N_{f}|J^{\alpha}(\vec{x}_{2}) J^{\beta}(\vec{x}_{1})|N_{i} \rangle
e^{i(p^{0}_{1}x^{0}_{1}+p^{0}_{2}x^{0}_{2})} e^{iq_{i}^{0}(x^{0}_{2}-x^{0}_{1})}
\nonumber\\
&&=-i\sum_{n} \frac{\langle N_{f}|J^{\beta}(\vec{x}_{2})|N_{n}\rangle\langle N_{n}|
J^{\alpha}(\vec{x}_{1}))|N_{i}}{E_{n}+p^{0}_{1}+q^{0}_{i}-E_{i}-i\epsilon}
\,
2\pi\delta(E_{f}+p^{0}_{1}+p^{0}_{2}-E_{i})
.
\end{eqnarray}
In Eqs.~(\ref{phases3}) and (\ref{phases4}) we used the relations
\begin{align}
\null & \null
\int_{-\infty}^{0}e^{i a x^{0}_{2}}
\,
dx^{0}_{2}
\to \int_{-\infty}^{0}e^{i (a-i\epsilon) x^{0}_{2}}
\,
dx^{0}_{2}
=\lim_{\epsilon\to 0}\frac{-i}{a-i\epsilon}
,
\label{phases11}
\\
\null & \null
\int^{-\infty}_{0}e^{i a x^{0}_{2}}
\,
dx^{0}_{2}
\to \int^{\infty}_{0}e^{i (a+i\epsilon) x^{0}_{2}}
\,
dx^{0}_{2}
=\lim_{\epsilon\to 0}\frac{i}{a+i\epsilon}
,
\label{phases12}
\end{align}
which are based on the assumption that at $t \to \pm \infty$ the interaction is turned off.

Taking into account all these relations,
we obtain
\begin{align}
\null & \null
\langle f|S^{2}|i \rangle
=
-2i
\left(\frac{G_{F}\cos\vartheta_{C}}{\sqrt{2}}\right)^{2}N_{p_1}N_{p_2}
\overline{u}(p_1)\gamma_{\alpha}\gamma_{\beta}(1+\gamma_{5})C \overline{u}^{T}(p_2)
\int
d^{3}x_{1}
d^{3}x_{1}
\nonumber
\\
\times
\null & \null
e^{-i\vec{p_{1}}\cdot\vec{x}_{1}-i\vec{p_{2}}\cdot\vec{x}_{2}}
\sum_{i}
U^{2}_{ei}
m_{i}
\frac{1}{(2\pi)^{3}}
\int
d^{3}q
\,
\frac{e^{i\vec{q}\cdot(\vec{x_{1}}-\vec{x_{2}})}}{q_{i}^{0}}
\,
2\pi\delta(E_{f}+p^{0}_{1}+p^{0}_{2}-E_{i})
\nonumber
\\
\times
\null & \null
\left[
\sum_{n}
\frac{
\langle N_{f}|J^{\alpha}(\vec{x}_{1})|N_{n}\rangle
\langle N_{n}|J^{\beta}(\vec{x}_{2})|N_{i}\rangle
}{E_{n}+p^{0}_{2}+q^{0}_{i}-E_{i}-i\epsilon}
+
\sum_{n}
\frac{
\langle N_{f}|J^{\beta}(\vec{x_{2}})|N_{n}\rangle
\langle N_{n}|J^{\alpha}(\vec{x_{1}})|N_{i}\rangle
}{E_{n}+p^{0}_{1}+q^{0}_{i}-E_{i}-i\epsilon}
\right]
.
\label{Smatelem2}
\end{align}
This equation gives the exact expression for the matrix
element of $\beta\beta_{0\nu}$ decay in the second order of
perturbation theory. 
In the following we consider the $0^{+} \to 0^{+}$ ground state to ground state
transitions of even-even nuclei as those in Tab.~\ref{tab:G0n}.
For these transitions the following
approximations are standard \cite{Doi:1985dx,Bilenky:1987ty}:

\begin{enumerate}

\item Small neutrino masses can be safely
neglected in the expression for the neutrino energy $q^{0}_{i}$.

In fact, from the uncertainty relation for the average neutrino momentum we have
$|\vec{q}| \simeq 1/r$, where $r$ is
the average distance between two nucleons in the nucleus.
Taking into account that $r \simeq 10^{-13}$ cm,
we obtain $|\vec{q}| \simeq 100$ MeV.
For the neutrino masses we have $m_{i} \lesssim 1$ eV.
Thus, $|\vec{q}|^{2} \gg m^{2}_{i}$ and we have
$q^{0}_{i}=\sqrt{\vec{q}^{2}+m^{2}_{i}} \simeq |\vec{q}|$.

\item Long-wave approximation.

For the two emitted electrons
$|\vec{p}_{k}\cdot\vec{x}_{k}| \leq |\vec{p}_{k}| R$ for $k=1,2$, where
$R \simeq 1.2 \, A^{1/3} \times 10^{-13}$ cm is the radius of the nucleus.
Thus, $|\vec{p}_{k}| R \simeq 10^{-2} A^{1/3} |\vec{p}_{k}| \mathrm{MeV}^{-1}$.
Taking into account that
$|\vec{p}_{k}| \lesssim 1$ MeV, we obtain $|\vec{p}_{k}\cdot\vec{x}_{k}| \ll 1$ and
$e^{-i\vec{p}_{k}\cdot\vec{x}_{k}} \simeq 1$.
Thus, the two electrons are emitted predominantly in the $S$-state.

\item Closure approximation.

The energy of the virtual neutrino ($|\vec{q}| \simeq 100$ MeV) is much larger than
the excitation energy of the intermediate states:
$|\vec{q}| \gg (E_{n}-E_{i})$.
Taking into account this inequality,
we can replace the energies of the intermediate states $E_{n}$
with the average energy $\overline{E}$.
With this approximation we can perform the sum over total system of intermediate states:
\begin{equation}\label{Smatelem31}
\sum_{n}
\frac{
\langle N_{f}|J^{\alpha}(\vec{x_{1}})|N_{n} \rangle
\langle N_{n}|J^{\beta}(\vec{x_{2}}))|N_{i} \rangle
}{E_{n}+p^{0}_{2}+q^{0}_{i}-E_{i}-i\epsilon}
\simeq
\frac{
\langle N_{f}|J^{\alpha}(\vec{x_{1}}) J^{\beta}(\vec{x_{2}}))|N_{i} \rangle
}{\overline{E}+p^{0}_{2}+|\vec{q}|-E_{i}-i\epsilon}
.
\end{equation}
Furthermore, neglecting the nuclear recoil, in the laboratory system we have
\begin{equation}
M_{i}=M_{f}+p^{0}_{2}+p^{0}_{1}
,
\label{norecoil}
\end{equation}
where $M_{i}$ and $M_{f}$ are masses of the initial and final nuclei.
Using these relations,
the energy denominators in Eq.~(\ref{Smatelem2}) become
\begin{equation}\label{edenom}
\overline{E}+ |\vec{q}| +p^{0}_{1,2}-M_{i}=\overline{E} + |\vec{q}| \pm
\frac{p^{0}_{1}-p^{0}_{2}}{2}-\frac{M_{i}+M_{f}}{2}\simeq
\overline{E}+ |\vec{q}| -\frac{M_{i}+M_{f}}{2},
\end{equation}
where we took into account that $(p^{0}_{1}-p^{0}_{2})/2 \ll |\vec{q}|$.

After all these approximations the matrix element of $\beta\beta_{0\nu}$ decay takes the form
\begin{align}
\null & \null
\langle f|S^{2}|i \rangle
=
-2i
\left(\frac{G_{F}\cos\vartheta_{C}}{\sqrt{2}}\right)^{2}
m_{\beta\beta}
N_{p_1}
N_{p_2}
\bar u(p_1)\gamma_{\alpha} \gamma_{\beta} (1+\gamma_{5}) C \bar u^{T}(p_2)
\nonumber
\\
\times
\null & \null
\int
d^{3}x_{1}
d^{3}x_{2}
\,
\frac{1}{(2\pi)^{3}}
\int d^{3}q
\,
\frac{e^{i\vec{q} \cdot (\vec{x}_{1}-\vec{x}_{2})}}
{|\vec{q}|\left(|\vec{q}|+\overline{E} -\frac{M_{i}+M_{f}}{2}\right)}
\,
\langle N_{f}|(J^{\alpha}(\vec{x}_{1})J^{\beta}(\vec{x}_{2})
\nonumber
\\
\null & \null
\hspace{2cm}
+
J^{\beta}(\vec{x}_{2})J^{\alpha}(\vec{x}_{1}))|N_{i}\rangle
\,
2\pi
\,
\delta(M_{f}+p^{0}_{1}+p^{0}_{2}-M_{i})
,
\label{Smatelem4}
\end{align}
with the effective Majorana neutrino mass
in Eq.~(\ref{effMaj}).

Let us stress that the fact that
the matrix element of $\beta\beta_{0\nu}$ decay is proportional to the
effective Majorana neutrino mass
is a consequence of the fact that only the left-handed neutrino fields
enter in the leptonic current.

The expression in Eq.~(\ref{Smatelem4}) of the matrix element of
$\beta\beta_{0\nu}$ decay can be further simplified.
In fact, we have
\begin{equation}\label{gamrel}
\gamma_{\alpha}\gamma_{\beta}
=
g_{\alpha\beta}
+
\frac{1}{2}
\left(
\gamma_{\alpha}\gamma_{\beta}
-
\gamma_{\beta}\gamma_{\alpha}
\right)
.
\end{equation}
Since the hadronic part of Eq.~(\ref{Smatelem4}) is symmetric under the exchange
$\alpha \leftrightarrows \beta$,
the second term of Eq.~(\ref{gamrel}) does not contribute to the matrix element of the process.
Thus, we can simplify Eq.~(\ref{gamrel}) to
\begin{align}
\langle f|S^{2}|i \rangle
=
\null & \null
-2i
m_{\beta\beta}
\left(\frac{G_{F}\cos\vartheta_{C}}{\sqrt{2}}\right)^{2}
N_{p_1}
N_{p_2}
\bar u(p_1)\gamma_{\alpha} \gamma_{\beta} (1+\gamma_{5}) C \bar u^{T}(p_2)
\nonumber
\\
\null & \null
\times
\int
d^{3}x_{1}
d^{3}x_{2}
\,
\frac{1}{(2\pi)^{3}}
\int d^{3}q
\,
\frac{e^{i\vec{q} \cdot (\vec{x}_{1}-\vec{x}_{2})}}
{|\vec{q}|\left(|\vec{q}|+\overline{E} -\frac{M_{i}+M_{f}}{2}\right)}
\nonumber
\\
\null & \null
\hspace{0.5cm}
\times
\langle N_{f}|(J^{\alpha}(\vec{x}_{1})J^{\beta}(\vec{x}_{2})|N_{i}\rangle
\,
2\pi
\,
\delta(M_{f}+p^{0}_{1}+p^{0}_{2}-M_{i})
.
\label{Smatelem5}
\end{align}

\item The impulse approximation.

In the impulse approximation the hadronic charged current has the form
\begin{equation}\label{impapr}
J^{\alpha}(\vec{x})
=
\sum_{n}
\delta(\vec{x}-\vec{r}_{n})
\,
\tau^{n}_{+}
\left(
g^{\alpha 0}J_{n}^{0}(q^{2})
+
g^{\alpha k}J_{n}^{k}(q^{2})
\right)
,
\end{equation}
where
\begin{equation}\label{impapr1}
J_{n}^{0}(q^{2})
=
g_{V}(q^{2})
,
\quad
\vec{J}_{n}(q^{2})
=
g_{A}(q^{2})
\vec{\sigma}_{n}
+
i g_{M}(q^{2})
\frac{\vec{\sigma}_{n} \times \vec{q}}{2M}
-
g_{P}(q^{2})
\frac{\vec{\sigma}_{n}\cdot\vec{q}}{2M}
\vec{q}
.
\end{equation}
Here $g_{V}(q^{2})$, $g_{A}(q^{2})$ and $g_{M}(q^{2})$ are CC vector, axial and magnetic form factors of the nucleon.
We have
$g_{V}(0)=1$,
$g_{A}(0)=g_{A}\simeq 1.27$
and
$g_{M}(0)=\mu_{p}-\mu_{n}$,
where
$\mu_{p}$ and $\mu_{n}$
are the anomalous magnetic moments of the proton and the neutron.
From PCAC it follows that the pseudoscalar form factor is given by
$g_{P}(q^{2}) = 2 M g_{A} / (q^{2}+m^{2}_{\pi})$.

\end{enumerate}

After the integration over $\vec{x}_{1}$ and $\vec{x}_{2}$, from Eqs.~(\ref{Smatelem5}) and (\ref{impapr})
we obtain
\begin{align}
\langle f|S^{2}|i \rangle
=
\null & \null
2i
m_{\beta\beta}
\left(\frac{G_{F}\cos\vartheta_{C}}{\sqrt{2}}\right)^{2}
\frac{1}{(2\pi)^{3}\sqrt{p^{0}_{1}p^{0}_{2}}}
\,
\bar u(p_1)(1+\gamma_{5})C \bar u^{T}(p_2)
\nonumber
\\
\null & \null
\times
\sum_{n,m}
\frac{1}{(2\pi)^{3}}
\int d^{3}q
\,
\frac{e^{i\vec{q} \cdot (\vec{r}_{n}-\vec{r}_{m})}}{|\vec{q}|(|\vec{q}|+\overline{E}-\frac{M_{i}+M_{f}}{2})}
\nonumber
\\
\null & \null
\times
\langle N_{f}|\sum_{n,m}\tau^{n}_{+}\tau^{m}_{+}(J_{n}^{0}J_{m}^{0}-\vec{J}_{n}\vec{J}_{m})|N_{i}\rangle
2\pi\delta(M_{f}+p^{0}_{1}+p^{0}_{2}-M_{i})
.
\label{Smatelem3}
\end{align}
In this expression we can perform the integration over the angles of the vector $\vec{q}$:
\begin{equation}\label{int1}
\frac{1}{(2\pi)^{3}}
\int
\frac{e^{i\vec{q}\cdot\vec{r}_{nm}}d^{3}q}{|\vec{q}|(|\vec{q}|+\overline{E}-\frac{1}{2}(M_{i}+M_{f}))}
=
\frac{1}{2\pi^{2} r_{nm}}
\int^{\infty}_{0}
\frac{\sin(|\vec{q}| r_{nm}) \, d|\vec{q}|}{|\vec{q}|+\overline{E}-\frac{1}{2}(M_{i}+M_{f})},
\end{equation}
where $\vec{r}_{nm}=\vec{r}_{n}-\vec{r}_{m}$.

Considering only the major contributions to the nuclear matrix elements given by the vector and axial terms\footnote{Smaller contributions of the pseudoscalar and magnetic terms are usually also taken into account in the calculations of the nuclear matrix elements (see Ref.~\citen{Simkovic:1999re}).}
in the hadronic charged current (\ref{impapr1}),
we obtain
\begin{align}
\langle f|S^{2}|i
=
\null & \null
-i
m_{\beta\beta}
\left(\frac{G_{F}\cos\vartheta_{C}}{\sqrt{2}}\right)^{2}
\frac{1}{(2\pi)^{3} \sqrt{p^{0}_{1}p^{0}_{2}}}
\,
\frac{1}{R}
\,
\bar{u}(p_1) (1+\gamma_{5}) C \bar{u}^{T}(p_2)
\nonumber
\\
\null & \null
\times
M^{0\nu}
\delta(p^{0}_{1}+p^{0}_{2}+M_{f}-M_{i})
,
\label{Smat6}
\end{align}
where $R$ is the radius of the nucleus and
\begin{equation}\label{NME}
M^{0\nu}=g^{2}_{A}
\left(
M^{0\nu}_{GT} - \frac{1}{g^{2}_{A}} \, M^{0\nu}_{F}
\right)
\end{equation}
is the nuclear matrix element.
Here,
\begin{equation}\label{GTme}
M^{0\nu}_{GT}= \langle
\Psi_{f}|
\sum_{n,m}
H(r_{n,m},\overline{E})
\,
\tau^{n}_{+}\tau^{m}_{+}
\,
\vec{\sigma}^{n}\cdot\vec{\sigma}^{m})|\Psi_{i}
\rangle
\end{equation}
is the (axial) Gamow-Teller matrix element and
\begin{equation}\label{Fme}
M^{0\nu}_{F}= \langle
\Psi_{f}|\sum_{n,m}H(r_{n,m},\overline{E})
\,
\tau^{n}_{+}\tau^{m}_{+}|\Psi_{i}
\rangle
\end{equation}
is the (vector) Fermi matrix element.
The function $H(r_{n,m},\overline{E}))$ is given by
\begin{equation}\label{2neuprop}
H(r_{n,m},\overline{E}))
=
\frac{2R}{\pi r_{nm}}
\int^{\infty}_{0} \frac{\sin(|\vec{q}| r_{nm}) \, d|\vec{q}|}{|\vec{q}|+\overline{E}-\frac{1}{2}(M_{i}+M_{f})}
,
\end{equation}
and
$|\Psi_{i}\rangle$ and $|\Psi_{i}\rangle$ are the wave functions of the initial and final nuclei.

Thus, the matrix element of $\beta\beta_{0\nu}$ decay is a
product of the effective Majorana mass $m_{\beta\beta}$, the electron
matrix element and the nuclear matrix element which includes the neutrino
propagator.
Taking into account that
$|\vec{q}| \gg \overline{E}-(M_{i}+M_{f})/2$,
for the function $H(r)$ (neutrino potential) we obtain the approximate expression
\begin{equation}\label{hfunc2}
H(r) \simeq \frac{2R}{\pi r} \int^{\infty}_{0} \frac{\sin(|\vec{q}| r)}{|\vec{q}|} \, d|\vec{q}|
=
\frac{R}{r}
.
\end{equation}
Let us now calculate the probability of $\beta\beta_{0\nu}$ decay.
We have
\begin{align}
\sum_{r_{1},r_{2}}
\left|
\bar{u}^{r_{1}}(p_1) (1+\gamma_{5}) C (\bar{u}^{r_{2}}(p_2))^{T}
\right|^{2}
=
\null & \null
\text{Tr}
\left[
(1+\gamma_{5})
( \gamma \cdot p_{2} + m_{e} )
(1-\gamma_{5})
( \gamma \cdot p_{1} - m_{e} )
\right]
\nonumber
\\
=
\null & \null
8 \, p_{1} \cdot p_{2}
.
\label{prob}
\end{align}
From Eqs.~(\ref{Smat6}) and (\ref{prob}), we find that the decay rate of
$\beta\beta_{0\nu}$ decay is given by
\begin{align}
d\Gamma^{0\nu}
=
\null & \null
|m_{\beta\beta}|^{2}
\,
|M^{0\nu}|^{2}
\,
\frac{4\left(G_{F}\cos\vartheta_{C}\right)^4}{(2\pi)^{5} R^{2}}
\left(
E_{1}E_{2} - p_{1}p_{2} \cos\theta \right)
\nonumber
\\
\null & \null
\times
F(E_{1},(Z+2))
F(E_{2},(Z+2))
|\vec{p}_{1}|
|\vec{p}_{2}|
\sin\theta
\,
d\theta
\,
dE_{1}
.
\label{decrate}
\end{align}
Here $E_{1,2} \equiv p^{0}_{1,2}$ are the energies of the two emitted electrons
(such that $E_{1}+E_{2}=M_{i}-M_{f}$)
and $\theta$ is the angle between the two electron
momenta $\vec{p}_{1}$ and $\vec{p}_{2}$.
The function $F(E,Z)$ describes final state electromagnetic interaction of the electron and the nucleus.
For a point-like nucleus it is given by the Fermi function
\begin{equation}\label{Ffunc}
F(E,Z)
\simeq
\frac{2\pi\eta}{1-e^{-2\pi\eta}}
,
\quad
\eta = Z \alpha \, \frac{m_{e}}{p}
.
\end{equation}
From Eq.~(\ref{decrate}) it follows that for the ultra relativistic electrons
the decay rate is proportional to $(1-\cos\theta)$
and is equal to zero at $\theta=0$.
This is due to the fact that the high-energy electrons
produced in weak decays have negative helicity.
If the two electrons are emitted in the same direction,
the projection of their total angular momentum on the direction of the momentum is equal to $-1$.
In $0^{+}\to0^{+}$ nuclear transitions
this configuration is forbidden by angular momentum conservation.

From Eq.~(\ref{decrate}),
for the half-life of $\beta\beta_{0\nu}$ decay we obtain
\begin{equation}
(T^{0\nu}_{1/2})^{-1}
=
\dfrac{\Gamma^{0\nu}}{\ln2}
=
|m_{\beta\beta}|^{2}
\,
|M^{0\nu}|^{2}
\,
G^{0\nu}(Q,Z)
,
\label{totrate}
\end{equation}
with the phase-space factor\footnote{
An additional factor 1/2 is due to the fact that in the final state we have two identical electrons.
}
\begin{equation}\label{Gfac}
G^{0\nu}(Q,Z)
=
\dfrac{\left(G_{F}\cos\vartheta_{C}\right)^4}{\ln2 \left(2\pi\right)^{5} R^{2}}
\int_{0}^{Q} \text{d}T_{1}
E_{1} |\vec{p}_{1}|
E_{2} |\vec{p}_{2}|
F(E_{1},Z')
F(E_{2},Z')
.
\end{equation}
Here $T_{1}=E_{1}-m_{e}$
is the kinetic energy of one of the two emitted electrons,
$Z' = Z+2$,
and
\begin{equation}
Q=M_{i}-M_{f}-2m_{e}
\label{Qvalue}
\end{equation}
is the $Q$-value of the process,
i.e. the
total released kinetic energy.

\begin{table}[b!]
\caption{\label{tab:G0n}
Double-$\beta$ decays for which $G^{0\nu}(Q,Z)$ has been calculated in Ref.~\citen{Kotila:2012zza}.
The columns give
the $\beta\beta^{-}$ nuclear decay,
$Q_{\beta\beta}$,
$G^{0\nu}(Q,Z)$ and
the natural abundance.
}
\centering
{
\begin{tabular}{cccc}
$\beta\beta^{-}$ decay	&$Q_{\beta\beta}$	&$G^{0\nu}(Q,Z)$				&nat.~abund.		\\
			&[keV]			&[$10^{-26}\,\text{y}^{-1}\,\text{eV}^{-2}$]	&[\%]			\\
\hline
$\nCa{48} \to\nTi{48} $	&$4272.26 \pm 4.04$	&9.501						&$0.187 \pm 0.021$	\\
$\nGe{76} \to\nSe{76} $	&$2039.061\pm 0.007$	&0.9049						&$7.73 \pm 0.12$	\\
$\nSe{82} \to\nKr{82} $	&$2995.12 \pm 2.01$	&3.891						&$8.73 \pm 0.22$	\\
$\nZr{96} \to\nMo{96} $	&$3350.37 \pm 2.89$	&7.881						&$2.80 \pm 0.09$	\\
$\nMo{100}\to\nRu{100}$	&$3034.40 \pm 0.17$	&6.097						&$9.82 \pm 0.31$	\\
$\nPd{110}\to\nCd{110}$	&$2017.85 \pm 0.64$	&1.844						&$11.72 \pm 0.09$	\\
$\nCd{116}\to\nSn{116}$	&$2813.50 \pm 0.13$	&6.396						&$7.49 \pm 0.18$	\\
$\nSn{124}\to\nTe{124}$	&$2286.97 \pm 1.53$	&3.462						&$5.79 \pm 0.05$	\\
$\nTe{128}\to\nXe{128}$	&$865.87 \pm 1.31$	&0.2251						&$31.74 \pm 0.08$	\\
$\nTe{130}\to\nXe{130}$	&$2526.97 \pm 0.23$	&5.446						&$34.08 \pm 0.62$	\\
$\nXe{136}\to\nBa{136}$	&$2457.83 \pm 0.37$	&5.584						&$8.8573\pm 0.0044$	\\
$\nNd{148}\to\nSm{148}$	&$1928.75 \pm 1.92$	&3.868						&$5.756 \pm 0.021$	\\
$\nNd{150}\to\nSm{150}$	&$3371.38 \pm 0.20$	&24.14						&$5.638 \pm 0.028$	\\
$\nSm{154}\to\nGd{154}$	&$1251.03 \pm 1.25$	&1.155						&$22.75 \pm 0.29$	\\
$\nGd{160}\to\nDy{160}$	&$1729.69 \pm 1.26$	&3.661						&$21.86 \pm 0.19$	\\
$\nPt{198}\to\nHg{198}$	&$1047.17 \pm 3.11$	&2.894						&$7.36 \pm 0.13$	\\
$\nTh{232}\to\nU{232} $	&$842.15 \pm 2.46$	&5.335						&$100$			\\
$\nU{238} \to\nPu{238}$	&$1144.98 \pm 1.25$	&12.87						&$99.274$		\\
\hline
\end{tabular}
}
\end{table}

The total rate of $\beta\beta_{0\nu}$ decay is the product of
three factors:
\begin{enumerate}

\item
The squared modulus of the effective Majorana mass $m_{\beta\beta}$ given in Eq.~(\ref{effMaj}).
The purpose of $\beta\beta_{0\nu}$-decay experiments
is to probe the Majorana nature of massive neutrinos
and, if the process is observed,
to extract the value of $|m_{\beta\beta}|$
from measurements of
$T^{0\nu}_{1/2}$
assuming that the other two factors as known.
In Section~\ref{bb6}
we discuss the implications for $|m_{\beta\beta}|$
of the values of the neutrino squared mass differences
and mixing angles
measured in neutrino oscillation experiments
(see Section~\ref{bb2}).

\item
The squared modulus of the nuclear matrix element $M^{0\nu}$,
which must be calculated on the basis of our knowledge of nuclear physics.
Unfortunately,
nuclear physics effects can be calculated only with nuclear models
which describe approximately the many-body interactions
of nucleons in nuclei.
Since different models have been developed
for the description of different aspects of nuclear physics
and the calculations rely on numerical approximations
of the many-body interactions and on truncation of the
large set of excitations of the nuclear states,
there are big differences between different calculations
which imply a large theoretical uncertainty for the nuclear matrix elements.
In the following Section~\ref{bb5}
we present a comparative review of the results of recent calculations.

\item
The phase-space factor $G^{0\nu}(Q,Z)$,
which must also be calculated.
However,
the calculation is much easier,
since $G^{0\nu}(Q,Z)$ is given by Eq.~(\ref{Gfac}).
Table~\ref{tab:G0n} gives the values of $G^{0\nu}(Q,Z)$ calculated
recently with high accuracy in Ref.~\citen{Kotila:2012zza}
for several nuclei.
The calculation was made considering
the exact Dirac wave functions of the electrons,
taking into account the finite nuclear size and electron screening.

\end{enumerate}

In this review we consider only
$\beta\beta_{0\nu}$ decay due to light neutrino masses
through the effective Majorana mass $m_{\beta\beta}$,
but there are many mechanisms in models beyond the Standard Model
which can generate $\beta\beta_{0\nu}$ decay
through new interactions and/or the exchange of new particles
(see Refs.~\citen{Faessler:1999zg,Choi:2002bb,Ibarra:2010xw,Tello:2010am,Rodejohann:2011mu,delAguila:2012nu,Deppisch:2012nb,deGouvea:2013zba}).
However,
it is important to realize that
in any case the existence of $\beta\beta_{0\nu}$ decay
implies that neutrinos are Majorana particles
\cite{Schechter:1982bd,Nieves:1984sn,Takasugi:1984xr}.
The reason is that
one can consider the $\beta\beta_{0\nu}$ elementary interaction process
$d d \to u u e^{-} e^{-}$
as generated by the black box depicted in Fig.~\ref{fig:blackbox},
which can include any mechanism.
Then, as shown in Fig.~\ref{fig:blackbox} \cite{Schechter:1982bd},
the external lines of the black box representing the interacting particles
can be arranged to form a diagram which generates
$\bar\nu_{e}\to\nu_{e}$
transitions.
This diagram contributes to the Majorana mass of the electron neutrino
through radiative corrections at some order of perturbation theory,
even if there is no tree-level Majorana neutrino mass term.
Furthermore,
unless there is a global symmetry which forbids a $\nu_{e}$ Majorana mass term,
it is very unlikely that this diagram and its variations
(for example, exchanging a photon between two lines \cite{Schechter:1982bd})
are canceled exactly by the contributions of other diagrams.
Since it can be proved that the existence of a global symmetry which forbids a $\nu_{e}$ Majorana mass term would forbid also $\beta\beta_{0\nu}$ decay
\cite{Nieves:1984sn,Takasugi:1984xr}
(see Section~14.3.2 of Ref.~\citen{Giunti:2007ry}),
the Majorana nature of neutrinos is a sufficient and necessary condition for
the existence of $\beta\beta_{0\nu}$ decay.
This is an important fact which must be kept in mind
to appreciate the power of $\beta\beta_{0\nu}$-decay experiments:
in spite of the nuclear physics uncertainty,
an unambiguous observation of $\beta\beta_{0\nu}$ decay
would represent a proof that neutrinos are Majorana particles!

It must be however clear that if
$\beta\beta_{0\nu}$ decay
is not generated by neutrino masses at the tree level
and a $\nu_{e}$ Majorana mass term is generated by radiative corrections
through the black-box mechanism,
the corresponding Majorana mass is much smaller than the value of $|m_{\beta\beta}|$
which would generate the same $\beta\beta_{0\nu}$ decay,
since the amplitude of the process in Fig.~\ref{fig:blackbox}
is suppressed at least by a factor $G_{F}^{2}$ with respect to the $\beta\beta_{0\nu}$-decay amplitude
corresponding to the black box.
Indeed,
the explicit calculation in Ref.~\citen{Duerr:2011zd}
shows that
a $\beta\beta_{0\nu}$ decay amplitude
corresponding to a value of $|m_{\beta\beta}|$ of the order of $10^{-1} \, \text{eV}$
generates radiatively a $\nu_{e}$ Majorana mass
of the order of $10^{-24} \, \text{eV}$,
which is many orders of magnitude smaller.

A simple positive measurement of $\beta\beta_{0\nu}$ decay
is not sufficient to determine if it is generated by light neutrino masses
with the rate is given by Eq.~(\ref{totrate}),
but more complex measurements may reveal
the dominant mechanism of $\beta\beta_{0\nu}$ decay.
Possible techniques which have been proposed are:

\begin{itemize}

\item
A comparison of the measured decay rates of different nuclei
\cite{Hirsch:1994es,Bilenky:2004um,Deppisch:2006hb,Gehman:2007qg}.
Obviously,
this method can succeed only
if the theoretical uncertainty of the corresponding nuclear matrix elements is not too large.

\item
The determination of the angular and energy
distribution of the two outgoing electrons
\cite{Doi:1983wv,Tomoda:1986yz,Ali:2007ec,Arnold:2010tu}.
In the recent past,
the
TGV\cite{Brudanin:2000in},
ELEGANT \cite{Ejiri:2001fx,Ogawa:2004fy}
and
NEMO \cite{Arnold:2005rz,Arnold:2013dha}
experiments
were able to make such measurements through spectroscopic observations of the two electrons.
Their sensitivity will be improved by the SuperNEMO experiment
\cite{Arnold:2010tu,Barabash:2011aa}.

\end{itemize}

Nevertheless,
we think that if $\beta\beta_{0\nu}$ decay is found with a rate
which corresponds to an effective Majorana mass $m_{\beta\beta}$
which is compatible with the predictions that can be obtained from the measured
neutrino masses and mixing
(see Section~\ref{bb6}),
then
it would be very likely that it is generated
by the neutrino masses,
which is the simplest and most natural mechanism given the known massive nature of neutrinos.
\section{Nuclear matrix elements}
\label{bb5}

The main purpose of neutrinoless double-$\beta$ decay experiments
is to prove the Majorana nature of massive neutrinos
and to measure the effective Majorana mass in Eq.~(\ref{effMaj})
if the $\beta\beta_{0\nu}$ decay is generated by the known light neutrino masses.
In this case,
the value of the effective Majorana mass
can be obtained from a measurement of the
neutrinoless double-$\beta$ decay rate in Eq.~(\ref{totrate})
only if the
nuclear matrix element (NME)
$|M^{0\nu}|$
is known with a sufficient accuracy.
Unfortunately,
since
$|M^{0\nu}|$
cannot be measured independently\footnote{
Note however that if the rates of the $\beta\beta_{0\nu}$ decays of at least two
different nuclei are measured,
the ratio of the corresponding NMEs can be determined from the data.
This could allow to test the nuclear model calculations of the NMEs
\cite{Bilenky:2002ga}.
},
it must be calculated.
This is a complicated nuclear many-body problem
whose discussion is beyond the scope of this review
(see Refs.~\citen{Tomoda:1990rs,Suhonen:1998ck,Faessler:1999zg,Vergados:2002pv,Caurier:2004gf,Vergados:2012xy}).
In this Section we present briefly the results of recent calculations
of the NMEs of several nuclei
and a discussion of the uncertainties of the values of the NMEs
which are important for the interpretation of the experimental data.

\begin{figure}[t!]
\begin{center}
\includegraphics*[width=\textwidth]{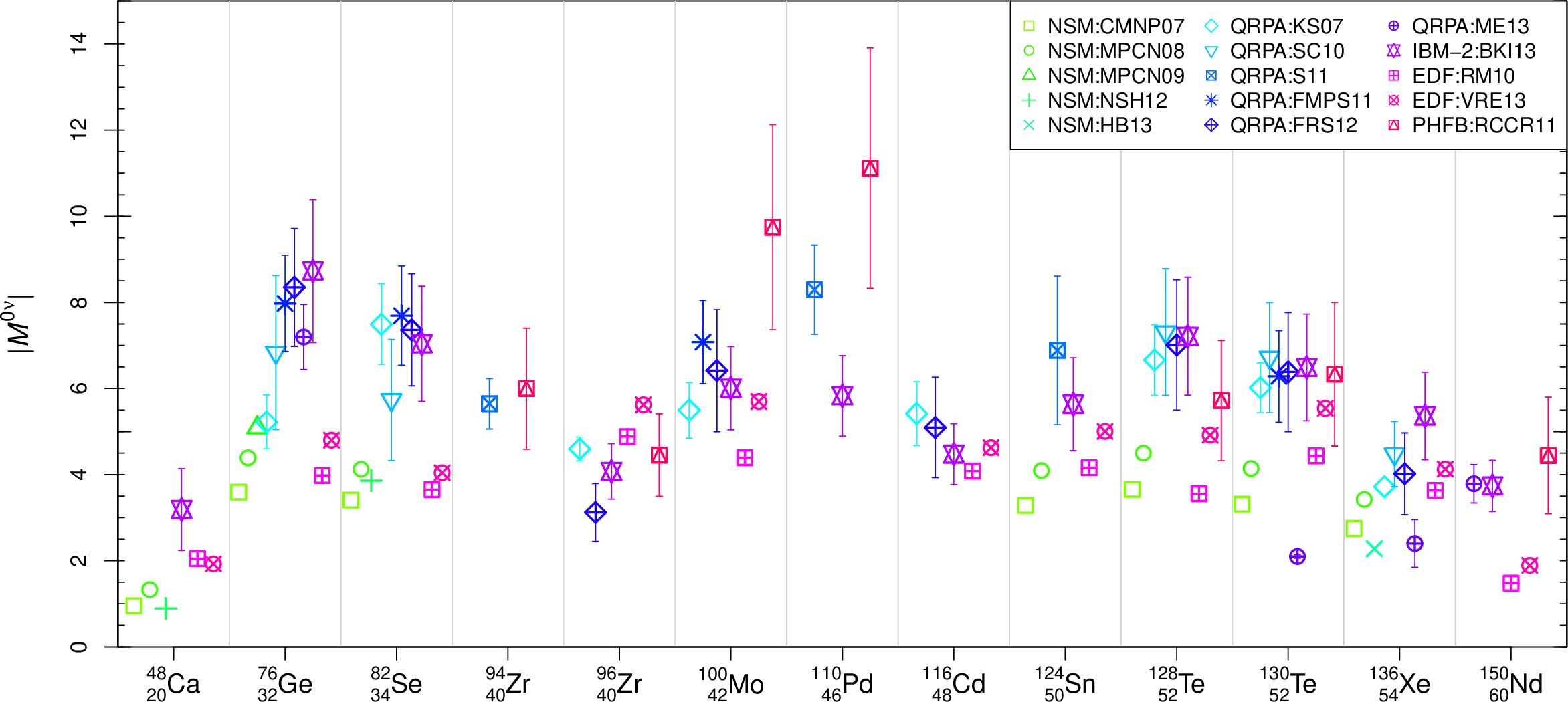}
\end{center}
\caption{\label{fig:mbg}
Values of the nuclear matrix element (NME)
$|M^{0\nu}|$
for several double-$\beta$ decaying nuclei of experimental interest
calculated with different methods:
Nuclear Shell Model (NSM):
CMNP07 \cite{Caurier:2007wq},
MPCN08 \cite{Menendez:2008jp},
MPCN09 \cite{Menendez:2009xa},
NSH12 \cite{Neacsu:2012id},
HB13 \cite{Horoi:2013jx};
Quasi-particle Random Phase Approximation (QRPA):
KS07 \cite{Kortelainen:2007rh,Kortelainen:2007mn},
SC10 \cite{Suhonen:2010zzc},
S11 \cite{Suhonen:2011zzd} (UCOM SRC);
Self-consistent Renormalized QRPA (SRQRPA):
FMPS11 \cite{Faessler:2011qw};
Renormalized QRPA (RQRPA):
FRS12 \cite{Faessler:2012ku} (Argonne and CD-Bonn SRC);
Deformed Skyrme QRPA (DSQRPA):
ME13 \cite{Mustonen:2013zu};
Interacting Boson Model-2 (IBM-2):
BKI13 \cite{Barea:2013bz};
Energy Density Functional (EDF):
RM10 \cite{Rodriguez:2010mn},
VRE13 \cite{Vaquero:2014dna};
Projected Hartree-Fock-Bogoliubov model (PHFB):
RCCR11 \cite{Rath:2010zz}.
The error bars do not have a statistical meaning,
but represent the range of values of $|M^{0\nu}|$ in the corresponding model
under variations of the model parameters and treatment of interactions.
}
\end{figure}

Five different methods
have been used for the calculation of
neutrinoless double-$\beta$ decay NMEs:

\begin{description}

\item[NSM:]
Nuclear Shell Model
\cite{Menendez:2008jp,Menendez:2009xa,Neacsu:2012id,Horoi:2013jx}
(see also
Ref.~\citen{Caurier:2004gf}).

\item[QRPA:]
Quasi-particle Random Phase Approximation
\cite{Simkovic:1999re,Bobyk:2000dw,Rodin:2003eb,Rodin:2005dp,Rodin:2006yk,Simkovic:2007vu,Kortelainen:2007rh,Kortelainen:2007mn,Simkovic:2009pp,Simkovic:2010zw,Suhonen:2011zzc,Suhonen:2011zzd,Faessler:2011qw,Faessler:2012ku,Suhonen:2012zz,Suhonen:2013dda,Suhonen:2013rca,Mustonen:2013zu}
(see also
Refs.~\citen{Tomoda:1990rs,Suhonen:1998ck,Faessler:1999zg,Suhonen-2007}).

\item[IBM:]
Interacting Boson Model
\cite{Barea:2009zz,Iachello:2011zz,Iachello:2011zzb,Barea:2013bz}
(see also
Refs.~\citen{Arima:1981hp,Iachello:1987zz}).

\item[EDF:]
Energy Density Functional
\cite{Rodriguez:2010mn,Rodriguez:2012rv,Vaquero:2014dna},
also called
Generating Coordinate Method
\cite{Bender:2003jk} (GCM).

\item[PHFB:]
Projected Hartree-Fock-Bogoliubov approach
\cite{Rath:2009dr,Rath:2010zz}.

\end{description}

\begin{figure}[t!]
\begin{center}
\includegraphics*[width=0.5\textwidth]{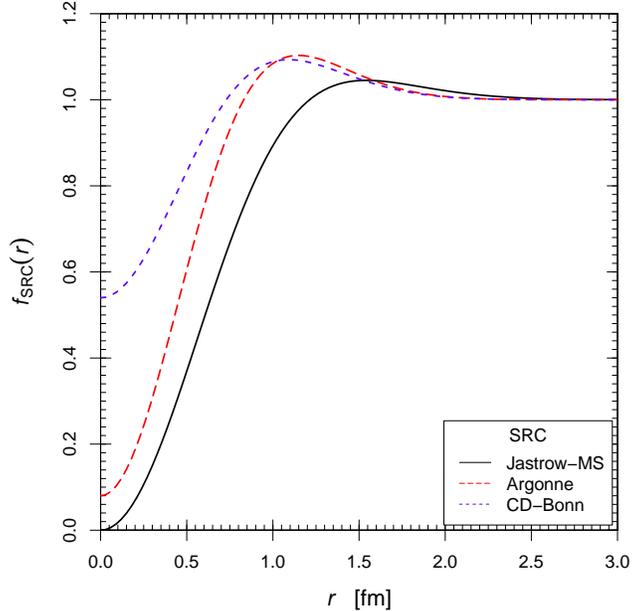}
\end{center}
\caption{ \label{fig:src}
The short-range correlation function in Eq.~(\ref{jastrow})
in the traditional
Jastrow-MS \cite{Jastrow:1955zz,Miller:1975hu}
approach
and in the parameterizations of Ref.~\citen{Simkovic:2009pp}
which describe approximately the
Argonne \cite{Wiringa:1984tg},
and
CD-Bonn \cite{Machleidt:1995km}
short-range correlations (SRC).
}
\end{figure}

Figure~\ref{fig:mbg}
shows the results of recent implementations of these methods.
The error bars in Fig.~\ref{fig:mbg} do not have a statistical meaning,
but represent the range of values of $|M^{0\nu}|$ in the corresponding model
under variations of the model parameters and treatment of interactions.
The most important are:

\begin{description}

\item[The axial coupling constant $g_{A}$.]
The value of $g_{A}$ considered by several authors is the traditional value $g_{A}=1.25$,
whereas the current value is
$g_{A} = 1.2723 \pm 0.0023$
\cite{PDG-2014}.
Moreover,
the strength of Gamow-Teller transitions must be ``quenched'' in order to fit the experimental data of
single $\beta$ decays, electron capture and $(p,n)$ charge-exchange reactions
(see Ref.~\citen{Osterfeld:1991ii,Caurier:2004gf,Suhonen:2013laa}).
Since the quenching depends on the model calculation,
different authors consider different effective values $g_{A}^{\text{eff}}$ of $g_{A}$.
Among the calculations considered in Fig~\ref{fig:mbg},
the EDF calculations
\cite{Rodriguez:2010mn,Vaquero:2014dna}
assume $g_{A}^{\text{eff}}=0.93$,
all the QRPA calculations
\cite{Kortelainen:2007rh,Kortelainen:2007mn,Suhonen:2011zzc,Suhonen:2011zzd,Faessler:2011qw,Faessler:2012ku,Mustonen:2013zu}
and the PHFB calculation
\cite{Rath:2010zz}
consider both $g_{A}^{\text{eff}}=1.25$ and $g_{A}^{\text{eff}}=1.0$.
All the NSM calculations
\cite{Caurier:2007wq,Menendez:2008jp,Menendez:2009xa,Neacsu:2012id,Horoi:2013jx}
assume the standard value
$g_{A}^{\text{eff}}=1.25$
and the IBM-2 calculation
\cite{Barea:2013bz}
uses $g_{A}^{\text{eff}}=1.269$,
which is the 2006 PDG value \cite{PDG-2006}.
It is also possible that $g_{A}^{\text{eff}}$ has a value much smaller than unity,
as discussed in Refs.~\citen{Faessler:2007hu,Suhonen:2013laa,Yoshida:2013jh}.

\item[The short-range correlations among nucleons.]
The short-range repulsion between nucleons is traditionally taken into account with the
Jastrow method
\cite{Jastrow:1955zz}
of multiplying the two-body wave function by a short-range correlation (SRC) function
$f_{\text{SRC}}(r)$,
where $r$ is the distance between the two nucleons\footnote{
The additional effects of the finite size of a nucleon
are taken into account with appropriate form factors in momentum space,
typically of dipole form.
}.
In the traditional parameterization of Miller and Spencer
\cite{Miller:1975hu},
\begin{equation}
f_{\text{SRC}}(r)
=
1 - c \, e^{-ar^2} \left( 1 - b \, r^2 \right)
,
\label{jastrow}
\end{equation}
with
$a = 1.1 \, \text{fm}^{-2}$,
$b = 0.68 \, \text{fm}^{-2}$,
$c = 1$
(Jastrow-MS).
In recent years,
several authors have adopted the newer
Argonne potential
(Argonne) \cite{Wiringa:1984tg},
or the
charge-dependent Bonn potential
(CD-Bonn) \cite{Machleidt:1995km},
or the
unitary correlation operator method (UCOM) \cite{Feldmeier:1997zh}.
In general,
in the framework of the same nuclear model,
the Jastrow-MS, UCOM, Argonne and CD-Bonn short-range correlations
give increasing values of $|M^{0\nu}|$
(see, for example, the comparisons in Refs.~\citen{Kortelainen:2007rh,Rath:2010zz,Suhonen:2011zzd,Faessler:2011qw,Faessler:2012ku,Barea:2013bz}).
The different approaches are discusses in details in Ref.~\citen{Simkovic:2009pp},
where the effects of the
Argonne and CD-Bonn potentials are approximated by the short-range correlation function in Eq.~(\ref{jastrow})
with
$a = 1.59 \, \text{fm}^{-2}$,
$b = 1.45 \, \text{fm}^{-2}$,
$c = 0.92$,
and
$a = 1.52 \, \text{fm}^{-2}$,
$b = 1.88 \, \text{fm}^{-2}$,
$c = 0.46$,
respectively.
Figure~\ref{fig:src}
shows the corresponding dependence of $f_{\text{SRC}}(r)$ on $r$
for the
Jastrow-MS, Argonne and CD-Bonn
parameterizations.
One can see that
larger values of $|M^{0\nu}|$
are associated with larger
short-range correlations for $r \lesssim 1 \, \text{fm}$.

\item[The $g_{pp}$ parameter in QRPA and its variants.]
\label{gpp}
This is a parameter of order one which renormalizes the particle-particle interaction
\cite{Cha:1983zz,Vogel:1986nj}
(see Refs.~\citen{Suhonen:1998ck,Faessler:1999zg})
and depends on the specific nuclear transition.
For each $\beta\beta_{0\nu}$ decay,
the value of $g_{pp}$
is fixed by the measured lifetime of the corresponding $\beta\beta_{2\nu}$ decay\footnote{
The available $\beta\beta_{2\nu}$ decay data are reviewed in Refs.~\citen{Tretyak:2002dx,Barabash:2010ie}.
}
\cite{Rodin:2003eb,Rodin:2005dp,Simkovic:2007vu},
or by that of the electron-capture and/or single $\beta$ decay of the intermediate nucleus
\cite{Suhonen:2004pw,Civitarese:2005jf,Civitarese:2005jb},
or both
\cite{Faessler:2007hu}.
Note that the value of $g_{pp}$ obtained with these methods is obviously correlated with that of $g_{A}^{\text{eff}}$
obtained from the same data
(see Refs.~\citen{Kortelainen:2007rh,Kortelainen:2007mn,Fang:2011da,Faessler:2011qw,Faessler:2012ku}).

\end{description}

From Fig.~\ref{fig:mbg}
one can see that the NSM values
\cite{Caurier:2007wq,Menendez:2008jp,Menendez:2009xa,Neacsu:2012id,Horoi:2013jx}
of the NME of each $\beta\beta_{0\nu}$ decay calculated in different publications do not differ much
and for some nuclei they are significantly smaller than those calculated with other methods.
Moreover,
the variation of the NSM values of
$|M^{0\nu}|$
for different nuclei is small\footnote{
Notice that the weak dependence of the value of the nuclear matrix element on the decaying nucleus in NSM calculations implies that
a measurement of the half-life of $\beta\beta_{0\nu}$-decay of a nucleus
${}^{A}_{Z}\text{X}$
allows to predict the $\beta\beta_{0\nu}$-decay half-life of another nucleus ${}^{A'}_{Z'}\text{X}$.
In fact, we have
$
T_{1/2}^{0\nu}({}^{A}_{Z}\text{X}) \simeq T_{1/2}^{0\nu}({}^{A'}_{Z'}\text{X}) G^{0\nu}(Q,Z) / G^{0\nu}(Q',Z')
$.
This relation could allow to test one of the main features of NSM calculations.
},
with the only
exception of the double-magic nucleus $\nCa{48}$,
for which the NME is significantly suppressed.

QRPA calculations
\cite{Kortelainen:2007rh,Kortelainen:2007mn,Suhonen:2011zzc,Suhonen:2011zzd,Faessler:2011qw,Faessler:2012ku,Mustonen:2013zu}
approximately agree for the nuclei in Fig.~\ref{fig:mbg},
except for the smaller and much smaller value of $|M^{0\nu}|$ for
$\nXe{136}$ and $\nTe{130}$
obtained in the recent
ME13 calculation \cite{Mustonen:2013zu}.

\begin{table}[b!]
\caption{\label{tab:nme}
Range of calculated values of $|M^{0\nu}|$,
ratio
$
|M^{0\nu}|_{\text{max}} / |M^{0\nu}|_{\text{min}}
$,
and
range of half-lives $T^{0\nu}_{1/2}$ for $m_{\beta\beta} = 0.1 \, \text{eV}$
for all the $\beta\beta^{-}$ decays of experimental interest (see Tab.~\ref{tab:exp})
and others for which more than one calculation of $|M^{0\nu}|$ exists.
}
\centering
{
\begin{tabular}{cccc}
$\beta\beta^{-}$ decay	&$|M^{0\nu}|$ &$\dfrac{|M^{0\nu}|_{\text{max}}}{|M^{0\nu}|_{\text{min}}}$ &$\displaystyle{T^{0\nu}_{1/2}(m_{\beta\beta} = 0.1 \, \text{eV}) \atop [10^{26}\,\text{y}]}$ \\
\hline
$\nCa{48} \to\nTi{48} $	&$0.89\div4.14$ &$4.6$ &$0.6\div13.3$ \\
$\nGe{76} \to\nSe{76} $	&$3.59\div10.39$ &$2.9$ &$1.0\div8.6$ \\
$\nSe{82} \to\nKr{82} $	&$3.41\div8.84$ &$2.6$ &$0.3\div2.2$ \\
$\nZr{96} \to\nMo{96} $	&$2.45\div5.62$ &$2.3$ &$0.4\div2.1$ \\
$\nMo{100}\to\nRu{100}$	&$4.39\div12.13$ &$2.8$ &$0.1\div0.8$ \\
$\nPd{110}\to\nCd{110}$	&$4.90\div13.91$ &$2.8$ &$0.3\div2.3$ \\
$\nCd{116}\to\nSn{116}$	&$3.77\div6.26$ &$1.7$ &$0.4\div1.1$ \\
$\nSn{124}\to\nTe{124}$	&$3.28\div8.61$ &$2.6$ &$0.4\div2.7$ \\
$\nTe{128}\to\nXe{128}$	&$3.55\div8.78$ &$2.5$ &$5.8\div35.2$ \\
$\nTe{130}\to\nXe{130}$	&$2.06\div8.00$ &$3.9$ &$0.3\div4.3$ \\
$\nXe{136}\to\nBa{136}$	&$1.85\div6.38$ &$3.4$ &$0.4\div5.2$ \\
$\nNd{150}\to\nSm{150}$	&$1.48\div5.80$ &$3.9$ &$0.1\div1.9$ \\
\hline
\end{tabular}
}
\end{table}

The IBM-2 calculation
\cite{Barea:2013bz}
covers almost all the nuclei in Fig.~\ref{fig:mbg}.
For several nuclei it is in approximate agreement with
QRPA calculations
(an exception is $\nPd{110}$).

The EDF
\cite{Rodriguez:2010mn,Vaquero:2014dna}
values of $|M^{0\nu}|$
are in approximate agreement with
QRPA and/or IBM-2 values
for some nuclei and more in agreement with the NSM values for others.
The recent calculation in Ref.~\citen{Vaquero:2014dna}
(VRE13)
improved that in Ref.~\citen{Rodriguez:2010mn}
(RM10)
by taking into account the effect of pairing fluctuations,
which gives a larger value of $|M^{0\nu}|$
for all the nuclei in Fig.~\ref{fig:mbg}, except $\nCa{48}$.

The PHFB
\cite{Rath:2010zz}
values of $|M^{0\nu}|$
tend to be large for some nuclei in Fig.~\ref{fig:mbg}
and intermediate for others.

Considering the range of values of $|M^{0\nu}|$ for each nucleus in Fig.~\ref{fig:mbg}
it is clear that there is a large theoretical uncertainty
which is quite unsatisfactory for future prospects to measure $m_{\beta\beta}$.
Table~\ref{tab:nme}
shows the range of calculated values of $|M^{0\nu}|$
and the ratio
$
|M^{0\nu}|_{\text{max}} / |M^{0\nu}|_{\text{min}}
$
for all the $\beta\beta^{-}$ decays of experimental interest (see Tab.~\ref{tab:G0n})
and others for which more than one calculation of $|M^{0\nu}|$ exists.
One can see that the minimum discrepancy among different calculations
is a factor
$1.7$
for \nCd{116}
and the discrepancy can be as large as a factor
$4.6$
for \nCa{48}!

The theoretical uncertainties of NMEs is an important problem which has been discussed
by several authors with different approaches
\cite{Bahcall:2004ip,Rodin:2005dp,Faessler:2008xj,Rath:2010zz,Faessler:2013hz}.
Since all the five methods described above have advantages and disadvantages
and for each method there are variants of the technique on which the method is based
(e.g. QRPA, RQRPA, SRQRPA, DSQRPA, etc.)
and the necessity to assume the values of some parameters
(e.g. $g_{A}$, $g_{pp}$, etc.)
and the treatment of some interactions
(e.g SRC),
there is no consensus among experts on a method that can be preferred over the others
and each method is subject to future developments.
In this unsettled situation,
the range of values of the NME
obtained with different methods
for each $\beta\beta_{0\nu}$ decay
is the only indication that we have on the NME uncertainty.
Therefore, in the following we consider this range as the
theoretical NME uncertainty.
However,
we think that this range is only indicative,
because the true NME may well be out of the range.
Contrary to some authors,
we think that
the theoretical NME uncertainty
should not be considered with a statistical meaning,
because theoretical calculations are not random events.
Hence, in the following, when we derive the value of a quantity which depends on
$|M^{0\nu}|$,
we give the results corresponding to the range of available
calculations of $|M^{0\nu}|$,
keeping in mind that it has no statistical meaning
and
that in the future it may collapse to a definite value or
to a much narrower range if one method of calculation will be proved to be
better than others.

In Tab.~\ref{tab:nme} we present also the range
of half-lives
$T^{0\nu}_{1/2}$
of the considered nuclei
for $m_{\beta\beta} = 0.1 \, \text{eV}$,
corresponding to the theoretical NME uncertainty.
Figure~\ref{fig:thalf}
shows the same total ranges of half-lives and those obtained with the
different methods
(NSM, QRPA, IBM-2, EDF, PHFB),
compared with the corresponding current experimental most stringent limits at 90\% CL
(which are listed in Tab.~\ref{tab:explim}).

The fractional uncertainty between about 2 and 3 of
the NME of most nuclei in Tab.~\ref{tab:nme}
implies larger fractional uncertainties
between about 4 and 9
of the corresponding half-lives,
which depend on
$|M^{0\nu}|^2$.
This is a problem for the planning of experiments,
if one wants to reach a sensitivity to a small value of
$|m_{\beta\beta}|$,
because the uncertainty on the corresponding half-life
is rather large,
as shown in Fig.~\ref{fig:thalf}.
The problem is exacerbated by the uncertainty in the quenching of $g_{A}$:
since the half-life is proportional to $g_{A}^4$ an uncertainty by a factor of 2 for $g_{A}$
implies an uncertainty of a factor 16 for $T^{0\nu}_{1/2}$!

On the other hand,
since for a given $T^{0\nu}_{1/2}$
the effective Majorana mass is inversely proportional to
$|M^{0\nu}|$,
\begin{equation}
|m_{\beta\beta}|
=
|M^{0\nu}|^{-1}
\left[
T^{0\nu}_{1/2} G^{0\nu}(Q,Z)
\right]^{-1/2}
,
\label{mbbdep}
\end{equation}
the fractional uncertainty on the extraction of $|m_{\beta\beta}|$
from a measured value or a limit of $T^{0\nu}_{1/2}$
is the same as that of $|M^{0\nu}|$,
i.e. a factor between about 2 and 3.
This is shown in Fig.~\ref{fig:exp},
in which we compared the limits on $|m_{\beta\beta}|$
obtained from recent experimental results
taking into account the uncertainty of $|M^{0\nu}|$.
\section{Effective Majorana mass}
\label{bb6}

\begin{figure}[t!]
\begin{center}
\includegraphics*[width=0.5\textwidth]{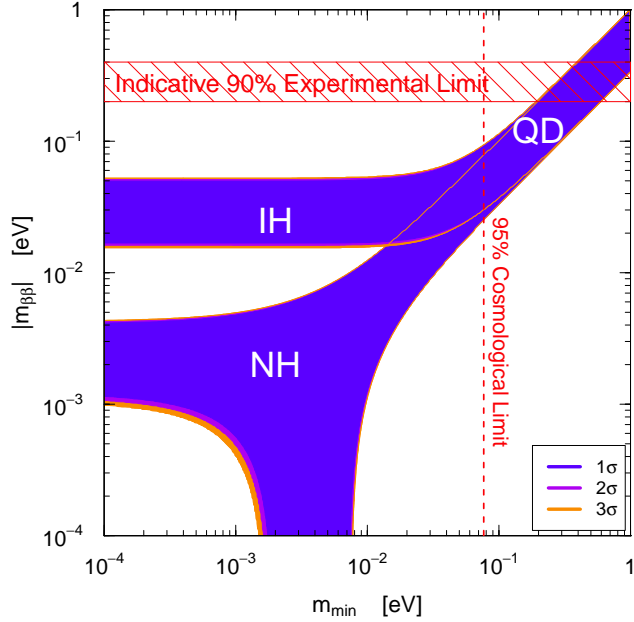}
\end{center}
\caption{\label{fig:mbb}
Value of the effective Majorana mass $|m_{\beta\beta}|$ as a
function of the lightest neutrino mass
in the three cases discussed in Section~\ref{sub:absolute}:
normal hierarchy (NH),
inverted hierarchy (IH)
and
quasi-degenerate spectra (QD).
}
\end{figure}

The existing atmospheric, solar, and long-baseline reactor and accelerator neutrino oscillation data
are perfectly described by the three neutrino mixing paradigm
(see Section~\ref{bb2}),
with
\begin{equation}\label{3numix}
\nu_{lL}=\sum^{3}_{i=1}U_{li}\nu_{i}
\qquad
(l=e,\mu,\tau)
.
\end{equation}
Using the standard parameterization in Eq.~(\ref{U}) of the three-neutrino mixing matrix,
the effective Majorana mass in $\beta\beta_{0\nu}$ decay can be written as
\begin{equation}\label{mbb3nu}
|m_{\beta\beta}|
=
|c^{2}_{13}c^{2}_{12}e^{2i\alpha_{1}}m_{1}+c^{2}_{13}s^{2}_{12}e^{2i\alpha_{2}}m_{2}+s^{2}_{13}m_{3}|,
\end{equation}
where $\alpha_{i}=\lambda_{i}+\delta$.

Since the values of the mixing angles and of the squared-mass differences are known from oscillation data
(see Table~\ref{tab:global}),
the value of $|m_{\beta\beta}|$ can be plotted as a
function of the lightest neutrino mass
$m_{\text{min}}=m_{1}$ in the normal spectrum
and
$m_{\text{min}}=m_{3}$ in the inverted spectrum,
as shown in Fig.~\ref{fig:mbb},
where we used the relations (\ref{mNS}) and (\ref{mIS}).
The largeness of the allowed bands are mainly due to our complete ignorance of the values of the two phases
$\alpha_{1}$
and
$\alpha_{2}$,
which can cause significant cancellations between the contributions to $|m_{\beta\beta}|$ in Eq.~(\ref{mbb3nu})
(see Refs.~\citen{Vissani:1999tu,Feruglio:2002af,Strumia:2006db,Xing:2014yka}).
Figure~(\ref{fig:mbb})
shows that a complete cancellation is not possible in the case of an inverted spectrum,
for which $|m_{\beta\beta}|$ is bounded to be larger than about $2\times10^{-2}\,\text{eV}$.
In the case of a normal spectrum a complete cancellation is not possible in the quasi-degenerate region,
but it is possible in the normal hierarchy region for $m_{\text{min}}=m_{1}$ in the approximate interval
$(2-7)\times10^{-3}\,\text{eV}$.

\begin{figure}[t!]
\begin{minipage}[t]{0.49\textwidth}
\begin{center}
\includegraphics*[width=\textwidth]{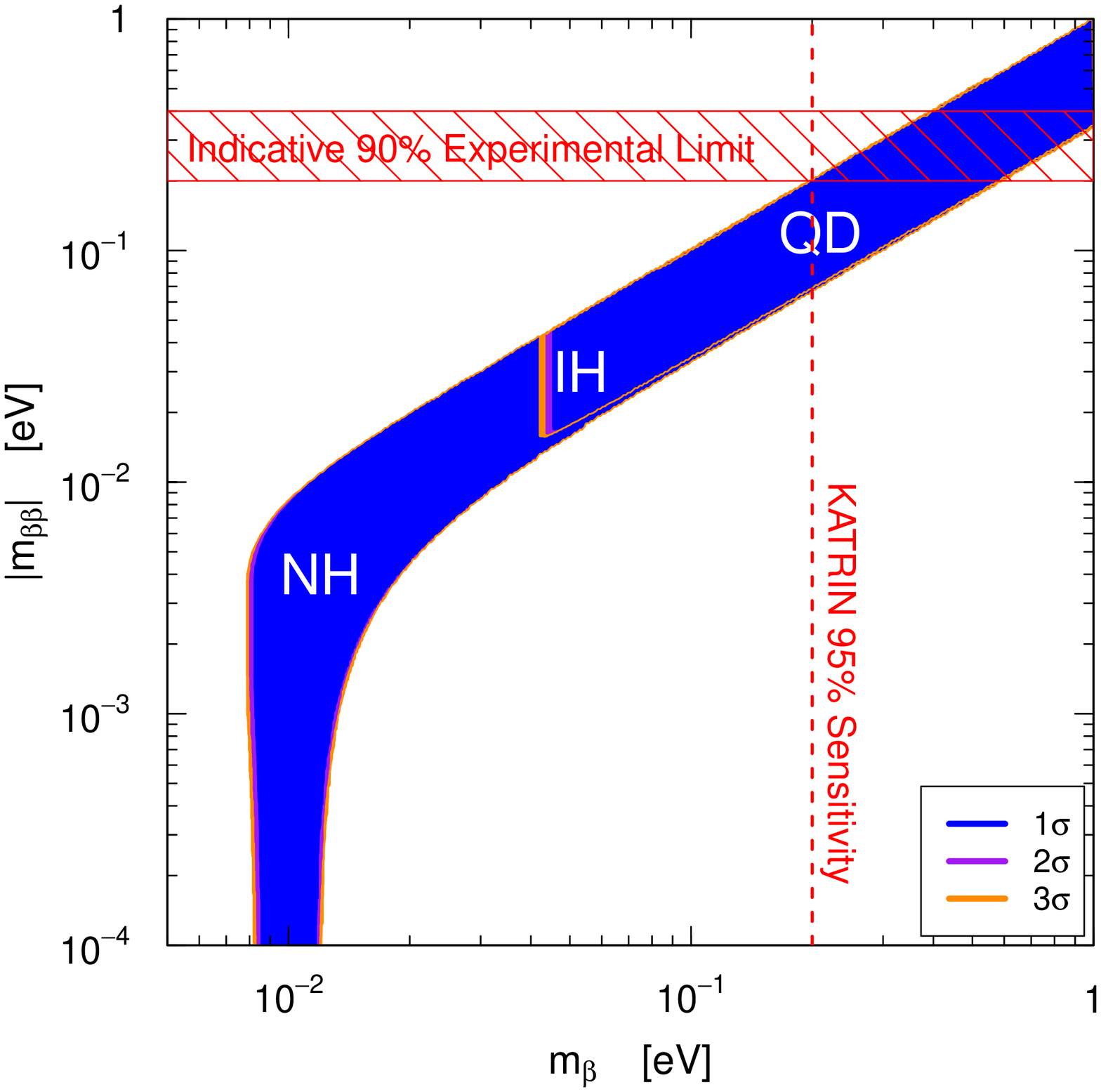}
\end{center}
\caption{\label{fig:mbt}
Value of the effective Majorana mass $|m_{\beta\beta}|$ as a
function of the effective electron neutrino mass (\ref{trimass}) in $\beta$ decay
in the three cases discussed in Section~\ref{sub:absolute}:
normal hierarchy (NH),
inverted hierarchy (IH)
and
quasi-degenerate spectra (QD).
}
\end{minipage}
\hfill
\begin{minipage}[t]{0.49\textwidth}
\begin{center}
\includegraphics*[width=\textwidth]{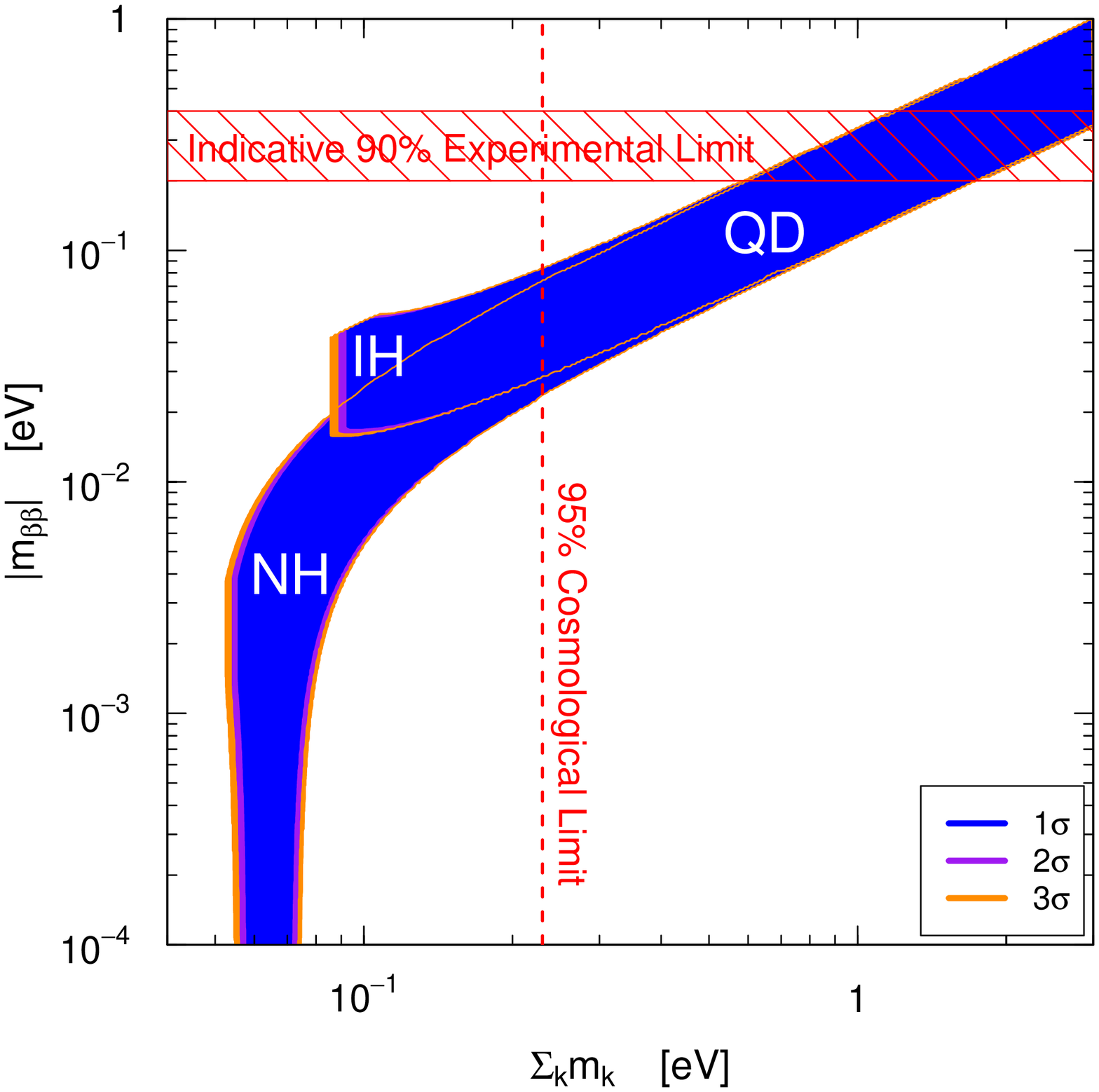}
\end{center}
\caption{\label{fig:flm}
Value of the effective Majorana mass $|m_{\beta\beta}|$ as a
function of the sum of neutrino masses
in the three cases discussed in Section~\ref{sub:absolute}:
normal hierarchy (NH),
inverted hierarchy (IH)
and
quasi-degenerate spectra (QD).
}
\end{minipage}
\end{figure}

The lower bound on $|m_{\beta\beta}|$ of about $2\times10^{-2}\,\text{eV}$
in the case of an inverted spectrum
provides a strong encouragement for the experimental searches of
$\beta\beta_{0\nu}$ decay in the near future,
with the aim of measuring $\beta\beta_{0\nu}$ decay if the neutrino masses have an inverted spectrum
or excluding the inverted spectrum if no signal is found.
Let us stress, however, that if $\beta\beta_{0\nu}$ decay is discovered in these experiments
the problem of the determination of the type of neutrino mass spectrum will remain unsolved.
In fact,
from Fig.~\ref{fig:mbb} one can see that the case of an inverted hierarchy
can be established only if it is known independently that
$m_{\text{min}} \lesssim 10^{-2} \, \text{eV}$.
Otherwise,
the neutrino mass spectrum can be either normal or inverted, with nearly quasi-degenerate masses.

Since it is difficult to measure directly the value of $m_{\text{min}}$
and the current and near-future measurements of the absolute values of neutrino masses
are done through the measurements
of the effective electron neutrino mass $m_{\beta}$ in $\beta$-decay experiments
and through the measurements of the sum
$\sum_k m_{k}$
of the neutrino masses in cosmological experiments
(see Section~\ref{sub:absolute}),
it is useful to plot the allowed interval of $|m_{\beta\beta}|$
as a function of these two quantities
\cite{Barger:1999na,Matsuda:2000iw,Barger:2002xm,Fogli:2004as},
as done\footnote{
Note that, contrary to some similar figures published in the literature,
Figs.~\ref{fig:mbt} and \ref{fig:flm}
take into account the uncertainties of
$m_{\beta}$
and of
$\sum_{k} m_{k}$
induced by the uncertainties of the neutrino oscillation parameters
(given in Tab.~\ref{tab:global}).
}
in Figs.~\ref{fig:mbt} and \ref{fig:flm}.
One can see that the allowed regions in the normal and inverted spectra have large overlaps.
Therefore,
if $|m_{\beta\beta}|$ is found to be larger than about $2\times10^{-2}\,\text{eV}$
it may be difficult to distinguish the normal and inverted spectra with
absolute neutrino mass measurements\footnote{
There are however several
accelerator \cite{Bian:2013saa,Abe:2014tzr},
reactor \cite{Li:2013zyd,Li:2014qca}
and
atmospheric \cite{Ghosh:2012px,Aartsen:2014oha}
neutrino oscillation experiments
aimed at a measurement of the neutrino mass hierarchy
in the near future
(see also Ref.~\citen{Cahn:2013taa}).
}.

In the following Subsections we consider in some details the three cases NH, IH and QD
(see also the recent discussions in Refs.~\citen{Dev:2013vxa,Haba:2013xwa,Dodelson:2014tga,Dell'Oro:2014yca}).

\subsection{Normal hierarchy of neutrino masses (NH)}
\label{sub:NH}

In this case the contribution to $|m_{\beta\beta}|$ of the first term in Eq.~(\ref{mbb3nu})
can be neglected, leading to
\begin{equation}\label{hierarchy1}
|m_{\beta\beta}|
\simeq
\left|
\cos^{2}\vartheta_{13} \sin^{2}\vartheta_{12} e^{2i\alpha_{2}} \sqrt{\Delta m^{2}_{S}}
+
\sin^{2}\vartheta_{13} \sqrt{\Delta m^{2}_{\text{A}}}
\right|
\end{equation}
The first term on the right-hand side is small because of the smallness of the solar squared-mass difference $\Delta m^{2}_{S}$
and
in the second term the contribution of the ``large'' atmospheric squared-mass difference $\Delta m^{2}_{\text{A}}$ is suppressed by the small factor $\sin^{2}\vartheta_{13}$.
Using the best fit values of the parameters we have
\begin{equation}\label{hierarchy2}
\cos^{2}\vartheta_{13} \sin^{2}\vartheta_{12} \sqrt{\Delta m^{2}_{S}}
\simeq
3 \times 10^{-3} \, \text{eV}
,
\quad
\sin^{2}\vartheta_{13} \sqrt{\Delta m^{2}_{\text{A}}}
\simeq
1 \times 10^{-3} \, \text{eV}
.
\end{equation}
Thus, the absolute values of the two terms in Eq.~(\ref{hierarchy1}) are of the same order of magnitude.
Taking into account the $3\sigma$ range of the mixing parameters in Tab.~\ref{tab:global},
we obtain the upper bound
\begin{equation}\label{hierarchy3}
|m_{\beta\beta}|
\leq
\cos^{2}\vartheta_{13} \sin^{2}\vartheta_{12} \sqrt{\Delta m^{2}_{S}}
+
\sin^{2}\vartheta_{13} \sqrt{\Delta m^{2}_{\text{A}}}
\lesssim
4 \times 10^{-3} \, \text{eV}
,
\end{equation}
in agreement with Fig.~\ref{fig:mbb}.
Unfortunately,
since
this upper bound is significantly smaller than the sensitivity of future planned experiments on the search for $\beta\beta_{0\nu}$ decay
(see Section~\ref{bb7}),
it will be very difficult to explore the NH with
future $\beta\beta_{0\nu}$ decay experiments.

On the other hand,
if there are light sterile neutrinos at the eV scale,
as suggested by anomalies found in short-baseline oscillation experiments
(see Refs.~\citen{Aguilar:2001ty,Abdurashitov:2005tb,Giunti:2010zu,Mention:2011rk,Kopp:2011qd,Conrad:2012qt,Giunti:2012tn,Kopp:2013vaa,Giunti:2013aea}),
their additional contribution to
$|m_{\beta\beta}|$
cannot be canceled by that of the standard three light neutrinos with a normal hierarchy
\cite{Goswami:2005ng,Goswami:2007kv,Barry:2011wb,Li:2011ss,Rodejohann:2012xd,Giunti:2012tn,Girardi:2013zra,Pascoli:2013fiz,Meroni:2014tba,Abada:2014nwa,Giunti-NNN-2014}.
In this case
the future planned $\beta\beta_{0\nu}$ decay experiments can find a signal.

\subsection{Inverted hierarchy of neutrino masses (IH)}
\label{sub:IH}

\begin{figure}[t!]
\begin{center}
\includegraphics*[width=\textwidth]{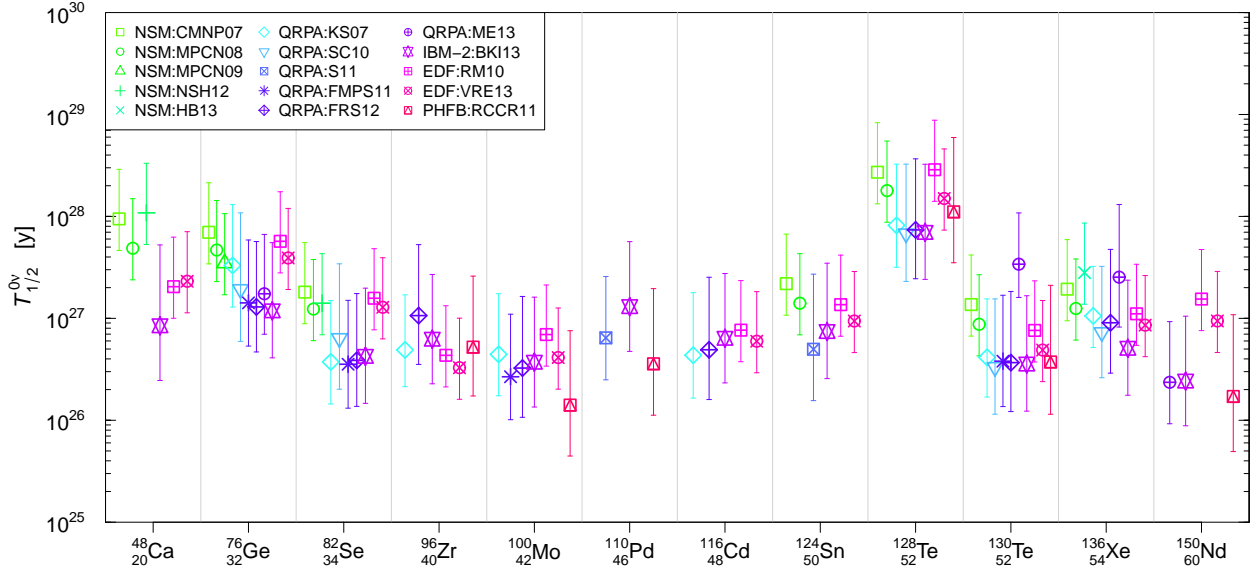}
\end{center}
\caption{\label{fig:tih}
Intervals of $T^{0\nu}_{1/2}$
obtained from Eq.~(\ref{TIH})
in the case of an inverted neutrino mass hierarchy
for several double-$\beta$ decaying nuclei of experimental interest
with the NME calculations in Fig.~\ref{fig:mbg}.
}
\end{figure}

In this case, the contribution of the small $m_{3}$,
which is suppressed by the small $\sin^{2}\vartheta_{13}$ coefficient,
can be neglected in Eq.~(\ref{mbb3nu}),
leading to
\begin{equation}\label{invhierarchy3}
|m_{\beta\beta}|
\simeq
\sqrt{\Delta m^{2}_{\text{A}}}
\,
\sqrt{1-\sin^{2}2\vartheta_{12}\sin^{2}\alpha}
,
\end{equation}
where the Majorana phase difference
$\alpha=\alpha_{2}-\alpha_{1}$
is the only unknown parameter.
Hence,
in this case $|m_{\beta\beta}|$ is bounded in the interval
\begin{equation}\label{invhierarchy4}
\sqrt{\Delta m^{2}_{\text{A}}}\cos2\vartheta_{12}
\lesssim
|m_{\beta\beta}|
\lesssim
\sqrt{\Delta m^{2}_{\text{A}}}
.
\end{equation}
The upper (lower) bound of (\ref{invhierarchy4}) corresponds to equal (opposite) CP parities of $\nu_{1}$ and $\nu_{2}$ if the lepton sector is CP-invariant\footnote{
In fact, from CP invariance it follows that $U_{ei}=U^{*}_{ei}\eta_{i}$,
where $\eta_{i}=\pm i$ is the CP parity of the Majorana neutrino with mass $m_{i}$.
From this condition we find $e^{2i\alpha_{i}}=\eta_{i}$.
Thus, we have
$e^{2i(\alpha_{2}-\alpha_{1})}=e^{2i\alpha}=\eta_{2}\eta^{*}_{1}$.
If $\eta_{2}=\eta_{1}$ we have $\alpha=0,\pi$
(the upper bound in Eq.~(\ref{invhierarchy4})),
and
if $\eta_{2}=-\eta_{1}$ we have $\alpha=\pm\pi/2$
(the lower bound in Eq.~(\ref{invhierarchy4})).
}.
Taking into account the $3\sigma$ ranges of $\Delta m^{2}_{\text{A}}$ and $\sin^{2}\vartheta_{12}$
in Tab.~\ref{tab:global},
we obtain
\begin{equation}\label{invhierarchy5}
2 \times 10^{-2}
\lesssim
|m_{\beta\beta}|
\lesssim
5 \times 10^{-2} \, \text{eV}
,
\end{equation}
in agreement with Fig.~\ref{fig:mbb}.
The $\beta\beta_{0\nu}$ decay experiments of the next generation
are aimed at exploring at least part of this interval of $|m_{\beta\beta}|$.
From Eqs.~(\ref{totrate}) and (\ref{invhierarchy4}),
the predicted half-lives of each nuclear
$\beta\beta_{0\nu}$ decay must lie in the range
\begin{equation}
\left[
\Delta m^{2}_{\text{A}}
|M^{0\nu}|^{2}
G^{0\nu}(Q,Z)
\right]^{-1}
\lesssim
T^{0\nu}_{1/2}
\lesssim
\left[
\Delta m^{2}_{\text{A}}
\cos2\vartheta_{12}
|M^{0\nu}|^{2}
G^{0\nu}(Q,Z)
\right]^{-1}
.
\label{TIH}
\end{equation}
Figure~\ref{fig:tih}
shows the values of this interval of $T^{0\nu}_{1/2}$,
predicted in the case of an inverted neutrino mass hierarchy,
for several double-$\beta$ decaying nuclei of experimental interest
obtained with the NME calculations in Fig.~\ref{fig:mbg}.

In the near future,
several
accelerator \cite{Bian:2013saa,Abe:2014tzr},
reactor \cite{Li:2013zyd,Li:2014qca}
and
atmospheric \cite{Ghosh:2012px,Aartsen:2014oha}
neutrino oscillation experiments are aimed at the determination of
the normal or inverted character of the neutrino mass spectrum
(see also Ref.~\citen{Cahn:2013taa}).
If it is found that the spectrum is inverted and if
future measurements of the absolute value of the neutrino masses
exclude the quasi-degenerate scenario
(see Section~\ref{sub:QD},
we will have an evidence in favor of the inverted hierarchy of the neutrino masses.
Then,
if massive neutrinos are Majorana particles,
neutrinoless double $\beta$-decay could be observed
in future experiments sensitive to $m_{\beta\beta}$
in the range (\ref{invhierarchy5}).
A measurement of the values of the half-lives of the $\beta\beta_{0\nu}$-decay of different nuclei and a comparison of these values with the predicted range (\ref{TIH})
could allow to check whether the Majorana mass mechanism is the only mechanism
generating neutrinoless double $\beta$-decay.
In particular,
in the extreme case in which
$\beta\beta_{0\nu}$ decay is not observed in the experiments sensitive to the lowest value of
$m_{\beta\beta}$
in Eq.~(\ref{invhierarchy5}),
it will be possible to conclude that
\begin{itemize}
\item
neutrinos with definite masses are Dirac particles, or
\item
in addition to the contribution of the three light neutrinos
there are contributions to the amplitude of $\beta\beta_{0\nu}$ decay
due to new particles
(light sterile neutrinos
\cite{Goswami:2005ng,Goswami:2007kv,Barry:2011wb,Li:2011ss,Rodejohann:2012xd,Giunti:2012tn,Girardi:2013zra,Pascoli:2013fiz,Meroni:2014tba,Abada:2014nwa,Giunti-NNN-2014},
or others
\cite{Rodejohann:2011mu,Vergados:2012xy,Deppisch:2012nb})
which cancel the three-neutrino contribution.
\end{itemize}

\subsection{Quasi-degenerate neutrino mass spectrum (QD)}
\label{sub:QD}

In this case,
neglecting  in Eq.~(\ref{mbb3nu}) the contribution of $m_{3}$,
which is suppressed by the small $\sin^{2}\vartheta_{13}$ coefficient,
and using the approximate common mass $m_{\nu}^{\text{QD}}$ in Eq.~(\ref{mQD}),
we obtain
\begin{equation}\label{quasi1}
|m_{\beta\beta}|
\simeq
m_{\nu}^{\text{QD}}
\,
\sqrt{1-\sin^{2}2\vartheta_{12}\sin^{2}\alpha}
,
\end{equation}
with
$\alpha=\alpha_{2}-\alpha_{1}$.
Thus, in the case of a quasi-degenerate neutrino mass spectrum
$|m_{\beta\beta}|$
depends on two unknown parameters:
the mass $m_{\nu}^{\text{QD}}$ and the Majorana CP phase difference $\alpha$.
From Eq.~(\ref{quasi1}) we find
\begin{equation}\label{quasi2}
m_{\nu}^{\text{QD}} \, \cos2\vartheta_{12}
\lesssim
|m_{\beta\beta}|
\lesssim
m_{\nu}^{\text{QD}}
.
\end{equation}
Taking into account that $\cos2\vartheta_{12}$ is large ($\simeq 0.38$),
we can conclude that if
the $\beta\beta_{0\nu}$ decay is observed with a large effective Majorana mass
(significantly larger than
$\sqrt{\Delta m^{2}_{\text{A}}} \simeq 5 \times 10^{-2} \, \text{eV}$),
it will be an evidence in favor of a quasi-degenerate neutrino mass spectrum.
In this case, $m_{\nu}^{\text{QD}}$ is bounded in the interval
\begin{equation}\label{quasi3}
|m_{\beta\beta}|
\lesssim
m_{\nu}^{\text{QD}}
\lesssim
\frac{|m_{\beta\beta}|}{\cos2\vartheta_{12}}
\simeq
2.6 \, |m_{\beta\beta}|
.
\end{equation}

\begin{figure}[t!]
\begin{center}
\includegraphics*[width=0.5\textwidth]{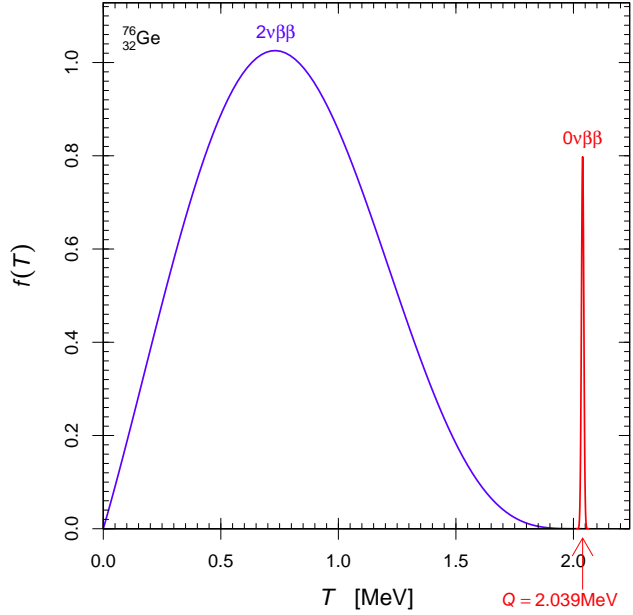}
\end{center}
\caption{\label{fig:spec}
Total electron kinetic energy spectra of the $\beta\beta_{2\nu}$ and $\beta\beta_{0\nu}$
decays of $\nGe{76}$.
}
\end{figure}

Information on $m_{\nu}^{\text{QD}}$ is given by the experiments sensitive to the absolute values of neutrino masses discussed in Section~\ref{sub:absolute}.

In the case of a quasi-degenerate neutrino mass spectrum the effective electron neutrino mass
in Eq.~(\ref{trimass}) becomes
\begin{equation}\label{quasi4}
m_{\beta} \simeq m_{\nu}^{\text{QD}}
,
\end{equation}
leading to a direct measurement of $m_{\nu}^{\text{QD}}$
in the experiments which measure with high precision the
end-point part of the electron spectrum in $\beta$ decays.
The sensitivity to $m_{\beta}$ of the KATRIN experiment \cite{Fraenkle:2011uu},
which is scheduled to start data taking in 2016,
is illustrated by the vertical dashed line in Fig.~\ref{fig:mbt}.

The analysis of cosmological data is sensitive to the sum of neutrino masses,
which in the case of a quasi-degenerate neutrino mass spectrum
is given by
\begin{equation}
\sum_{i}m_{i} \simeq 3 m_{\nu}^{\text{QD}}
.
\label{QD8}
\end{equation}
Hence, the cosmological bound in Eq.~(\ref{cosmo})
implies the rather stringent limit given by the vertical dashed line in Fig.~\ref{fig:flm},
which excludes most of the QD region.
Let us however note that the results of analyses of cosmological data are
model-dependent and must be confirmed by direct measurements.
\section{Experimental status of $\beta\beta_{0\nu}$ decay}
\label{bb7}

\begin{figure}[t!]
\begin{center}
\includegraphics*[width=\textwidth]{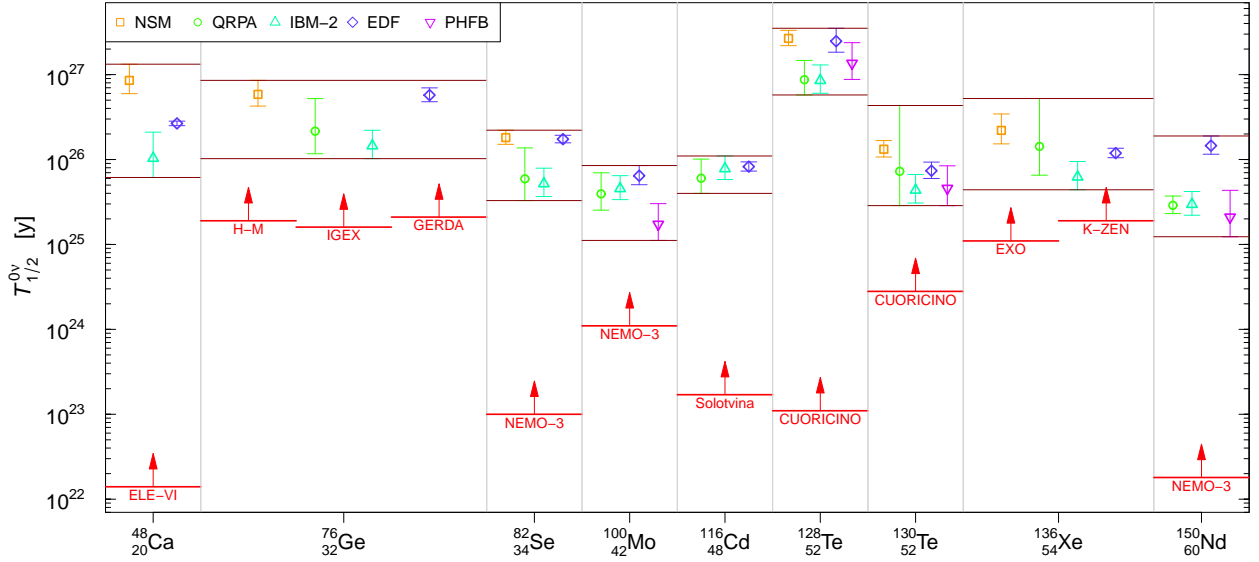}
\end{center}
\caption{\label{fig:thalf}
Comparison of the
90\% CL experimental lower limits on $T^{0\nu}_{1/2}$
of the experiments listed in Tab.~\ref{tab:explim}
confronted with the corresponding range of theoretical predictions
for $m_{\beta\beta} = 0.1 \, \text{eV}$
(see Tab.~\ref{tab:nme}).
}
\end{figure}

\begin{table}[b!]
\caption{\label{tab:exp}
Recent past, current and future neutrinoless double-$\beta$ decay experiments.
See Ref.~\citen{Tretyak:2002dx} for a complete list of previous experiments and results.
}
\centering
{
\begin{tabular}{llll}
$\beta\beta^{-}$ decay	&past exp.						&current exp.			&future exp.		\\
\hline
\multirow{2}{*}{
$\nCa{48} \to\nTi{48} $
}			&ELEGANT-VI\cite{Ogawa:2004fy}				&				&CANDLES\cite{Ogawa:2012gh}\\
			&TGV\cite{Brudanin:2000in}				&				&\\
\hline
\multirow{2}{*}{
$\nGe{76} \to\nSe{76} $
}			&Heidelberg-Moscow\cite{Klapdor-Kleingrothaus:2001yx}	&GERDA\cite{Agostini:2013mzu}	&Majorana\cite{Abgrall:2013rze}\\
			&IGEX\cite{Aalseth:2002rf}				&				&\\
\hline
\multirow{2}{*}{
$\nSe{82} \to\nKr{82} $
}			&NEMO-3\cite{Arnold:2005rz}				&				&SuperNEMO\cite{Shitov:2010nt}\\
			&							&				&LUCIFER\cite{Cardani:2012gf}\\
\hline
\multirow{1}{*}{
$\nZn{70} \to\nGe{70} $
}
			&Solotvina\cite{Danevich:2004tj}&COBRA\cite{Bloxham:2007aa}	&\\
\hline
\multirow{3}{*}{
$\nMo{100}\to\nRu{100}$
}			&ELEGANT-V\cite{Ejiri:2001fx}				&				&MOON\cite{Fushimi:2010zzb}\\
			&NEMO-3\cite{Arnold:2013dha}				&				&AMoRE\cite{Bhang:2012gn}\\
			&							&				&LUMINEU\cite{Barabash:2014una}\\
\hline
\multirow{1}{*}{
$\nCd{116}\to\nSn{116}$
}			&Solotvina\cite{Danevich:2003ef}			&COBRA\cite{Bloxham:2007aa}	&\\
\hline
\multirow{1}{*}{
$\nTe{128}\to\nXe{128}$
}			&CUORICINO\cite{Arnaboldi:2002te}			&COBRA\cite{Bloxham:2007aa}	&\\
\hline
\multirow{2}{*}{
$\nTe{130}\to\nXe{130}$
}			&CUORICINO\cite{Arnaboldi:2002te}			&COBRA\cite{Bloxham:2007aa}	&CUORE\cite{Andreotti:2010vj}\\
			&							&				&SNO+\cite{Biller:2014eha}\\
\hline
\multirow{2}{*}{
$\nXe{136}\to\nBa{136}$
}			&Gotthard\cite{Luscher:1998sd}				&EXO\cite{Albert:2014awa}		&NEXT\cite{Gomez-Cadenas:2013lta}\\
			&							&KamLAND-Zen\cite{Gando:2012zm} &XMASS\cite{Abe:2013tc}\\
\hline
\multirow{1}{*}{
$\nNd{150}\to\nSm{150}$
}			&NEMO-3\cite{Argyriades:2008pr}				&				&DCBA\cite{Ishihara:2012gg}\\
\hline
\end{tabular}
}
\end{table}

Table~\ref{tab:exp}
presents a list of past, current and future experiments and the corresponding double-$\beta$ nuclear transitions.
Detailed reviews of neutrinoless double-$\beta$ decay experiments
have been given recently in Refs.~\citen{GomezCadenas:2011it,Schwingenheuer:2012zs,Giuliani:2012zu,Cremonesi:2013vla}.

The experiments searching for $\beta\beta_{0\nu}$ decay
observe the two electrons emitted in the process (\ref{bb0-}).
As illustrated in Fig.~\ref{fig:spec},
in $\beta\beta_{0\nu}$ decay the sum of the energies of the two emitted electrons
is equal to the $Q$-value of the process given by Eq.~(\ref{Qvalue}).
This signature allows to distinguish $\beta\beta_{0\nu}$ decay
events from $\beta\beta_{2\nu}$ decay events,
whose electrons have the continuous spectrum depicted in Fig.~\ref{fig:spec}.

From the experimental point of view there are two main requirements for the search of
$\beta\beta_{0\nu}$ decay:
1)
a sufficiently low background in an appropriate search energy window around the $Q$-value,
which depends on the energy resolution of the detector;
2)
a sufficient number of $\beta\beta_{0\nu}$-decaying nuclei which allow to observe
a significant signal excess over the background.

Background reduction is achieved by placing the
detector in a deep underground laboratory
in order to reduce the cosmic-ray background,
by shielding the detector from
environmental radioactivity and by constructing
the detector with low-radioactivity material.

\begin{figure}[t!]
\begin{center}
\includegraphics*[width=\textwidth]{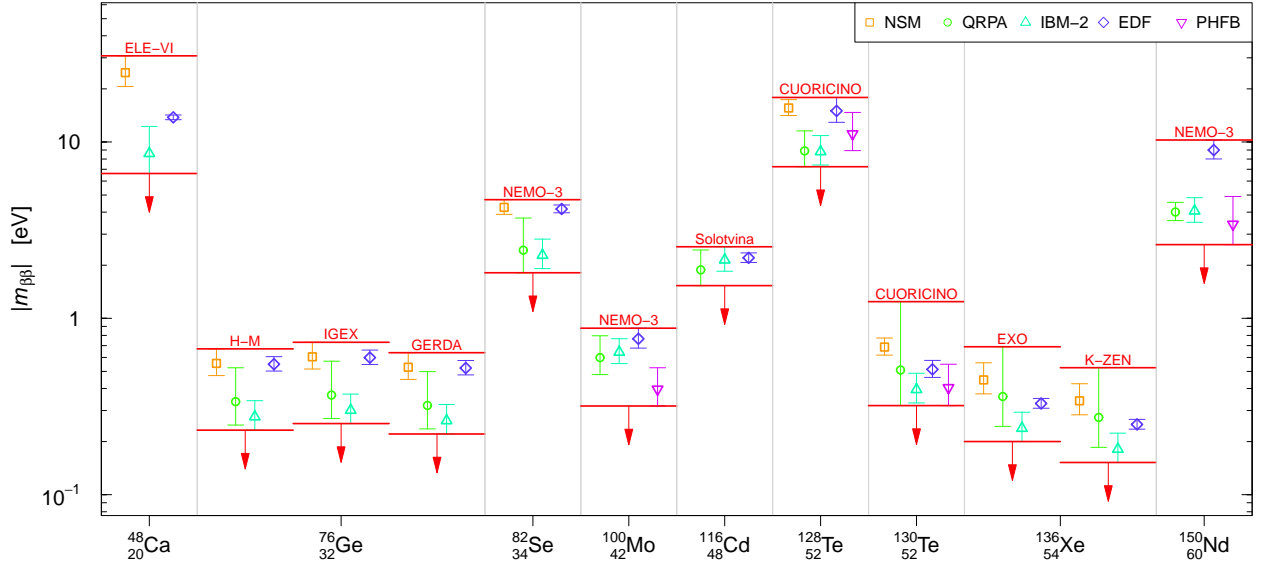}
\end{center}
\caption{\label{fig:exp}
Comparison of the
the limits on $|m_{\beta\beta}|$
obtained from the
90\% CL experimental lower limits on $T^{0\nu}_{1/2}$
in Fig.~\ref{fig:thalf}
taking into account the uncertainties of the theoretical
nuclear matrix element calculations.
}
\end{figure}

Table~\ref{tab:explim}
gives the most stringent 90\% CL experimental lower limits on the $\beta\beta_{0\nu}$
half-lives of several nuclei
established by recent experiments
and the corresponding upper bounds for $m_{\beta\beta}$ taking into account the NME theoretical uncertainties
in Tab.~\ref{tab:nme}.
One can see that the most stringent upper limits on $m_{\beta\beta}$ are between about
$0.2 \, \text{eV}$
and
$0.6 \, \text{eV}$,
depending on the NME theoretical uncertainties.
The same limits on $T^{0\nu}_{1/2}$ are illustrated in Fig.~\ref{fig:thalf},
where they are compared with the corresponding range of the
NSM, QRPA, IBM-2, EDF and PHFB
theoretical predictions
for $m_{\beta\beta} = 0.1 \, \text{eV}$.
Figure~\ref{fig:exp} illustrates the experimental
upper bounds for $m_{\beta\beta}$ obtained with the
NSM, QRPA, IBM-2, EDF and PHFB
calculation of the nuclear matrix elements.
One can see that the $\nGe{76}$ and $\nXe{136}$ experiments
give the strongest upper limit on $m_{\beta\beta}$,
which is close to about
$0.2 \, \text{eV}$
according to the IBM-2 and some QRPA calculations.
According to the NSM and EDF calculations,
for $\nGe{76}$ the upper limit on $m_{\beta\beta}$
is close to about
$0.6 \, \text{eV}$,
whereas for $\nXe{136}$ it is close to about
$0.3-0.4 \, \text{eV}$.

\begin{table}[b!]
\caption{\label{tab:explim}
Recent 90\% CL experimental lower limits on $T^{0\nu}_{1/2}$
and corresponding upper bounds for $m_{\beta\beta}$ taking into account the NME theoretical uncertainties
in Tab.~\ref{tab:nme}.
See Ref.~\cite{Tretyak:2002dx} for a complete list of previous experiments and results.
}
\centering
{
\begin{tabular}{llcc}
$\beta\beta^{-}$ decay	&experiment						&$T^{0\nu}_{1/2}$ [y]	&$m_{\beta\beta}$ [eV]	\\
\hline
$\nCa{48} \to\nTi{48} $	&ELEGANT-VI\cite{Ogawa:2004fy}				&$>1.4\times10^{22}$	&$<6.6-31$	\\
\hline
\multirow{3}{*}{
$\nGe{76} \to\nSe{76} $
}			&Heidelberg-Moscow\cite{Klapdor-Kleingrothaus:2001yx}	&$>1.9\times10^{25}$	&$<0.23-0.67$		\\
			&IGEX\cite{Aalseth:2002rf}				&$>1.6\times10^{25}$	&$<0.25-0.73$		\\
			&GERDA\cite{Agostini:2013mzu}				&$>2.1\times10^{25}$	&$<0.22-0.64$		\\
\hline
$\nSe{82} \to\nKr{82} $	&NEMO-3\cite{Arnold:2005rz}				&$>1.0\times10^{23}$	&$<1.8-4.7$	\\
\hline
$\nMo{100}\to\nRu{100}$	&NEMO-3\cite{Arnold:2013dha}				&$>2.1\times10^{25}$	&$<0.32-0.88$	\\
\hline
$\nCd{116}\to\nSn{116}$	&Solotvina\cite{Danevich:2003ef}			&$>1.7\times10^{23}$	&$<1.5-2.5$	\\
\hline
$\nTe{128}\to\nXe{128}$	&CUORICINO\cite{Arnaboldi:2002te}			&$>1.1\times10^{23}$	&$<7.2-18$	\\
\hline
$\nTe{130}\to\nXe{130}$	&CUORICINO\cite{Andreotti:2010vj}			&$>2.8\times10^{24}$	&$<0.32-1.2$	\\
\hline
\multirow{2}{*}{
$\nXe{136}\to\nBa{136}$
}			&EXO\cite{Albert:2014awa}				&$>1.1\times10^{25}$	&$<0.2-0.69$	\\
			&KamLAND-Zen\cite{Gando:2012zm}				&$>1.9\times10^{25}$	&$<0.15-0.52$	\\
\hline
$\nNd{150}\to\nSm{150}$	&NEMO-3\cite{Argyriades:2008pr}				&$>2.1\times10^{25}$	&$<2.6-10$	\\
\hline
\end{tabular}
}
\end{table}

\begin{figure}[t!]
\begin{center}
\includegraphics*[width=0.5\textwidth]{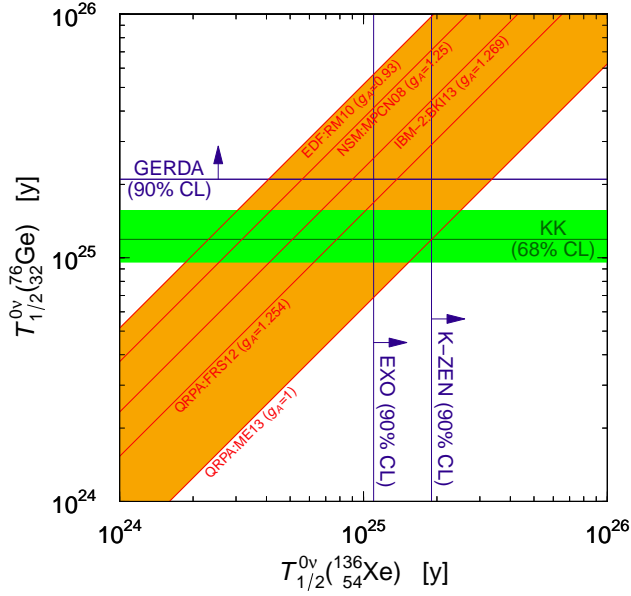}
\end{center}
\caption{\label{fig:xege}
Comparison of the alleged
\protect\cite{KlapdorKleingrothaus:2004wj}
observation of
$T^{0\nu}_{1/2}(\protect\nGe{76}) = 1.19 {}^{+0.37}_{-0.23} \times 10^{25} \, \text{y}$
(green 68\% CL band with the label ``KK'')
with the 90\% CL lower limit on $T^{0\nu}_{1/2}(\protect\nGe{76})$ of the GERDA experiment
\protect\cite{Agostini:2013mzu}
and with the
90\% CL lower limits on $T^{0\nu}_{1/2}(\protect\nXe{136})$ of the
EXO \protect\cite{Albert:2014awa}
and
KamLAND-Zen \protect\cite{Gando:2012zm}
experiments.
The diagonal orange band gives the correlation between
$T^{0\nu}_{1/2}(\protect\nGe{76})$
and
$T^{0\nu}_{1/2}(\protect\nXe{136})$
in the nuclear matrix element calculations discussed in Section~\ref{bb5}.
The cases of some specific calculations are shown with diagonal red lines.
}
\end{figure}

A question which has been debated in the literature in the last decade
concerns the validity of the claim of observation of
$\beta\beta_{0\nu}$ decay of $\nGe{76}$
presented in Refs.~\citen{KlapdorKleingrothaus:2004wj,KlapdorKleingrothaus:2006ff}
by part of the Heidelberg-Moscow collaboration
and its compatibility with other data
(see Refs.~\citen{Elliott:2004hr,Aalseth:2004hb,Strumia:2006db,Schwingenheuer:2012zs}).
Here we consider the result
$T^{0\nu}_{1/2}(\nGe{76}) = 1.19 {}^{+0.37}_{-0.23} \times 10^{25} \, \text{y}$
presented in Ref.~\citen{KlapdorKleingrothaus:2004wj},
because the larger value
($T^{0\nu}_{1/2}(\nGe{76}) = 2.23 {}^{+0.44}_{-0.31} \times 10^{25} \, \text{y}$)
presented in Ref.~\citen{KlapdorKleingrothaus:2006ff}
has been severely criticized in Ref.~\citen{Schwingenheuer:2012zs}.
Figure~\ref{fig:xege}
shows that the value
$T^{0\nu}_{1/2}(\nGe{76}) = 1.19 {}^{+0.37}_{-0.23} \times 10^{25} \, \text{y}$
is strongly disfavored by the recent results of the GERDA experiment
\cite{Agostini:2013mzu}.
Moreover,
Fig.~\ref{fig:xege}
shows that it is also disfavored by the results on $T^{0\nu}_{1/2}(\nXe{136})$ of the
EXO \cite{Albert:2014awa}
and
KamLAND-Zen \cite{Gando:2012zm}
experiments
with most of the nuclear matrix element calculations discussed in Section~\ref{bb5}.

Let us now briefly discuss the
general strategy of the design of future $\beta\beta_{0\nu}$-decay experiments.
The aim of an experiment is to observe
$\beta\beta_{0\nu}$-decay
by measuring a number of
events in the search energy window
which is larger than the expected background fluctuation
at a given statistical confidence level (CL).
Therefore,
in order to quantify and compare the performances of different $\beta\beta_{0\nu}$-decay
experiments it is useful to consider a detector sensitivity
$S_{0\nu}$,
which is defined as the process half-life
which generates a number of events which is equal to the expected background fluctuation
$\Delta{N}_{B}$
at $1\sigma$ (68.27\% CL).
In other words,
if the process half-life is larger than the sensitivity
$S_{0\nu}$
the $\beta\beta_{0\nu}$ signal can be hidden at $1\sigma$ by background fluctuations.
Experiments with larger $S_{0\nu}$
are sensitive to larger values of the half-life $T^{0\nu}_{1/2}$,
which correspond to smaller values of the effective Majorana mass $|m_{\beta\beta}|$.

In order to determine the expression of the sensitivity
$S_{0\nu}$
let us first note that in a detector with a number $N_{\text{nuc}}$ of nuclei which decay
with a rate $\Gamma^{0\nu}$,
the expected number of measured $\beta\beta_{0\nu}$ events in a time $T_{\text{obs}}$
is
\begin{equation}
N_{\beta\beta_{0\nu}}
=
\Gamma^{0\nu}
N_{\text{nuc}}
T_{\text{obs}}
\epsilon
=
\frac{\ln2}{T^{0\nu}_{1/2}}
\,
\frac{x \eta N_{A} M_{\text{det}}}{w_{\text{mol}}}
\,
T_{\text{obs}}
\epsilon
,
\label{N0nu}
\end{equation}
where
$\epsilon$ is the detection efficiency.
In the second equality we have taken into account that
the number of decaying nuclei $N_{\text{nuc}}$
in a detector made of molecules
containing atoms of type $\text{X}$
which have a $\beta\beta$-decaying isotope ${}^{A}\text{X}$
is given by the product of
the number $x$ of $\text{X}$ atoms in a molecule,
the isotopic abundance $\eta$ of ${}^{A}\text{X}$,
the Avogadro number $N_{A}$,
and
the detector mass $M_{\text{det}}$
divided by the molecular weight $w_{\text{mol}}$.

On the other hand,
since the background follows the Poisson statistics,
the expected $1\sigma$ fluctuation of the number $N_{B}$ of background events in the time $T_{\text{obs}}$ is given by
\begin{equation}
\Delta{N}_{B}
=
\sqrt{N_{B}}
=
\sqrt{R_{B} M_{\text{det}} T_{\text{obs}} \Delta_{E}}
,
\label{DNB}
\end{equation}
where
$\Delta_{E}$ is the search energy window
and
$R_{B}$ is the rate of background events per unit mass, time and energy.

Equating $N_{\beta\beta_{0\nu}}$ to $\Delta{N}_{B}$,
we obtain
\begin{equation}
S_{0\nu}
=
T^{0\nu}_{1/2}(N_{\beta\beta_{0\nu}}=\Delta{N}_{B})
=
\ln2
\,
\frac{x \eta N_{A} \epsilon}{w_{\text{mol}}}
\,
\sqrt{\frac{M_{\text{det}} T_{\text{obs}}}{R_{B} \Delta_{E}}}
.
\label{S0nu}
\end{equation}
This expression of $S_{0\nu}$ is useful to understand which quantities are important for
reaching a high sensitivity.
It is clear that an improvement of the enrichment of the material with the $\beta\beta$-decaying isotope,
which increases $\eta$,
is important because $S_{0\nu}$ is linear in $\eta$,
but its effect is limited, because $\eta \leq 1$.
The same applies to the efficiency $\epsilon$.
Unfortunately the quantities which can be improved without an obvious limit
are under the square root,
so that an improvement of $S_{0\nu}$ by a factor of two requires
a detector which is four times more massive, etc.

Note that the above definition of the sensitivity $S_{0\nu}$ applies to the experiments with a background.
However some future experiments are planned in order to have a ``zero background'' (ZB),
i.e. an expected number of background events
which is less than about one in all the life of the experiment.
In this case,
the sensitivity can be defined as the highest value of the half-life which is excluded
at $1\sigma$ if no signal event is measured.
Hence,
$\Delta{N}_{B}$ is replaced by the average number of signal events which can be excluded
if no event is measured,
which is of the order of one,
leading to
\begin{equation}
S_{0\nu}^{\text{ZB}}
=
T^{0\nu}_{1/2}(N_{\beta\beta_{0\nu}}=1)
=
\ln2
\,
\frac{x \eta N_{A} \epsilon}{w_{\text{mol}}}
\,
M_{\text{det}}
T_{\text{obs}}
.
\label{S0nuZB}
\end{equation}
It is clear that this situation is advantageous for experiments which can be
constructed with large detector masses
and
which can take data for long times,
since $S_{0\nu}^{\text{ZB}}$ is proportional to
$M_{\text{det}}$
and
$T_{\text{obs}}$.
On the other hand,
the sensitivity of these experiments cannot be improved by
improving the energy resolution
(which reduces the energy window $\Delta_{E}$)
and reducing the background
(which is already ``zero'')
as the experiments with background for which Eq.~(\ref{S0nu}) applies.

Hence,
the planning strategy of new $\beta\beta_{0\nu}$ experiments
(see the reviews in Refs.~\citen{GomezCadenas:2011it,Schwingenheuer:2012zs,Giuliani:2012zu,Cremonesi:2013vla})
is quite different in the case of experiments with background,
which can reach high sensitivities by combining
a large detector mass,
a long detection time,
a low background and
a high energy resolution,
but the improvement increases only as the square root,
and the experiments with zero background,
which can reach high sensitivities only with
a high detector mass and a long detection time,
but with a faster linear improvement
(see the discussion in Ref.~\citen{Biassoni:2013ywa}).
When it is possible to reach a zero-background level,
this approach is convenient for detectors made of substances which are not too expensive.
\section{Conclusions}
\label{bb8}

Neutrinoless double-$\beta$ decay
is the most promising process which could allow near-future experiments
to reveal the Majorana nature of massive neutrinos which is expected from
the physics beyond the Standard Model
(see Section~\ref{bb3}).

The analysis of the data of neutrino oscillation experiments
have determined with accuracies between about 3\% and 11\%
(see Tab.~\ref{tab:global})
the values of the neutrino oscillation parameters in the framework of three-neutrino mixing:
the two solar and atmospheric mass differences,
$\Delta{m}^2_{\text{S}}$
and
$\Delta{m}^2_{\text{A}}$,
and the three mixing angles
$\vartheta_{12}$,
$\vartheta_{23}$,
$\vartheta_{13}$.
The experiments on the study of neutrino oscillations enter now
a new stage of high-precision measurement of the
neutrino oscillation parameters,
of determination of the neutrino mass ordering
(normal or inverted),
and
of measurement of the CP-violating phase $\delta$,
which is the last unknown parameter of the three-neutrino mixing matrix in the case of Dirac neutrinos.
If massive neutrinos are Majorana particles,
there are two additional ``Majorana phases''
that are observable in processes which violate the total lepton number,
as neutrinoless double-$\beta$ decay.
In fact,
the effective Majorana neutrino mass
$m_{\beta\beta}$
in $\beta\beta_{0\nu}$ decay
depends on the Majorana phases,
which can cause cancellations among the
contributions of the three neutrino masses
(see Section~\ref{bb6}).
Hence, even assuming a value for the absolute scale of neutrino masses,
there are large uncertainties on the
possible value of $m_{\beta\beta}$,
which depend on the normal or inverted character of the neutrino mass spectrum
(see Figs.~\ref{fig:mbb}, \ref{fig:flm} and \ref{fig:mbt}).

Up to now no $\beta\beta_{0\nu}$ decay has been observed,
with the exception of the controversial claim
in Ref.~\citen{KlapdorKleingrothaus:2004wj}
of observation of $\beta\beta_{0\nu}$ decay of $\nGe{76}$,
which is currently strongly disfavored by
the direct lower limit on $T^{0\nu}_{1/2}(\nGe{76})$ of the GERDA experiment
\cite{Agostini:2013mzu}
and indirectly by the lower limits on
$T^{0\nu}_{1/2}(\nXe{136})$ of the
EXO \cite{Albert:2014awa}
and
KamLAND-Zen \cite{Gando:2012zm}
experiments
(see Fig.~\ref{fig:xege}).

The current experimental upper bound on $m_{\beta\beta}$
is between about 0.2 and 0.6 eV
(see Tab.~\ref{tab:explim} and Fig.~\ref{fig:exp}),
depending on the uncertainties in the theoretical calculations
of the nuclear matrix elements which determine the rates of
nuclear $\beta\beta_{0\nu}$ decays
(see Section~\ref{bb4}).

The problem of the calculation of the nuclear matrix elements
(see Section~\ref{bb5})
is open and so far unresolved.
There are several methods of calculation whose results differ
by a factor between about two and four for different nuclei
(see Tab.~\ref{tab:nme}).
It is clear that this situation is deeply unsatisfactory
and it requires an effort of clarification of the
nuclear physics community.
A promising improvement could come from tests of the nuclear matrix elements
with heavy-ion double charge-exchange reactions
(NUMEN project\cite{Agodi-NOW-2014})
which have a nuclear transition operator with the same spin-isospin mathematical structure
of that of $\beta\beta_{0\nu}$ decay
and proceed through the same excited levels of the intermediate nucleus.
It is interesting that also the quenching of $g_{A}$ is expected to be similar,
because both processes happen in the same nuclear medium.
Information on the NMEs can be obtained also in
charged-current neutrino-nucleus interactions
\cite{Volpe:2005iy}.

The next generation of $\beta\beta_{0\nu}$ decay experiments
is aimed at the exploration of values of $m_{\beta\beta}$
below 0.1 eV,
with the purpose of reaching the inverted hierarchy interval between about
0.02 and 0.05 eV.
This is a very important goal,
which will allow either to discover $\beta\beta_{0\nu}$ decay
or to exclude the inverted hierarchy.

In the first case, we will know that massive neutrinos are Majorana particles
and that there must be physics beyond the Standard Model which explains
the small Majorana masses (see Section~\ref{bb3}).
However,
it is not possible to know
from the measurement of $\beta\beta_{0\nu}$ decay alone
if the scheme of neutrino masses is inverted or normal
and it may be difficult to do it also with independent measurements
of the sum of the neutrino masses (see Fig.~\ref{fig:flm})
and
of the effective electron neutrino mass in $\beta$ decay (see Fig.~\ref{fig:mbt}).

On the other hand,
in the second case the lack of observation of $\beta\beta_{0\nu}$ decay
with $m_{\beta\beta} \gtrsim 0.2 \, \text{eV}$
will exclude the inverted hierarchy,
leaving only the normal hierarchy as a viable possibility.
In this case, the Majorana nature of massive neutrinos will be still uncertain
and it may be possible that the cancellations among the massive neutrino contributions
make $m_{\beta\beta}$ so small that it will be very difficult to measure it
in a foreseeable future.

Let us however remark that the implications of these two possibilities
for the scheme of the masses of the three light neutrinos would be radically changed
if future experiments will confirm the current indications
in favor of the existence of a sterile neutrino at the eV scale
(see Refs.~\citen{Goswami:2005ng,Goswami:2007kv,Barry:2011wb,Li:2011ss,Rodejohann:2012xd,Giunti:2012tn,Girardi:2013zra,Pascoli:2013fiz,Meroni:2014tba,Abada:2014nwa,Giunti-NNN-2014}).

Let us finally note that although in this review we considered only the neutrinoless double-$\beta^{-}$ decay
process in Eq.~(\ref{bb0-}),
if the total lepton number $L$ is violated some nuclei can undergo the neutrinoless double-$\beta^{+}$ decay
process
\begin{equation}
\beta\beta_{0\nu}^{+}:
\quad
{}^{A}_{Z}\text{X} \to {}^{\phantom{-2}A}_{Z-2}\text{X} + 2 e^{+}
,
\label{bb0+}
\end{equation}
which has the same expression (\ref{totrate}) for the decay rate
and can be used to measure the effective Majorana mass $|m_{\beta\beta}|$.
However there are only six nuclei with low natural isotopic abundances which can undergo $\beta\beta_{2\nu}^{+}$
(${}^{A}_{Z}\text{X} \to {}^{\phantom{-2}A}_{Z-2}\text{X} + 2 e^{+} + 2 \nu_{e}$)
and
$\beta\beta_{0\nu}^{+}$ decays and the corresponding decay rates
are suppressed by small phase space and unfavorable Fermi function
(see Ref.~\citen{Frekers:2005ze}).
If the total lepton number $L$ is violated,
also the following
neutrinoless positron-emitting electron capture
and
neutrinoless double electron capture
processes
\cite{Winter:1955zz,Eramzhian:1982bi}
obtained by crossing symmetry from Eq.~(\ref{bb0+})
are possible:
\begin{align}
\text{EC}\beta_{0\nu}:
\quad
\null & \null
e^{-} + {}^{A}_{Z}\text{X} \to {}^{\phantom{-2}A}_{Z-2}\text{X} + e^{+}
,
\label{ECbb0nu}
\\
\text{EC}\text{EC}_{0\nu}:
\quad
\null & \null
e^{-} + e^{-} + {}^{A}_{Z}\text{X} \to {}^{\phantom{-2}A}_{Z-2}\text{X}
.
\label{ECEC0nu}
\end{align}
It has been shown in Ref.~\citen{Bernabeu:1983yb} that the
$\text{EC}\text{EC}_{0\nu}$
process (\ref{ECEC0nu})
can occur through the capture of two atomic electrons
in a resonance transition of the initial atom
into an excited nuclear and/or atomic state of the final atom.
A resonant
$\text{EC}\text{EC}_{0\nu}$
transition takes place if the masses of the initial and final atoms are equal
(for recent discussions of this process see Refs.~\citen{Sujkowski:2003mb,Frekers:2005ze,Lukaszuk:2006ua,Krivoruchenko:2010ng}).
Experimental searches found lower bounds for the resonant
$\text{EC}\text{EC}_{0\nu}$
half-life of several atoms
($\nSe{74}$ \cite{Barabash:2006qx},
$\nRu{96}$ \cite{Belli:2009zz},
$\nCd{106}$ \cite{Rukhadze:2010zz},
$\nSn{112}$ \cite{Barabash:2008wj,Dawson:2007re,Dawson:2008kj,Kidd:2008zz,Barabash:2009ja,Barabash:2011zza},
$\nCe{136}$ \cite{Belli:2009zza};
see also the review in Ref.~\citen{Barabash:2011fg}).

\section*{Acknowledgments}

The work of S.M.B. is supported by the Alexander von Humboldt Stiftung, Bonn, Germany (contract Nr. 3.3-3-RUS/1002388),
by RFBR Grant N 13-02-01442 and by the Physics Department E15 of the Technical University Munich.
The work of C.G. is supported by the research grant {\sl Theoretical Astroparticle Physics} number 2012CPPYP7 under the program PRIN 2012 funded by the Ministero dell'Istruzione, Universit\`a e della Ricerca (MIUR).

\end{document}